\documentclass[twocolumn,showpacs,superscriptaddress,amsmath,amssymb,prc]{revtex4-1}
\usepackage{graphicx}% Include figure files
\usepackage{dcolumn}% Align table columns on decimal point
\usepackage{bm}% bold math
\usepackage{ifpdf}
\usepackage{epstopdf}
\usepackage{slashed}
\usepackage{amsfonts}
\usepackage{mathrsfs}
\usepackage{amssymb}
\usepackage{multirow}
\usepackage{xcolor}

\begin{document}
%\begin{CJK*}{UTF8}{song}

%\preprint{APS/123-QED}

\title{Quantitative analysis of tensor effects in the relativistic Hartree-Fock theory}

\author{Zhiheng Wang}
\affiliation{School of Nuclear Science and Technology, Lanzhou University, Lanzhou 730000, China}
\affiliation{Joint department for nuclear physics, Lanzhou University and Institute of Modern Physics,
Chinese Academy of Sciences, Lanzhou 730000, China}
\affiliation{Faculty of Pure and Applied Sciences, University of Tsukuba, Tsukuba 305-8571, Japan}
\affiliation{RIKEN Nishina Center, Wako 351-0198, Japan}

\author{Qiang Zhao}
\affiliation{School of Nuclear Science and Technology, Lanzhou University, Lanzhou 730000, China}
\affiliation{Joint department for nuclear physics, Lanzhou University and Institute of Modern Physics,
Chinese Academy of Sciences, Lanzhou 730000, China}
\affiliation{Institut de Physique Nucl\'{e}aire de Lyon, Universit\'{e} Claude Bernard Lyon 1, F-69622 Villeurbanne Cedex, France}

\author{Haozhao Liang}%
\email{haozhao.liang@riken.jp}
\affiliation{RIKEN Nishina Center, Wako 351-0198, Japan}
\affiliation{Department of Physics, Graduate School of Science, The University of Tokyo,
  Tokyo 113-0033, Japan}

\author{Wen Hui Long}
\email{longwh@lzu.edu.cn}
\affiliation{School of Nuclear Science and Technology, Lanzhou University, Lanzhou 730000, China}
\affiliation{Joint department for nuclear physics, Lanzhou University and Institute of Modern Physics,
Chinese Academy of Sciences, Lanzhou 730000, China}

\date{\today}% It is always \today, today,

\begin{abstract}
Tensor force is identified in each meson-nucleon coupling in the relativistic Hartree-Fock theory.
It is found that all the meson-nucleon couplings, except the $\sigma$-scalar one, give rise to the tensor force.
The effects of tensor force on various nuclear properties can now be investigated quantitatively, which allows fair and direct comparisons with the corresponding results in the non-relativistic framework.
The tensor effects on nuclear binding energies and the evolutions of the $Z,\,N = 8,\,20$, and $28$ magic gaps are studied.
The tensor contributions to the binding energies are shown to be tiny in general.
The $Z,\,N = 8$ and $20$ gaps are sensitive to the tensor force, but the $Z,\,N = 28$ gaps are not.
\end{abstract}


\maketitle
%\end{CJK*}

\section{Introduction}

The tensor force is one of the most important components of the nucleon-nucleon interaction \cite{YUKAWA1935Proc.Phys.Math.Soc.Japan17.48, Fayache1997Phys.Rep.290.201, Sagawa2014Prog.Part.Nucl.Phys.76.76}.
At early stage of nuclear physics, the tensor force was recognized to be responsible for the deuteron binding \cite{Rarita1941Phys.Rev.59.436} and electric quadrupole moment \cite{Gerjuoy1942Phys.Rev.61.138}.
With the advance of radioactive-ion-beam facilities around the world, much progress has been made in the study of the structure of exotic nuclei. From the $\beta$-stability valley towards the drip lines, the shell evolution, particularly the disappearance of the traditional magic numbers and the emergence of new ones, is of great interest \cite{Sorlin2008Prog.Part.Nucl.Phys.61.602, Wienholtz2013Nature498.346, Steppenbeck2013Nature502.207}.
It has been pointed out by Otsuka \textit{et al.} \cite{Otsuka2001Phys.Rev.Lett.87.082502, Otsuka2005Phys.Rev.Lett.95.232502, Otsuka2006Phys.Rev.Lett.97.162501, Otsuka2010Phys.Rev.Lett.104.012501} in the scheme of nuclear shell model that the tensor force plays a critical role in the shell evolution in exotic nuclei.

Among the state-of-the-art nuclear methodologies, the nuclear density functional theory (DFT) \cite{Bender2003Rev.Mod.Phys.75.121, Meng2006Prog.Part.Nucl.Phys.57.470, Nakatsukasa2016Rev.Mod.Phys.88.045004, Xia2018At.DataNucl.DataTables121.1} is the only approach that can cover almost the whole nuclear chart, in particular, the exotic nuclei, now and in the near future.
The first study of the role of tensor force in the shell evolution can be traced back 40 years ago \cite{Stancu1977Phys.Lett.B68.108} in the Hartree-Fock (HF) theory using the Skyrme \cite{Skyrme1958NuclearPhys.9.615} interaction.
However, in that study, minor improvements or even, in some cases, some deteriorations were predicted in the description of single-particle energies and spin-orbit splittings.
Actually, for decades, the tensor force had been neglected in the Skyrme HF theory.
In the non-relativistic Gogny HF theory, the tensor force is also not included in the widely used versions \cite{Decharge1980Phys.Rev.C21.1568, Berger1991Comput.Phys.Commun.63.365}.
The same applies to the relativistic framework.
In the widely used relativistic mean-field (RMF) theory \cite{Walecka1974Ann.Phys.83.491, RING1996PROG.PART.NUCL.PHYS.37.193, NIKSIC2011PROG.PART.NUCL.PHYS.66.519, VRETENAR2005PHYS.REP.409.101, meng2016}, the tensor force is not included because only the Hartree terms are taken into account.
Note that in the scheme of DFT, the tensor effects refer to those effects of the tensor force acting on the system wave function as a single Slater determinant.
The higher-order effects of tensor force \cite{Kamada2001Phys.Rev.C64.044001}, e.g., the two-particle-two-hole effect \cite{Myo2005Prog.Theor.Phys.113.763, Myo2007Phys.Rev.C76.024305, Myo2017Phys.Lett.B769.213, Myo2017Phys.Rev.C95.044314, Myo2017Prog.Theor.Exp.Phys.2017.111D01}, are supposed to be implicitly absorbed in the effective interactions.
%In contrast, such two-particle-two-hole tensor effects are studied explicitly in the \textit{ab initio} calculations \cite{Kamada2001Phys.Rev.C64.044001}, the tensor-optimized shell model \cite{Myo2005Prog.Theor.Phys.113.763, Myo2007Phys.Rev.C76.024305}, the tensor-optimized antisymmetrized molecular dynamics \cite{Myo2017Phys.Lett.B769.213, Myo2017Phys.Rev.C95.044314}, the high-momentum antisymmetrized molecular dynamics \cite{Myo2017Prog.Theor.Exp.Phys.2017.111D01}, and so on.

Such a situation of the study of tensor force has been dramatically changed since the experimental data on the shell evolution of nuclei far from the stability line, such as the energy differences between the $1h_{11/2}$ and $1g_{7/2}$ single-proton states along the $Z = 50$ isotopes, the energy differences between the $1i_{13/2}$ and $1h_{9/2}$ single-neutron states along the $N = 82$ isotones \cite{Schiffer2004Phys.Rev.Lett.92.162501}, and the energy differences between the $2s_{1/2}$ and $1d_{5/2}$ single-neutron states along the $Z=20$ isotopes \cite{Cottle1998Phys.Rev.C58.3761}.

This bloomed a series of works focused on the tensor effects on the shell evolution in both the non-relativistic \cite{Otsuka2006Phys.Rev.Lett.97.162501, Brown2006Phys.Rev.C74.061303, Brink2007Phys.Rev.C75.064311, Colo2007Phys.Lett.B646.227,  Sugimoto2007Phys.Rev.C76.054310,  Zou2008Phys.Rev.C77.014314, Anguiano2012Phys.Rev.C86.054302, Nakada2013Phys.Rev.C87.014336} and relativistic \cite{Long2007Phys.Rev.C76.034314, Long2008Europhys.Lett.82.12001, Wang2013Phys.Rev.C87.047301, Marcos2014Phys.At.Nucl.77.299, Lopez-Quelle2018Nucl.Phys.A971.149} DFT.
Readers are referred to Ref.~\cite{Sagawa2014Prog.Part.Nucl.Phys.76.76} for a recent review.
In particular, the comparisons between the relativistic and non-relativistic frameworks were carried out in Ref.~\cite{Tarpanov2008Phys.Rev.C77.054316} for the proton $1d$ spin-orbit splitting and neutron $2p_{3/2}$-$1f_{7/2}$ gap, and in Ref.~\cite{Moreno-Torres2010Phys.Rev.C81.064327} for the $Z,\,N = 8,\,20$, and $28$ magic gaps.
Agreements between the relativistic and non-relativistic results were found in a qualitative way.
However, quantitative analysis of tensor effects in the relativistic framework was still missing \cite{Tarpanov2008Phys.Rev.C77.054316, Moreno-Torres2010Phys.Rev.C81.064327}.

In the non-relativistic framework, the tensor force is included explicitly and its strengths are fitted basically in two ways.
One is to add the tensor force onto a given existing effective interaction perturbatively, and adjust only the tensor strengths so as to reproduce at best the shell evolution along the isotopic or isotonic chains \cite{Brown2006Phys.Rev.C74.061303, Colo2007Phys.Lett.B646.227}.
Another is to fit the strengths of the tensor force, e.g., $\alpha_T$ and $\beta_T$ in Skyrme \cite{Lesinski2007Phys.Rev.C76.014312}, $f_G$ in Gogny \cite{Otsuka2006Phys.Rev.Lett.97.162501}, on an equal footing with the other components of the effective interaction.
In both cases, the tensor force is isolated from the other components, and thus its effects can be identified clearly.

In the relativistic framework, first of all, to include the tensor force, the Fock terms must be taken into account.
This is the relativistic Hartree-Fock (RHF) theory \cite{Bouyssy1987Phys.Rev.C36.380}.
By adopting the density-dependent meson-nucleon couplings, the RHF theory \cite{Long2006Phys.Lett.B640.150, Long2007Phys.Rev.C76.034314, Long2005PhD.Thesis} achieved for the first time the quantitative description of the ground-state properties of many nuclear systems on the same level as RMF.
It has been also shown that the Fock terms play very important roles in the nucleon effective mass splitting \cite{Long2006Phys.Lett.B640.150}, symmetry energies \cite{Sun2008Phys.Rev.C78.065805, Zhao2015J.Phys.G:Nucl.Part.Phys.42.095101, Liu2018Phys.Rev.C97.025801}, pseudospin and spin symmetries \cite{LONG2006Phys.Lett.B639.242, Liang2010Eur.Phys.J.A44.119, Liang2015Phys.Rep.570.1}, halo and bubble-like structures \cite{Long2010Phys.Rev.C81.031302, Li2016Phys.Rev.C93.054312}, deformation \cite{Ebran2011Phys.Rev.C83.064323}, superheavy elements \cite{LI2014Phys.Lett.B732.169}, new magic numbers \cite{Li2016Phys.Lett.B753.97, Li2018arXiv:1807.10000v1}, Coulomb effects and isospin-symmetry breaking \cite{Liang2009Phys.Rev.C79.064316, Gu2013Phys.Rev.C87.041301R}, spin-isospin resonances \cite{Liang2008Phys.Rev.Lett.101.122502, Liang2012Phys.Rev.C85.064302, Liang2012Phys.Rev.C86.021302R, Niu2017Phys.Rev.C95.044301}, $\beta$-decay half-lives \cite{NIU2013Phys.Lett.B723.172}, and the properties of neutron stars \cite{Long2012Phys.Rev.C85.025806, Li2018Phys.Lett.B783.234, Li2018Eur.Phys.J.A54.133}. It is, however, not straightforward to identify the tensor effects in the RHF theory, because the tensor force is mixed together with other components, such as the central and spin-orbit ones. For example, simply excluding the pion-nucleon coupling, which is known as the most important carrier of the tensor force, leads to substantial changes also in the central part of the mean field.

Within the RHF theory, there have been several attempts to identify the tensor force and evaluate its effects on the shell evolution.
In Refs.~\cite{Long2008Europhys.Lett.82.12001, Wang2013Phys.Rev.C87.047301}, it is found that the tensor components of nuclear interaction arising from the $\pi$-pseudo-vector ($\pi$-PV) and $\rho$-tensor ($\rho$-T) couplings play an essential role in the self-consistent description of the relevant shell evolutions. Moreover, both the $\pi$-PV and $\rho$-T couplings are found to be essential in triggering the new magicity $N=32$ in $^{52}$Ca \cite{Li2016Phys.Lett.B753.97}. It is also recognized that the interaction matrix elements from the Fock terms and their contributions to the spin-orbit splittings show characteristic spin dependence, and such a spin-dependent feature can be extracted almost completely by the proposed relativistic tensor formalism, see Fig.~2 in Ref.~\cite{Jiang2015Phys.Rev.C91.034326}. In particular, more distinct effects were found in the isoscalar channels, namely the $\sigma$-scalar ($\sigma$-S) and $\omega$-vector ($\omega$-V) couplings, rather than the isovector ones ($\pi$-PV and $\rho$-T) \cite{Jiang2015Phys.Rev.C91.034326, Jiang2015Phys.Rev.C91.025802, Zong2018Chin.Phys.C42.024101}, since this spin-dependent feature originates not only from the tensor force, but also from the exchange parts of the central force \cite{Colo2007Phys.Lett.B646.227}.
Nevertheless, with these attempts, a fair and direct comparison between the relativistic and non-relativistic schemes  about the tensor force and its effects remains an open question.

In this work, we will perform the non-relativistic reduction for the relativistic two-body interactions in the RHF theory, and obtain the corresponding non-relativistic reduced operators. These operators, which are expanded in a systematic way in the powers of $1/M$, are nothing but the central, spin-orbit, and tensor forces, etc., in the conventional non-relativistic sense. We can then evaluate the tensor effects on various nuclear properties in a quantitative way, and eventually compare with the non-relativistic results.

This paper is organized as follows. In Sec.~\ref{Sec:Theory}, the RHF theory is briefly introduced, and the identification of the tensor forces in RHF is shown with the formalism of non-relativistic reduction. More details are in Appendixes~\ref{App:Reduction}, \ref{App:Reduction*}, and \ref{APP:Matrix_tensor}. The sum rule of the two-body matrix elements of tensor force is verified, and the tensor effects on nuclear binding energies and shell evolutions are studied in Sec.~\ref{Sec:results}. Summary and perspectives are given in Sec.~\ref{Sec:summary}.

\section{Theoretical Framework}\label{Sec:Theory}

\subsection{Relativistic Hartree-Fock theory}

In the relativistic framework, the nucleon-nucleon interaction is mediated by the exchange of mesons \cite{meng2016}.
The starting point of the RHF theory is the effective Lagrangian density $\mathscr L$.
It is constructed with the degrees of freedom associated with the nucleon field $\psi$, two isoscalar meson fields $\sigma$ and $\omega$, two isovector meson fields $\pi$ and $\rho$, and the photon field $A$.
It is composed of the free parts of the nucleon, meson, and photon fields as well as the interaction parts between nucleons and mesons (photons) \cite{Bouyssy1987Phys.Rev.C36.380, Long2006Phys.Lett.B640.150, Long2007Phys.Rev.C76.034314},
\begin{equation}\label{Lagrangian}
  \mathscr L =\, \mathscr L_0 + \mathscr L_I,
\end{equation}
where
\begin{subequations}
\begin{align}
%\mathscr L=& \mathscr L_0 + \mathscr L_I,\\
\mathscr L_0=\,&\bar\psi\left(i\gamma_\mu\partial^\mu - M\right)\psi\nonumber\\
&+ \frac{1}{2}\partial_\mu\sigma\partial^\mu\sigma - \frac{1}{2} m_\sigma^2\sigma^2 +\frac{1}{2} m_\omega^2 \omega_\mu\omega^\mu - \frac{1}{4} \Omega_{\mu\nu}\Omega^{\mu\nu}\nonumber\\
&+ \frac{1}{2} m_\rho^2\vec \rho_\mu\cdot\vec\rho^\mu -\frac{1}{4}\vec R_{\mu\nu}\cdot\vec R^{\mu\nu} + \frac{1}{2}\partial_\mu\vec\pi\cdot\partial^\mu\vec\pi \nonumber\\
&-\frac{1}{4} F_{\mu\nu}F^{\mu\nu},\label{Lfree}\\
\mathscr L_I =\,& -\bar\psi\left[g_\sigma\sigma + g_\omega\gamma^\mu\omega_\mu + g_\rho \gamma^\mu\vec\tau\cdot\vec\rho_\mu - \frac{f_\rho}{2M}\sigma^{\mu\nu} \vec\tau\cdot\partial_\nu\vec\rho_\mu\right.\nonumber\\
&\left.\qquad\quad + \frac{f_\pi}{m_\pi}\gamma_5\gamma^\mu\vec\tau\cdot\partial_\mu\vec\pi
+ e\gamma^\mu\frac{1-\tau_3}{2} A_\mu\right]\psi, \label{LI}
\end{align}
\end{subequations}
with the nucleon mass $M$, the meson masses $m$, the meson-nucleon coupling strengths $g$ and $f$, the field tensors
\begin{subequations}
\begin{align}
\Omega^{\mu\nu} & \equiv \,\partial^\mu\omega^\nu - \partial^\nu\omega^\mu, \\
\vec R^{\mu\nu} & \equiv \,\partial^\mu\vec\rho^\nu - \partial^\nu\vec\rho^\mu, \\
F^{\mu\nu} & \equiv \,\partial^\mu A^\nu - \partial^\nu A^\mu,
\end{align}
\end{subequations}
and $\sigma^{\mu\nu}=\frac{i}{2}[\gamma^\mu, \gamma^\nu]$.
In this paper, the isovectors are denoted by arrows and the space vectors are in bold type.

The Hamiltonian can be derived through the Legendre transformation and further presented with the nucleon degree of freedom as
\begin{align}\label{Hamil}
    H=&\,\int d^3x\,\bar{\psi}(x)[-i\bm\gamma\cdot\bm{\nabla}+M]\psi(x)\nonumber\\
    &\,+\frac{1}{2}\sum_{\phi}\iint d^3x\, d^4y\,
    \bar{\psi}(x)\bar{\psi}(y)
    \Gamma_\phi(x,y)D_\phi(x,y)\nonumber\\
    &\qquad\qquad\qquad\qquad\times\psi(y)\psi(x),
\end{align}
where $\phi$ denotes the meson-nucleon couplings, i.e., the Lorentz $\sigma$-scalar ($\sigma$-S), $\omega$-vector ($\omega$-V), $\rho$-vector ($\rho$-V), $\rho$-tensor ($\rho$-T), $\rho$-vector-tensor ($\rho$-VT), and $\pi$-pseudovector ($\pi$-PV) couplings, as well as the photon-vector ($A$-V) coupling.
To make no confusion, the capital letter ``T'' here means the Lorentz tensor coupling.
In contrast, the small letter ``t'' will be used later to denote the word ``tensor'' in the relevant contexts of tensor force.
The interaction vertices $\Gamma_\phi(x,y)$ in the Hamiltonian~\eqref{Hamil} read
\begin{subequations}\label{eq:vertex}
\begin{align}
\Gamma_{\sigma\text{-S}} =&\, -[g_\sigma]_x[g_\sigma]_y, \\
\Gamma_{\omega\text{-V}} =&\, +\left[ g_\omega \gamma_\mu\right]_x \left[g_\omega\gamma^\mu\right]_y, \\
\Gamma_{\rho\text{-V}} =&\, +\left[g_\rho\gamma_\mu\vec\tau \right]_x \cdot \left[ g_\rho\gamma^\mu\vec\tau\right]_y, \\
\Gamma_{\rho\text{-T}} =&\, +\left[\frac{f_\rho}{2M} \sigma_{\mu\nu}\vec\tau \partial^\nu\right]_x\cdot \left[\frac{f_\rho}{2M} \sigma^{\mu\lambda}\vec\tau \partial_\lambda\right]_y, \label{rhot}\\
\Gamma_{\rho\text{-VT}} =&\, +\left[\frac{f_\rho}{2M}\sigma_{\mu\nu}\vec\tau\partial^\mu\right]_x\cdot\left[g_\rho \gamma^\nu\vec\tau\right]_y + (x \leftrightarrow y), \label{rhvt}\\
\Gamma_{\pi\text{-PV}} =&\, -\left[\frac{f_\pi}{m_\pi}\vec\tau\gamma_5\gamma_\mu\partial^\mu\right]_x \cdot \left[\frac{f_\pi}{m_\pi}\vec\tau\gamma_5\gamma_\nu\partial^\nu \right]_y, \label{pi-sv}\\
\Gamma_{A\text{-V}} =&\, +  \left[e\gamma_\mu\frac{1-\tau_3}2\right]_x \left[e\gamma^\mu\frac{1-\tau_3}2\right]_y.
\end{align}
\end{subequations}
The propagators $D_\phi(x,y)$ read
\begin{align}\label{pro}
  D_\phi(x,y)=\,-\int \frac{d^4k}{(2\pi)^4}e^{-ik(x-y)}\frac{1}{k^2-m_\phi^2}.\\\nonumber
\end{align}
When the retardation effect is neglected \cite{Bouyssy1987Phys.Rev.C36.380}, the meson and photon propagators become the standard Yukawa and Coulomb forms,
\begin{align}
D_\phi =\, \frac{1}{4\pi} \frac{e^{-m_\phi\left|\bm r_1 -\bm r_2\right|}}{\left|\bm r_1 -\bm r_2\right|},\qquad
D_{A\text{-V}} =\, \frac{1}{4\pi} \frac{1}{\left|\bm r_1 -\bm r_2\right|},
\end{align}
respectively.
Hereafter, we use $\bm{r}_1$ and $\bm{r}_2$ to denote the space coordinates at vertices $x$ and $y$, and the indices ``1'' and ``2'' are always used to denote the vertices.

The nucleon-field operators $\psi(x)$ and $\psi^\dagger(x)$ can be expanded on the set of creation and annihilation operators defined by a complete set of Dirac spinors $\{\varphi_\alpha(\bm{x})\}$,
\begin{subequations}\label{psi}
\begin{align}
  \psi(x)&=\,\sum_\alpha \varphi_\alpha(\bm{x})e^{-i\varepsilon_\alpha t}c_\alpha,\\
  \psi^\dag(x)&=\,\sum_\alpha \varphi_\alpha^\dag(\bm{x}) e^{i\varepsilon_\alpha t}c^\dag_\alpha,
\end{align}
\end{subequations}
where $c_\alpha$ and $c_\alpha^\dag$ represent the annihilation and creation operators for the nucleon in state $|\alpha\rangle$ with the single-particle energy $\varepsilon_\alpha$.
The trial Hartree-Fock ground-state wave function of a nucleus with $A$ particles is constructed as
\begin{align}\label{HF}
   |\Phi_0\rangle=\,\prod_{\alpha}^A c_\alpha^\dag |0\rangle.
\end{align}
The no-sea approximation \cite{Walecka1974Ann.Phys.83.491} indicates that the index $\alpha$ runs over only the occupied states in the Fermi sea.

The expectation energy of the Hamiltonian~\eqref{Hamil} on  the trial ground state, excluding the rest mass, can be derived as
\begin{widetext}
\begin{align}\label{HFEDF}
    E=&\,\langle\Phi_0|H|\Phi_0\rangle - A M
    =\,E^{\text{K}}+\sum_\phi(E^\text{D}_\phi+E^\text{E}_\phi)\nonumber\\
     =&\,\sum_{\alpha}\int d\bm{r}\,\bar{\varphi}_\alpha(\bm{r})(-i\bm{\gamma}\cdot\bm{\nabla}+M)\varphi_\alpha(\bm{r}) -AM+\frac{1}{2}\sum_{\phi,\alpha\beta}\left\{\iint d{\bm r_1}\,d\bm r_2\,
                 \bar{\varphi}_\alpha(\bm{r}_1)\bar{\varphi}_\beta(\bm{r}_2)\Gamma_\phi(\bm r_1,\bm r_2) D_\phi (\bm r_1,\bm r_2)
      {\varphi}_{\alpha}(\bm{r}_1)\varphi_{\beta}(\bm{r}_2)\right.\nonumber\\
      &\left.\quad\quad-\iint d{\bm r_1}\,d\bm r_2\,      \bar{\varphi}_\alpha(\bm{r}_1)\bar{\varphi}_\beta(\bm{r}_2)\Gamma_\phi(\bm r_1,\bm r_2) D_\phi(\bm r_1,\bm r_2) {\varphi}_{\beta}(\bm{r}_1)\varphi_{\alpha}(\bm{r}_2)\right\},
\end{align}
\end{widetext}
where $E^{\text{K}}$ denotes the kinetic energy, and $E^\text{D}_\phi$ and $E^\text{E}_\phi$ correspond to the energy contributions from the direct (Hartree) and exchange (Fock) terms, respectively.

Adopting the spherical symmetry, the single-particle states are specified by a set of quantum numbers $\alpha \equiv (a, m_\alpha) \equiv  (\tau_a, n_a, l_a, j_a, m_\alpha)$.
Note that because of the spherical symmetry, here we use $a$ to represent the other quantum numbers apart from the magnetic one $m_\alpha$.
For the isospin, $\tau_a=1/2 $ corresponds to the neutron state and $\tau_a=-1/2$ to the proton state.
The Dirac spinors of nucleon are explicitly written as
\begin{align}\label{DSSN}%Dirac spinor of spherical nuclei
  \varphi_\alpha(\bm r)
= \, \left(
  \begin{array}{c}
  \xi_\alpha(\bm r) \\
  \zeta_\alpha(\bm r)\\
  \end{array}
 \right)
=\,\frac{1}{r}\left(
 \begin{array}{c}
 iG_a(r) \\
 F_a(r)\hat{\bm \sigma}\cdot \hat {\bm r}\\
 \end{array}
  \right)
  \mathscr Y_\alpha(\hat {\bm r})\chi_\frac{1}{2}(\tau_a),
\end{align}
where $\xi_\alpha(\bm r)$ and $\zeta_\alpha(\bm r)$ are the upper and lower components of the Dirac spinor, and $G_a(r)$ and $F_a(r)$ are their radial parts, respectively.
$\chi_\frac{1}{2}(\tau_a)$ are the isospin spinors, and $\mathscr Y_\alpha(\hat {\bm r})$ are the tensor spherical harmonics defined through the coupling of the spherical harmonics and the spin spinors,
\begin{align}
  \mathscr Y_\alpha(\hat {\bm r})=\,\sum_{u s}C_{l_au\frac{1}{2}s}^{j_am_\alpha}Y_{l_au}(\hat {\bm r})\chi_\frac{1}{2}(s),
\end{align}
with $\hat {\bm r}\equiv{\bm r}/{r}$.

The variational principle,
{\begin{equation}
  \delta\left[E-\sum_{\alpha}\varepsilon_\alpha\int d \bm r\, \varphi_\alpha^\dag(\bm r) \varphi_\alpha(\bm r)\right]=\,0,
\end{equation}}
leads to the Hartree-Fock equation for the single-particle states $\{\varphi_\alpha(\bm{r})\}$.
The Lagrangian multipliers $\varepsilon_\alpha \equiv e_\alpha + M$ can be verified to be the single-particle energies, including the rest mass of nucleon.
The corresponding Hartree-Fock equation for the radial part of the wave functions reads
\begin{widetext}
 \begin{subequations}\label{eq:HF}
 \begin{align}
  \varepsilon_aG_a(r)=&\,-\left[\frac{d}{dr}-\frac{\kappa_a}{r}-\Sigma_T(r)\right]F_a(r)+[M+\Sigma_S(r)+\Sigma_0(r)]G_a(r)+Y_a(r),\\\label{HFE4}
  \varepsilon_aF_a(r)=&\,+\left[\frac{d}{dr}+\frac{\kappa_a}{r}+\Sigma_T(r)\right]G_a(r)-[M+\Sigma_S(r)-\Sigma_0(r)]F_a(r)+X_a(r),
\end{align}
\end{subequations}
\end{widetext}
with $\kappa_a \equiv (2j_a+1)(l_a-j_a)$.
$\Sigma_S$, $\Sigma_0$, and $\Sigma_T$ are the contributions to the self-energy from the direct terms.
$X$ and $Y$ denote the contributions from the exchange terms.
See all the detailed expressions, e.g., in Refs.~\cite{Long2005PhD.Thesis, Long2006Phys.Lett.B640.150, Long2007Phys.Rev.C76.034314, Sun2008Phys.Rev.C78.065805}.
Note that the density-dependence in the meson-nucleon coupling strengths leads to the contributions of the rearrangement terms to the self-energy.

\subsection{Tensor force in relativistic Hartree-Fock theory}\label{Sec:Tensor}

\subsubsection{Non-relativistic reduction}\label{SSec:Reduction}

To identify various components embraced in the relativistic meson-exchange picture, such as the central, spin-orbit, and tensor forces in the nucleon-nucleon interactions in the conventional non-relativistic sense, we will perform the non-relativistic reduction for the relativistic two-body interactions.

In the RHF theory, the relativistic meson-exchange two-body interactions,
\begin{equation}
  \hat{V}_\phi(\bm{r}_1,\bm{r}_2) = \,\gamma_0(\bm{r}_1)\gamma_0(\bm{r}_2)\Gamma_\phi(\bm{r}_1,\bm{r}_2)D_\phi(\bm{r}_1,\bm{r}_2),
\end{equation}
include those provided by the $\sigma$, $\omega$, $\rho$, and $\pi$ mesons.
The corresponding two-body interaction matrix elements read
\begin{align}\label{eq:Vabcd}
 &V_{\phi,\alpha\beta\gamma\delta}\nonumber\\
 =\,& \langle {\varphi}_\alpha{\varphi}_\beta | \hat{V}_\phi | {\varphi}_{\gamma} \varphi_{\delta} \rangle\nonumber\\
 =\,& \iint d \bm{r}_1\, d \bm{r}_2\, {\varphi}^\dagger_\alpha(\bm{r}_1){\varphi}^\dagger_\beta(\bm{r}_2) \hat{V}_\phi(\bm{r}_1,\bm{r}_2) {\varphi}_{\gamma}(\bm{r}_1) \varphi_{\delta}(\bm{r}_2).
\end{align}
Hereafter, we will use the indices ``$\alpha\beta\gamma\delta$'' or ''$abcd$'' to denote the single-particle states.
The non-relativistic reduction of $\hat{V}_\phi(\bm{r}_1,\bm{r}_2)$ leads to the non-relativistic two-body interaction $\hat{\mathcal{V}}_\phi(\bm{r}_1,\bm{r}_2)$ that satisfies \cite{Foldy1950Phys.Rev.78.29}
\begin{align}\label{eq:NRdef}
 V_{\phi,\alpha\beta\gamma\delta}
 =\, \langle {\varphi}_\alpha{\varphi}_\beta | \hat{\mathcal{V}}_\phi \Pi_+ | {\varphi}_{\gamma} \varphi_{\delta} \rangle
 =\,     \langle {\xi}_\alpha{\xi}_\beta | \hat{\mathcal{V}}_\phi | {\xi}_{\gamma} \xi_{\delta} \rangle,
\end{align}
where $\Pi_+$ is the projector to the upper components of the Dirac spinors, i.e., $\hat{\mathcal{V}}$ only acts on the upper components of the single-particle wave functions.
To make a clear distinction, hereafter we use the math calligraphic font $\hat{\mathcal{V}}$ to present the non-relativistic reduced two-body interactions.
In principle, $\hat{\mathcal{V}}$ can be expanded in the powers of $1/M$.

First, we discuss a specific case that the single-particle wave functions are the plane waves in the vacuum, i.e., in the zero-density limit.
The corresponding plane waves read
\begin{equation}\label{eq:Plane}
  {\varphi}_{\bm{p}_a}(\bm{r}) =\, u_{\bm{p}_a}\, e^{i \bm p_a \cdot \bm r},
\end{equation}
where
  \begin{align}\label{eq:upa}
 u_{\bm{p}_a}=\,\sqrt{\frac{M+\varepsilon_a}{2\varepsilon_a}}
  \left(
   \begin{array}{c}
    1 \\
    \frac{\bm{\sigma}\cdot\bm{p}_a}{M+\varepsilon_a} \\
     \end{array}
  \right)\chi_\frac{1}{2}(s_a)\chi_\frac{1}{2}(\tau_a),
\end{align}
for the positive-energy states in the Fermi sea.
Putting these expressions in Eq.~\eqref{eq:Vabcd}, we obtain
\begin{align}\label{eq:Vphi_abcd}
V_{\phi,abcd} = \,\bar u_{\bm{p}_a}(1)\bar u_{\bm{p}_b}(2)\frac{1}{m_\phi^2+\bm{q}^2}\Gamma_\phi{(1,2)}u_{\bm{p}_c}(1)u_{\bm{p}_d}(2),
\end{align}
where $\bm{q} \equiv \bm{p}_a - \bm{p}_c = \bm{p}_d - \bm{p}_b$ in the Yukawa propagator is the momentum transfer.
Here the expressions for the vertices $\Gamma_\phi{(1,2)}$ acting on the plane waves are shown in Eqs.~\eqref{vertexp} in Appendix~\ref{App:Reduction}.

\begin{table}
\caption{Expressions of $\mathscr{F}_{0,\phi}$ in Eq.~\eqref{eq:V_t0phi} for each meson-nucleon coupling in the zero-density limit.
The ratios to the $\pi$\text{-PV} coupling are evaluated by $\left(\frac{\mathscr{F}_{0,\phi}}{m_\phi^2+\bm{q}^2}\right) \left/ \left(\frac{\mathscr{F}_{0,\pi\text{-PV}}}{m_\pi^2+\bm{q}^2} \right)\right.$ with $\bm{q}=0$ using the bare interaction Bonn A~\cite{Machleidt1989Adv.Nucl.Phys.19.189} and the effective interaction PKA1~\cite{Long2007Phys.Rev.C76.034314}.}
\label{Table:F0}
\begin{ruledtabular}
\begin{tabular}{llll}
\vspace{0.3em} Coupling & $\mathscr{F}_{0,\phi}$ & \multicolumn{2}{c}{ Ratio to $\pi$-PV} \\
 & & Bonn A\footnote{The corresponding form factors are also taken into account.} & PKA1 \\
\hline
\vspace{0.3em} $\omega$\text{-V}&$\displaystyle\frac{g_\omega g_\omega}{4M^2}$& $-0.02$\footnote{This value is only for the $nn$ or $pp$ channel, whereas $0$ for the $np$ channel.} & $-0.02$\footnote{Same as $^{\text{b}}$.}\\
\vspace{0.3em} $\pi$\text{-PV}&$\displaystyle -\vec{\tau}\cdot\vec{\tau}\frac{f_{\pi}f_{\pi}}{m^2_\pi}$ & ~~\,$1$ & ~~\,$1$\\
\vspace{0.3em} $\rho$\text{-V}&$\vec{\tau}\cdot\vec{\tau}\displaystyle\frac{g_\rho g_\rho}{4M^2}$& $-0.0009$ & $-0.002$\\
\vspace{0.3em} $\rho$\text{-T}&$\vec{\tau}\cdot\vec{\tau}\displaystyle\frac{f_\rho f_\rho}{4M^2}$& $-0.03$ & $-0.02$\\
\vspace{0.3em} $\rho$\text{-VT}&$\vec{\tau}\cdot\vec{\tau}\displaystyle\frac{f_\rho g_\rho}{2M^2}$& $-0.01$ & $-0.01$\\
\end{tabular}
\end{ruledtabular}
\end{table}

As a result, the non-relativistic reduced two-body interactions $\hat{\mathcal{V}}_{0,\phi}$ provided by each meson-nucleon coupling are expressed up to the $1/M^2$ order in Eqs.~\eqref{Vabcd2}.
It is seen that all the couplings, except the $\sigma\text{-S}$ one, give rise to the tensor force.
This is in agreement with the realistic Bonn nucleon-nucleon interactions in the one-boson-exchange picture \cite{Machleidt1989Adv.Nucl.Phys.19.189}.
Explicitly, the tensor components of the non-relativistic reduced two-body interactions read
\begin{equation}\label{eq:V_t0phi}
  \hat{\mathcal{V}}^{\text t}_{0,\phi} = \,\frac{1}{m_\phi^2+\bm{q}^2} \mathscr{F}_{0,\phi} S_{12},
\end{equation}
where
\begin{align}\label{S12}
 S_{12}\equiv\,(\bm{\sigma}_1\cdot \bm q)(\bm{\sigma}_2\cdot \bm q)-\frac{1}{3}(\bm\sigma_1\cdot\bm\sigma_2)q^2,
 \end{align}
and $\mathscr{F}_{0,\phi}$ in each meson-nucleon coupling are shown in Table~\ref{Table:F0}.
See Appendix~\ref{App:Reduction} for detailed derivations.

To have ideas on the relative strengths of the tensor component generated from different couplings, Table~\ref{Table:F0} also shows their ratios to the $\pi$\text{-PV} coupling, which are evaluated by $\left(\frac{\mathscr{F}_{0,\phi}}{m_\phi^2+\bm{q}^2}\right) \left/ \left(\frac{\mathscr{F}_{0,\pi\text{-PV}}}{m_\pi^2+\bm{q}^2} \right)\right.$ with $\bm{q}=0$, by taking the bare interaction Bonn A \cite{Machleidt1989Adv.Nucl.Phys.19.189} and the effective interaction PKA1 \cite{Long2007Phys.Rev.C76.034314} as examples.
It is seen that the largest tensor contribution comes from the pion exchange, while all other couplings have opposite but negligible contributions in the zero-density limit.

For general single-particle wave functions, the two-body interaction matrix elements can be formally expressed as
\begin{align}\label{eq:V_abcd-expand}
  &\langle {\varphi}_\alpha{\varphi}_\beta | \hat{V}_\phi | {\varphi}_{\gamma} \varphi_{\delta} \rangle\nonumber\\
  =\,& \sum_{\bm{p}_a \bm{p}_b \bm{p}_c \bm{p}_d} \langle {\varphi}_\alpha | {\varphi}_{\bm{p}_a} \rangle \langle {\varphi}_\beta | {\varphi}_{\bm{p}_b} \rangle
  \langle {\varphi}_{\bm{p}_c} | {\varphi}_\gamma \rangle \langle {\varphi}_{\bm{p}_d} | {\varphi}_\delta \rangle \nonumber\\
  &\qquad\qquad \times \langle {\varphi}_{\bm{p}_a}{\varphi}_{\bm{p}_b} | \hat{V}_\phi | {\varphi}_{\bm{p}_c} \varphi_{\bm{p}_d} \rangle \nonumber\\
  \approx\,& \sum_{\bm{p}_a \bm{p}_b \bm{p}_c \bm{p}_d\in F} \langle {\varphi}_\alpha | {\varphi}_{\bm{p}_a} \rangle \langle {\varphi}_\beta | {\varphi}_{\bm{p}_b} \rangle
  \langle {\varphi}_{\bm{p}_c} | {\varphi}_\gamma \rangle \langle {\varphi}_{\bm{p}_d} | {\varphi}_\delta \rangle \nonumber\\
  &\qquad\qquad \times \langle {\varphi}_{\bm{p}_a}{\varphi}_{\bm{p}_b} | \hat{V}_\phi | {\varphi}_{\bm{p}_c} \varphi_{\bm{p}_d} \rangle\nonumber\\
  =\,&\sum_{\bm{p}_a \bm{p}_b \bm{p}_c \bm{p}_d\in F} \langle {\varphi}_\alpha | {\varphi}_{\bm{p}_a} \rangle \langle {\varphi}_\beta | {\varphi}_{\bm{p}_b} \rangle
  \langle {\varphi}_{\bm{p}_c} | {\varphi}_\gamma \rangle \langle {\varphi}_{\bm{p}_d} | {\varphi}_\delta \rangle \nonumber\\
  &\qquad\qquad \times \langle {\varphi}_{\bm{p}_a}{\varphi}_{\bm{p}_b} | \hat{\mathcal{V}}_{0,\phi} \Pi_+ | {\varphi}_{\bm{p}_c} \varphi_{\bm{p}_d} \rangle.
\end{align}
Because the non-relativistic reduction performed for the plane waves at the last step in Eq.~\eqref{eq:V_abcd-expand} is valid only for the positive-energy states in the Fermi sea, we have to make a truncation $\bm{p}_a \bm{p}_b \bm{p}_c \bm{p}_d\in F$ before that.
Such a truncation introduces an approximation in the non-relativistic reduction for general cases, such as the single-particle wave functions in finite nuclei.

For finite nuclei, the Hartree-Fock equation~\eqref{eq:HF} shows that the ratio between the upper and lower components can be evaluated as
\begin{align}
  F(r) &\sim \,\frac{\frac{d}{dr}+\frac{\kappa}{r}}{\varepsilon + M +\Sigma_S(r) -\Sigma_0(r)} G(r) \nonumber\\
  &\sim\, \frac{\frac{d}{dr}+\frac{\kappa}{r}}{2 [M +\Sigma_S(r)]} G(r).
\end{align}
In the central region of nuclei, the nuclear density is around the saturation density $\rho_{\rm sat.}$, and \cite{Long2007Phys.Rev.C76.034314}
\begin{equation}
  M + \Sigma_S \sim \,0.6 M .
\end{equation}
In comparison, the ratio between the upper and lower components of the plane waves in the zero-density limit is around $\bm{\sigma}\cdot\bm{p}/(2M)$ as shown in Eq.~\eqref{eq:upa}.
Therefore, within the truncation $\bm{p}_a \bm{p}_b \bm{p}_c \bm{p}_d\in F$ adopted in Eq.~\eqref{eq:V_abcd-expand}, it is not an optimal choice to perform non-relativistic reduction for the single-particle wave functions in finite nuclei by expanding on the plane waves in the vacuum.

Following the spirit of the local density approximation (LDA), at each position $\bm r$ with finite density $\rho(\bm r)$ in nuclei, we seek for the corresponding properties of homogeneous nuclear matter with the same density $\rho$.

\begin{table}
\caption{Expressions for $\mathscr{F}_{\phi}$ in Eq.~\eqref{eq:Vtphi_abcd} for each meson-nucleon coupling with finite density.
The ratio to the $\pi$\text{-PV} coupling is evaluated by $\left(\frac{\mathscr{F}_{\phi}}{m_\phi^2+\bm{q}^2}\right) \left/ \left(\frac{\mathscr{F}_{\pi\text{-PV}}}{m_\pi^2+\bm{q}^2} \right)\right.$ with $\bm{q}=0$ and $\rho = \rho_{\text{sat.}}$ using the effective interaction PKA1.}
\label{Table:F}
\begin{ruledtabular}
\begin{tabular}{lll}
\vspace{0.3em} Coupling & $\mathscr{F}_{\phi}$ & Ratio to $\pi$\text{-PV} \\
\hline
\vspace{0.3em} $\omega$\text{-V}&$\displaystyle\frac{g_\omega(1) g_\omega(2)}{4M^*(1)M^*(2)}$& $-0.74$\footnote{This value is only for the $nn$ or $pp$ channel, whereas $0$ for the $np$ channel.} \\
\vspace{0.3em} $\pi$\text{-PV}&$\displaystyle -\vec{\tau}\cdot\vec{\tau}\frac{f_{\pi}(1)f_{\pi}(2)}{m^2_\pi}$ & ~~\,$1$\\
\vspace{0.3em} $\rho$\text{-V}&$\vec{\tau}\cdot\vec{\tau}\displaystyle\frac{g_\rho(1) g_\rho(2)}{4M^*(1)M^*(2)}$ & $-0.03$\\
\vspace{0.3em} $\rho$\text{-T}&$\vec{\tau}\cdot\vec{\tau}\displaystyle\frac{f_\rho(1) f_\rho(2)}{4M^2}$ & $-0.25$ \\
\vspace{0.3em} $\rho$\text{-VT}&$\vec{\tau}\cdot\vec{\tau}\displaystyle\frac{f_\rho(1) g_\rho(2)}{4MM^*(2)}+(1\leftrightarrow2)$& $-0.16$ \\
\end{tabular}
\end{ruledtabular}
\end{table}

In the relativistic framework, the single-particle plane waves in a homogeneous system generally read
\begin{equation}
  {\varphi}_{\bm{p}^*_a}(\bm{r}) =\, u_{\bm{p}^*_a}\, e^{i \bm p_a \cdot \bm r},
\end{equation}
where
  \begin{align}\label{eq:upa*}%\label{U}
 u_{\bm{p}^*_a}=\,\sqrt{\frac{M^*+\varepsilon^*_a}{2\varepsilon^*_a}}
  \left(
   \begin{array}{c}
    1 \\
    \frac{\bm{\sigma}\cdot\bm{p}^*_a}{M^*+\varepsilon^*_a} \\
     \end{array}
  \right)\chi_\frac{1}{2}(s_a)\chi_\frac{1}{2}(\tau_a).
\end{align}
In the RHF theory, the starred quantities are defined as \cite{Bouyssy1987Phys.Rev.C36.380}
\begin{subequations}
\begin{align}
    \bm{p}^*& \equiv \, \bm{p}+\hat{\bm{p}}{\Sigma}_V(p),\\
    M^*(p) & \equiv\, M + {\Sigma}_S(p),\\
    \varepsilon^*(p) & \equiv\, \varepsilon(p)-{\Sigma}_0(p),
\end{align}
\end{subequations}
with the momentum-dependent self-energies.
$M^*$ is the so-called Dirac mass.
As a result, the corresponding tensor components of the non-relativistic reduced two-body interactions become
\begin{equation}\label{eq:Vtphi_abcd}
  \hat{\mathcal{V}}^{\text t}_{\phi} =\, \frac{1}{m_\phi^2+\bm{q}^2} \mathscr{F}_{\phi} S_{12}.
\end{equation}
Table~\ref{Table:F} shows the explicit expressions of $\mathscr{F}_{\phi}$ and the ratios to the $\pi$\text{-PV} coupling, which are evaluated by $\left(\frac{\mathscr{F}_{\phi}}{m_\phi^2+\bm{q}^2}\right) \left/ \left(\frac{\mathscr{F}_{\pi\text{-PV}}}{m_\pi^2+\bm{q}^2} \right)\right.$ with $\bm{q}=0$ and $\rho = \rho_{\text{sat.}}$ using the PKA1 effective interaction.
Here the $M^*$ is evaluated with the Fermi momentum $M^*(p_F)$, due to its weak momentum dependence.
See Appendix~\ref{App:Reduction*} for detailed derivations and the relevant discussions.

On the one hand, similar to the case of plane waves in the zero-density limit, the largest tensor contribution comes from the pion exchange, while all other couplings have opposite contributions.
On the other hand, now these contributions, except $\rho$-V, become comparable with the $\pi$-PV one.
This is mainly due to the density-dependent behaviors of the coupling strengths as well as the Dirac mass $M^*$.
First, comparing with the other meson-nucleon coupling strengths, $f_\pi$ quenches more significantly as the nuclear density increases.
Second, factors of $M/{M^*}$ and even $M^2/{M^*}^2$ enhance the tensor components in the VT and V couplings, respectively.
In addition, it is noted that the ratios shown in Table~\ref{Table:F} are evaluated with $\bm{q}=0$.
These values will become larger with finite momentum transfer, because in typical cases $|\bm{q}| \sim m_\pi$ but $|\bm{q}| \ll m_\omega,\,m_\rho$.

Then, we will use the non-relativistic reduced two-body interactions in Eq.~\eqref{eq:Vtphi_abcd} to evaluate the contributions of the tensor component in finite nuclear systems, as we will present in the following.

\subsubsection{Evaluation of tensor contribution}

Based on the above discussions, the tensor contribution to the two-body interaction matrix elements $V_{\phi,\alpha\beta\gamma\delta}$ in each meson-nucleon coupling, denoted as $V^{\text{t}}_{\phi,\alpha\beta\gamma\delta}$, can be evaluated by
\begin{align}\label{eq:VTabba}
V^{\text{t}}_{\phi,\alpha\beta\gamma\delta}
 =\, \langle {\xi}_\alpha{\xi}_\beta | \hat{\mathcal{V}}_\phi^{\text{t}} | {\xi}_{\gamma} \xi_{\delta} \rangle.
\end{align}
In the coordinate representation, $\hat{\mathcal{V}}_\phi^t(\mathfrak r)$ is expressed as
\begin{align}\label{eq:ten_coor}
\hat{\mathcal{V}}_\phi^{\text{t}}(\mathfrak r) =\,
-\mathscr{F}_{\phi}(1,2)\,\frac{m_\phi^2 e^{-m_\phi \mathfrak r}}{ 4\pi\mathfrak{r}}\left(1+\frac{3}{m_\phi \mathfrak{r}}+\frac{3}{m_\phi^2 \mathfrak r^2}\right)S_{12}(\hat{\mathfrak r}),
\end{align}
where $S_{12}$ reads
\begin{align}\label{S12r}
S_{12}(\hat{\mathfrak r})\equiv\,{(\bm{\sigma}_1\cdot \hat{\mathfrak{r}})(\bm{\sigma}_2\cdot \hat{\mathfrak{r}})}-\frac{1}{3}\bm\sigma_1\cdot\bm\sigma_2,
\end{align}
with $\mathfrak{r}\equiv|\bm r_1-\bm r_2|$ and $\hat{\mathfrak{r}}\equiv({\bm r_1-\bm r_2})/{|\bm r_1-\bm r_2|}$.

In the RHF theory, only the exchange terms give rise to the non-vanishing matrix elements of tensor interaction, because $\bm q = 0$ in the direct terms.
By using the spherical symmetry, the matrix elements are evaluated by
\begin{align}\label{eq:VTabbasum}
& \sum_{m_\alpha = -j_a}^{j_a} \sum_{m_\beta = -j_b}^{j_b} {V}^{\text{t}}_{\phi,\alpha\beta\beta\alpha}\nonumber\\
=\,& \sum_{m_\alpha m_\beta} \langle {\xi}_\alpha{\xi}_\beta | \hat{\mathcal{V}}_\phi^{\text{t}} | {\xi}_{\beta} \xi_{\alpha} \rangle\nonumber\\
=\,& \sum_{m_\alpha m_\beta} \iint d\bm r_1\, d\bm r_2\, {\xi}_\alpha^\dagger(\bm r_1) {\xi}_\beta^\dagger(\bm r_2) \hat{\mathcal{V}}_\phi^{\text{t}}(\mathfrak r) {\xi}_\beta(\bm r_1) {\xi}_\alpha(\bm r_2).
\end{align}
The corresponding spherically averaged matrix elements are defined as
\begin{equation}
  {V}^{\text{t}}_{\phi,abba} \equiv \,\frac{1}{\hat j^2_a \hat j^2_b} \sum_{m_\alpha m_\beta} {V}^{\text{t}}_{\phi,\alpha\beta\beta\alpha},
\end{equation}
where $\hat j^2 \equiv 2j+1$ is the degeneracy of the orbital.
In principle, the above integrals can be carried out directly.
Nevertheless, it will be very difficult to decompose analytically the radial and angular parts of the tensor interaction in Eq.~\eqref{eq:ten_coor}.
In practice, we take the advantage of the existing RHF formalism and subroutines to calculate this integral in an alternative way.
See Appendix~\ref{APP:Matrix_tensor} for details.

\begin{figure}
\includegraphics[width=0.45\textwidth]{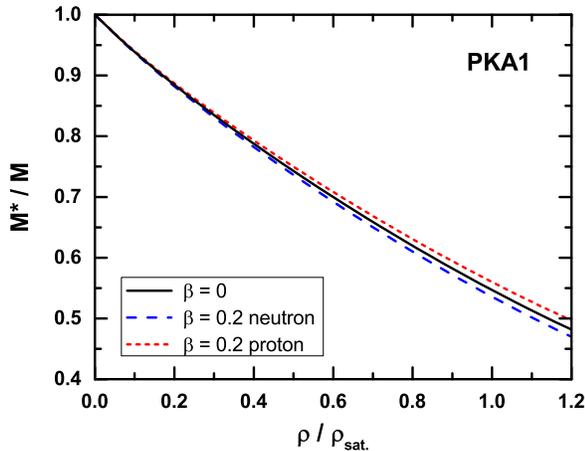}% Here is how to import EPS art
\caption{(Color online) Dirac mass $M^*$ at the Fermi momentum as a function of matter density $\rho$ calculated by the RHF theory with the PKA1 effective interaction.
The $M^*$ of the symmetric nuclear matter are shown with the solid line, while the $M^*_n$ and $M^*_p$ of the asymmetric matter with $\beta=0.2$ are shown with the dashed and short-dashed lines, respectively.
}\label{Fig:Mrho}
\end{figure}

Before ending this section, let us discuss the properties of the Dirac mass $M^*$ appearing in $\mathscr{F}_{\phi}(1,2)$ for the integral~\eqref{eq:VTabbasum}.
First of all, as discussed in Appendix~\ref{App:Reduction*}, for a given nuclear matter density the momentum dependence of $M^*$ is rather weak, and thus its value is evaluated with the Fermi momentum $M^*(p_F)$.
Second, according to the spirit of LDA, at vertices 1 and 2 with densities $\rho(\bm r_1)$ and $\rho(\bm r_2)$, we take
\begin{equation}
  M^*(\bm r_i) = \,M^*(\rho(\bm r_i)),\qquad i=1,2,
\end{equation}
i.e., their values in the corresponding homogeneous nuclear matter with the same densities, respectively.
In Fig.~\ref{Fig:Mrho}, the values of $M^*$ are shown as a function of matter density $\rho$ with the solid line for the symmetric nuclear matter.
An obvious density-dependent behavior is seen, and $M^*(\rho_{\text{sat.}})=0.55M$ at the saturation density.
Third, for a given matter density, in principle $M^*$ also depends on the isospin asymmetry $\beta \equiv (\rho_n - \rho_p)/\rho$ and appears the isospin splitting.
Nevertheless, at the central region of nuclei the isospin asymmetry $\beta$ is small, in contrast, at the surface region $\beta$ increases for neutron-rich nuclei while the density becomes small.
By taking the case of $\beta=0.2$ as an example, $M_n^*$ and $M_p^*$ are shown as a function of matter density $\rho$ in Fig.~\ref{Fig:Mrho}.
It is seen that such an isospin splitting is generally small.
Therefore, we will always adopt the $M^*$ values associated with the symmetric nuclear matter in the following calculations.

\section{Results and Discussion}\label{Sec:results}

\subsection{Sum rule of matrix elements}

\begin{table}%[htbp]\centering
\caption{Two-body matrix elements $V_{abba}^\text{t}$ of the tensor force in the $\omega$-V coupling in $^{208}$Pb.
The results are calculated by RHF with PKA1, and for each spin doublets the radial wave function of the spin-up state is adopted.
All units are in $10^{-2}$~MeV.}
\label{Tab:SumRule_Omega}
\begin{ruledtabular}
\begin{tabular}{crrr}
    $a$  & \multicolumn{3}{c}{$b$ }  \\
    & $\nu1p_{1/2}$ & $\nu1f_{7/2}$  & $\nu1h_{11/2}$   \\
\hline
$\nu1p_{ 3/2}$ & $-0.795260$ & $0.525426$   & $0.355980$ \\
$\nu1p_{ 1/2}$ & $1.590520$ & $-1.050852$ &  $-0.711960$ \\
Sum            & $0.000000$	& $0.000000$	& $0.000000$	\\
\hline
$\nu2f_{ 7/2}$ & $-1.131623$ &  $0.705647$ & $0.447433$ \\
$\nu2f_{ 5/2}$ &  $1.508830$ & $-0.940863$ & $-0.596577$ \\
Sum      &  $0.000000$	& $0.000000$	& $0.000000$ \\
\end{tabular}
\end{ruledtabular}
\end{table}

\begin{table}%[htbp]\centering
\caption{Same as Table~\ref{Tab:SumRule_Omega}, but for the two-body matrix elements $V_{abba}^\text{t}$ of the tensor force from all the couplings.}
\label{Tab:SumRule_total}
\begin{ruledtabular}
\begin{tabular}{crrr}
 $a$  &\multicolumn{3}{c}{$b$ }  \\
 & $\nu1p_{1/2}$ & $\nu1f_{7/2}$  & $\nu1h_{11/2}$   \\
\hline
$\nu1p_{ 3/2}$ & $-0.361054$    & $0.389579$ & $0.303729$ \\
$\nu1p_{ 1/2}$ & $0.722108$ & $-0.779159$     &  $-0.607458$ \\
Sum            & $0.000000$	& $0.000000$	  & $0.000000$	\\
\hline
$\nu2f_{ 7/2}$ & $-1.034539$    &  $0.623290$ & $0.254924$ \\
$\nu2f_{ 5/2}$ & $1.379386$ & $-0.831054$      & $-0.339899$ \\
Sum            & $0.000000$	& $0.000000$   & $0.000000$ \\
\end{tabular}
\end{ruledtabular}
\end{table}

As a benchmark, one of the most important properties of the tensor force is the following sum rule of the two-body interaction matrix elements \cite{Otsuka2005Phys.Rev.Lett.95.232502},
\begin{align}\label{Eq:SumRule}
\hat j^2_{a} V_{a b b a}^\text{t} + \hat j^2_{\tilde a} V_{\tilde a b b \tilde a}^\text{t} =\,0,
\end{align}
where $b$ is an arbitrary state, and the spin-up state $a$ with $j_> = l+1/2$ and the spin-down state $\tilde a$ with $j_<=l-1/2$ are a pair of spin doublets.
This sum rule is exactly satisfied on the condition that the radial wave functions of spin doublets are identical to each other.

We carry out such a benchmark for the tensor forces extracted above in the RHF theory.
Table~\ref{Tab:SumRule_Omega} shows the values of the two-body matrix elements $V^\text{t}_{abba}$ of the tensor force generated by the $\omega$-V coupling and the corresponding sum-rule values, by taking the several single-neutron states in $^{208}$Pb as examples.
First of all, it is seen that the matrix elements are positive between the $j_>$ and $j'_>$ ($j_<$ and $j'_<$) states, whereas they are negative between the $j_>$ and $j'_<$ ($j_<$ and $j'_>$) states.
This property is opposite to that emphasized in Ref.~\cite{Otsuka2005Phys.Rev.Lett.95.232502}, because the tensor forces generated by the $\omega$-V and $\pi$-PV couplings have different signs.
Note that it is the two-body matrix elements with a minus sign, $-V^\text{t}_{abba}$, that contribute to the single-particle and total energies, because they are the Fock terms.
To testify the sum rule, for each spin doublets, we make the radial wave function of the spin-down $j_<$ state identical to its spin-up $j_>$ counterpart.
It is confirmed that the sum rule is fulfilled with more than six digits, for both the nodal and non-nodal states with low and high angular momenta.

Individually, this sum rule is satisfied for the two-body matrix elements of the tensor forces generated by each meson-nucleon coupling.
As a result, the total values of the tensor matrix elements satisfy the sum rule at the same accuracy, as shown in Table~\ref{Tab:SumRule_total}.

\subsection{Tensor effects on binding energy}

\begin{table*}
\caption{Contributions to the total energy $E$ from different couplings in $^{48}$Ca and $^{208}$Pb calculated by RHF with PKA1.
$E^{\text{K}}$, $E^{\text{D}}$, $E^{\text{E}}$, and $E^{\text{CM}}$ are the kinetic, Hartree, Fock, and center-of-mass correction energies, respectively.
All units are in MeV.}
\label{Tab:Energy}
\begin{ruledtabular}
\begin{tabular}{ccrrrrrr}
 & & \multicolumn{3}{c}{$^{48}$Ca}              &      \multicolumn{3}{c}{$^{208}$Pb } \\
  & coupling & \multicolumn{1}{c}{neutron}  & \multicolumn{1}{c}{proton}& \multicolumn{1}{c}{total} & \multicolumn{1}{c}{neutron}& \multicolumn{1}{c}{proton} & \multicolumn{1}{c}{total} \\
\hline
$E^{\text{K}}$ &          & $392.398$    & $220.074$    & $612.471$  & $1596.568$    & $907.899$      & $2504.467$ \\
\hline
$E^{\text{D}}$
   &  $\sigma\text{-S}$  & $-2940.535$  & $-2253.310$ & $-5193.845$& $-14320.935$ & $-10022.547$ &$-24343.482$    \\
   &  $\omega\text{-V}$  & $2357.686$   & $1805.927$  & $4163.613$ & $11520.882$  &  $7962.551$  & $19483.433$\\
   &  $\rho\text{-V}$    & $15.595$     & $-11.161$   & $4.434$    & $98.419$     & $-65.123$    & $33.296$\\
   &  $\rho\text{-T}$    &  $-0.285$    & $0.134$     & $-0.150$   & $-0.308$     & $0.210$      & $-0.099$\\
   &  $\rho\text{-VT}$   &  $-0.997$    & $0.714$     & $-0.283$   & $-1.650$     & $1.077$      & $-0.571$\\
   &  $\pi\text{-PV}$    &   ------  & ------ & ------ & ------    & ------   & ------ \\
   &  $A\text{-V}$       &   ------  & $79.354 $   & $79.354$   & ------    & $827.640$   & $827.640$\\
%\cline{2-11}
   &   Total             &  $-568.536$	&$-378.342$	&$-946.877$	&$-2703.591$	&$-1296.192$ & $-3999.783$\\
\hline
$E^{\text{E}}$
   &  $\sigma\text{-S}$  & $709.344$    & $431.461$    & $1140.805$   &  $3503.261$   & $1774.293$    & $5277.554$ \\
   &  $\omega\text{-V}$  & $-515.922$ & $-306.724$ & $-822.646$ & $-2451.231$ & $-1265.209$ & $-3716.441$\\
   &  $\rho\text{-V}$    & $-59.121$  & $-48.867$  & $-107.987$ & $-266.194$  & $-210.099$  & $-476.293$\\
   &  $\rho\text{-T}$    & $-151.829$ & $-123.868$ & $-275.698$ & $-687.900$  & $-531.942$  & $-1219.843$\\
   &  $\rho\text{-VT}$   & $23.852$     & $19.370$     & $43.222$     & $122.508$     & $89.155$      & $211.663$ \\
   &  $\pi\text{-PV}$    & $-23.449$  & $-20.397$  & $-43.846$  & $-103.254$  & $-79.752$   & $-183.006$ \\
   &  $A\text{-V}$       & ------  & $-7.201$   & $-7.201$   & ------    & $-29.021$   & $-29.021$\\
%\cline{2-11}
   &   Total             & $-17.124$  & $-56.226$  & $-73.351$	& $117.190$	      &$-252.575$	&$-135.387$\\
   \hline
{ $E^{\text{CM}}$ }&   &    &           &{$-8.617$}  &      &          &{$-6.260$}\\
\hline
\multicolumn{2}{c}{Total energy $E$ }   &    &           &{$-416.373$}  &      &          &{$-1636.961$}\\
\end{tabular}
\end{ruledtabular}
\end{table*}

\begin{table*}
\caption{Contributions to the total energy from the tensor forces in different couplings in $^{48}$Ca and $^{208}$Pb calculated by RHF with PKA1.
All units are in MeV.}
\label{Tab:Energy_tensor}
\begin{ruledtabular}
\begin{tabular}{ccrrrrrr}
 & & \multicolumn{3}{c}{$^{48}$Ca (MeV)}                                                 &      \multicolumn{3}{c}{$^{208}$Pb (MeV)} \\
  & coupling & \multicolumn{1}{c}{neutron}  & \multicolumn{1}{c}{proton}& \multicolumn{1}{c}{total} & \multicolumn{1}{c}{neutron}& \multicolumn{1}{c}{proton} & \multicolumn{1}{c}{total} \\
\hline
$E^\text{t}$
   &  $\omega\text{-V}$          & $-1.148$ & $-0.006$ & $-1.155$ & $-1.769$ & $-1.090$ & $-2.859$  \\
   &  $\rho\text{-V}$   & $-0.060$ & $-0.005$ & $-0.065$ & $-0.215$ & $-0.179$ & $-0.394$  \\
   &  $\rho\text{-T}$   & $-0.202$ & $-0.01$2 & $-0.214$ & $-0.69$1 & $-0.553$ & $-1.243$  \\
   &  $\rho\text{-VT}$  & $-0.216$ & $-0.016$ & $-0.232$ & $-0.760$ & $-0.620$ & $-1.380$ \\
   &  $\pi\text{-PV}$   & $1.142$    & $0.037$    & $1.179$    & $3.566$   & $2.941$    & $6.507$   \\
  % \cline{2-11}
   &   Total            & $-0.486$	   & $-0.002$	  & $-0.488$	 & $0.131$    &  $0.499$   & $0.629$  \\
\end{tabular}
\end{ruledtabular}
\end{table*}

With the tensor contributions to the matrix elements from all the meson-nucleon couplings, we can evaluate the tensor contributions to the total energies of finite nuclei.

Let us first give an overview on the effective interaction PKA1 by showing the contributions to the total energy from the kinetic, Hartree, and Fock terms, as well as the center-of-mass correction in Table~\ref{Tab:Energy} for the nuclei $^{48}$Ca and $^{208}$Pb.
It is seen that the total energy is mainly determined by the delicate balance among the kinetic term, the $\sigma$-S, and the $\omega$-V couplings, in particular, their Hartree terms.
This is consistent with the original idea of the Walecka model \cite{Walecka1974Ann.Phys.83.491}.
Among other Hartree terms, the Coulomb interaction becomes more important as the proton number increases, and the $\rho$-V coupling contributes in neutron-rich nuclei for the proper isovector properties.
In contrast, the $\rho$-T and $\rho$-VT couplings give basically no contribution, and the $\pi$-PV coupling does not contribute at all due to the violation of parity conservation.
For the Fock terms, on the one hand, the biggest contributions still come from the $\sigma$-S and $\omega$-V couplings, but they are in general smaller than their Hartree counterparts by around a factor of $5$, and have opposite signs.
On the other hand, via the Fock terms, the $\rho$-V, $\rho$-T, $\rho$-VT, and $\pi$-PV couplings give much more important contributions to the total energy comparing with their Hartree counterparts, in particular, the $\rho$-T one.

The tensor contributions to the total energy are embraced in the Fock terms.
The corresponding values are shown in Table~\ref{Tab:Energy_tensor}.
It is noted that in general the tensor forces of all the couplings give very small contributions to the total energy.
In particular, for the protons in $^{48}$Ca which are spin saturated, the tensor contributions are especially small due to the sum rule \eqref{Eq:SumRule}.
Such a tiny contribution to the total energy is one of the most important reasons why the tensor forces had been neglected for many years in most of popular effective interactions.
Even if the tensor forces were included, their proper strengths were not well in control by fitting to the data such as nuclear masses.

Traditionally, the $\pi$-PV and $\rho$-T couplings are considered as the main carriers of the tensor force.
In the RHF theory with the effective interaction PKA1, the tensor force in the $\pi$-PV coupling makes nuclei less bound, while the tensor forces in all the other couplings give opposite contributions and largely cancel the $\pi$-PV one.
The present calculations show that for the completely spin-unsaturated system, e.g., $^{208}$Pb, the $\pi$-PV tensor contribution can reach around $0.4\%$ of the total energy, while the $\rho$-T coupling contributes less than $0.1\%$.
Furthermore, it is remarkable that the tensor contribution from the $\rho$-VT coupling is indeed comparable with the $\rho$-T one, and the tensor contribution from the $\omega$-V coupling is larger and even comparable with the $\pi\text{-PV}$ one.
Among all the couplings which can give rise to the tensor force, the contribution from the $\rho\text{-V}$ coupling is the smallest, mainly because of its small coupling strength around the saturation density.
These conclusions can also be understood by the guidance of Table~\ref{Table:F}.

Note that in the present scheme, the tensor effects on the total energy correspond to the expectation value of the tensor force on the system wave function as a Slater determinant.
As a result, these effects are in general tiny, while the higher-order effects of tensor force, e.g., the two-particle-two-hole effect, are supposed to be implicitly absorbed in the effective interactions.
In contrast, if the tensor effects on the total energy refer to the expectation value of the bare tensor force on the fully correlated system wave function, the corresponding effects are in general profound.
For example, it contributes about $-68$~MeV in $^4$He in various \textit{ab initio} calculations \cite{Kamada2001Phys.Rev.C64.044001}.
The two-particle-two-hole tensor effects are also studied explicitly in the tensor-optimized shell model \cite{Myo2005Prog.Theor.Phys.113.763, Myo2007Phys.Rev.C76.024305}, the tensor-optimized antisymmetrized molecular dynamics \cite{Myo2017Phys.Lett.B769.213, Myo2017Phys.Rev.C95.044314}, and the high-momentum antisymmetrized molecular dynamics \cite{Myo2017Prog.Theor.Exp.Phys.2017.111D01}.

\subsection{Tensor effects on shell evolution}

Even though there is only a tiny effect of the tensor force on nuclear binding energy, the tensor force plays a significant role in the shell evolution \cite{Otsuka2005Phys.Rev.Lett.95.232502, Sagawa2014Prog.Part.Nucl.Phys.76.76}, in particular, the emergence of new magic numbers far from the nuclear $\beta$-stability line \cite{Wienholtz2013Nature498.346, Steppenbeck2013Nature502.207}.

In Ref.~\cite{Moreno-Torres2010Phys.Rev.C81.064327}, the tensor effects on the shell evolution were investigated by comparing the non-relativistic Skyrme and Gogny Hartree-Fock theories as well as the relativistic Hartree-Fock theory.
Particular attention was paid to the evolution of the magic gaps along the $Z,\,N = 8,\,20$, and $28$ isotopes and isotones.
To our knowledge, this is the only literature so far that carries out such systematic comparisons of the tensor effects among these three types of the most successful nuclear DFT.
On the non-relativistic side, the effective interactions GT2 \cite{Otsuka2006Phys.Rev.Lett.97.162501} with tensor and its counterpart GT2$_{\text{nT}}$ without tensor were used for the Gogny calculations, and the SLy5 without tensor and its counterpart SLy5$_{\text{wT}}$ \cite{Colo2007Phys.Lett.B646.227} with tensor were used for the Skyrme calculations.
On the relativistic side, however, the results by PKA1 \cite{Long2007Phys.Rev.C76.034314} with tensor were compared to the results by a very different effective interaction DD-ME2 \cite{Lalazissis2005Phys.Rev.C71.024312} without tensor (and even without the Fock terms).
In principle, one should perform similar calculations as in the Skyrme and Gogny cases, where the tensor forces are switched on and off without changing the rest of the interaction.
Nevertheless, as mentioned in Ref.~\cite{Moreno-Torres2010Phys.Rev.C81.064327}, an explicit evaluation of the tensor effects in the relativistic framework was very difficult at that time.
Simply setting $f_\pi = f_\rho = 0$ would lead to huge changes also in the central part of the mean field and in most cases the mean-field calculations would not even converge.

Now, with the newly-developed formalism in this work, we can finally make a quantitative analysis of the tensor effects on the shell evolution in the relativistic framework.
Let us re-examine the evolution of the magic gaps along the $Z,\,N = 8,\,20$, and $28$ isotopes and isotones.

Following the procedure in Ref.~\cite{Moreno-Torres2010Phys.Rev.C81.064327}, the theoretical gaps are calculated as the differences of the HF single-particle energies.
The empirical gaps are approximately evaluated via the nuclear mass as adopted in Review~\cite{Sorlin2008Prog.Part.Nucl.Phys.61.602}:
For the proton gaps at $Z_{\text{mag.}}$, we calculate the single-particle energies of the last occupied and the first unoccupied orbitals, $e_b$ and $e_a$, as
\begin{subequations}
\begin{align}
  e_b(Z_{\text{mag.}}, N) &=\, E(Z_{\text{mag.}}, N) - E(Z_{\text{mag.}}-1, N),\\
  e_a(Z_{\text{mag.}}, N) &=\, E(Z_{\text{mag.}}+1, N) - E(Z_{\text{mag.}}, N).
\end{align}
\end{subequations}
Note that here $E$ are the total energies.
The energy of the magic gap is then evaluated as
\begin{equation}
  E_{\text{gap}}(Z_{\text{mag.}}, N) =\, e_a - e_b.
\end{equation}
The same procedure is followed for the evaluation of the empirical neutron magic gaps $E_{\text{gap}}(Z, N_{\text{mag.}})$.
All the experimental masses are taken from AME2016 \cite{Wang2017Chin.Phys.C41.030003}.

In the Appendix of Ref.~\cite{Sorlin2008Prog.Part.Nucl.Phys.61.602}, some warnings were provided about the use of this approximation to evaluate the empirical single-particle energies.
The separation energies are supposed to be similar to the single-particle energies only if one assumes that the proton or neutron magic core remains almost unchanged when one nucleon is added to or removed from it, which will be contaminated by various beyond-mean-field effects.
In particular, for the $N=Z$ nuclei, an extra beyond-mean-field correlation has been pointed out, which has led to intensive discussions on the so-called Wigner term in nuclear mass models.
We use a prescription introduced in the Skyrme Hartree-Fock-Bogoliubov mass model \cite{Goriely2002Phys.Rev.C66.024326},
\begin{equation}
  E_W =\,V_W \exp \left[ -\lambda \left( \frac{N-Z}{A}\right)^2 \right],
\end{equation}
with $V_W = -2.327$~MeV and $\lambda = 400$.
The empirical gaps after taking into account these Wigner corrections will be shown in the following figures.
Once again, we stress that we consider here the empirical values just as indications to provide qualitative (and not
precise) empirical trends to compare with the theoretical results.
See more relevant discussions in Ref.~\cite{Moreno-Torres2010Phys.Rev.C81.064327}.

\subsubsection{$Z=8$ isotopes and $N=8$ isotones}

The proton gap $Z=8$ is determined by the difference of the HF single-particle energies between the proton $1d_{5/2}$ and $1p_{1/2}$ states, which belong to the spin-up $j_>$ state of the $1d$ spin doublets and the spin-down $j_<$ state of the $1p$ spin doublets, respectively.
Going from $^{16}$O to $^{22}$O, the spin-up $j'_>$ neutron state $1d_{5/2}$ is occupied, and thus the proton $1d_{5/2}$ ($1p_{1/2}$) state is pushed upward (downward) by the $\pi$-PV tensor force \cite{Otsuka2005Phys.Rev.Lett.95.232502}.
As a result, the tensor effect is expected to enhance the $Z=8$ gap.

\begin{figure}
\includegraphics[width=0.45\textwidth]{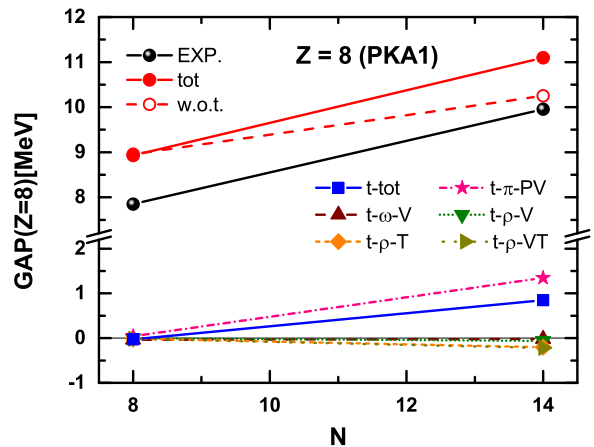}% Here is how to import EPS art
\caption{(Color online) Proton gap $Z=8$ in the O isotopes and the contributions from the tensor force in each meson-nucleon coupling as a function of neutron number $N$, calculated by the RHF theory with the PKA1 effective interaction.
The results with and without the tensor contributions are shown with filled and open circles, respectively.
The total tensor contributions are shown with filled squares, while the contributions from each coupling are denoted with different symbols.
See the text for the details of the empirical values.}
\label{Fig:gap-Z=8}
\end{figure}

We calculate the proton gap $Z=8$ in $^{16}$O and $^{22}$O by the RHF theory with the PKA1 effective interaction, and also separate the tensor effects generated by each meson-nucleon coupling through the tensor contributions to the corresponding single-particle energies.
The results are shown in Fig.~\ref{Fig:gap-Z=8} as a function of the neutron number $N$, and the corresponding empirical values are also given for an qualitative comparison.
Note that, same as in Ref.~\cite{Moreno-Torres2010Phys.Rev.C81.064327}, only several selected sub-shell-closure nuclei are investigated without pairing correlation, and thus the lines in the figures are plotted only to show the trends from one nucleus to the other more clearly.
From $^{16}$O to $^{22}$O, the $Z=8$ gap calculated by PKA1 increases by around $2$~MeV, which is in a nice agreement with the empirical trend.
Comparing the results with and without tensor, it is seen that the tensor force produces an enhancement of about $1$~MeV, which is also in agreement with the mechanism in Ref.~\cite{Otsuka2005Phys.Rev.Lett.95.232502}.

In Ref.~\cite{Moreno-Torres2010Phys.Rev.C81.064327}, it was shown that the results by PKA1 with tensor and those by DD-ME2 without tensor give less difference on the gap evolution.
Nevertheless, PKA1 is used within the RHF scheme, whereas DD-ME2 is used within the RMF scheme.
Their differences not only lie in the tensor interactions but also exist in all other effects coming from the Fock terms, such as the central, two-body spin-orbit interactions, etc.
Therefore, such a comparison cannot give us a clean conclusion about to what extent the tensor force in the relativistic framework influences the gap evolution.
But now, with the present newly-developed formalism, we can eventually identify the properties of the tensor force embraced in the effective interaction PKA1.
Due to such tensor properties, the tendency of gap evolution coincides with the empirical trend.

Let us look into the details of each meson-nucleon coupling.
As shown in Fig.~\ref{Fig:gap-Z=8}, for $^{16}$O with $Z=N=8$, all the couplings give almost vanishing tensor contributions, because both neutrons and protons are spin saturated, in which case the tensor contributions from all the states are basically canceled out by those from their spin partners.
The same feature was also seen in the Skyrme calculations with SLy5 and SLy5$_{\text{wT}}$ and the Gogny calculations with GT2$_{\text{nT}}$ and GT2 \cite{Moreno-Torres2010Phys.Rev.C81.064327}.
Going from $^{16}$O to $^{22}$O, the tensor contribution from the $\pi$-PV coupling increases by about $1.3$~MeV, and the  tensor contributions from the $\rho$-T and $\rho$-VT couplings compromise the $\pi$-PV one by around $0.2$~MeV each.
In contrast, there is no tensor contribution to the $Z=8$ gap from the $\omega\text{-V}$ coupling since the isoscalar $\omega$ meson cannot mediate the interaction between neutrons and protons.
The tensor contribution from the $\rho\text{-V}$ coupling is negligible mainly due to its small coupling strength.

\begin{figure}
\includegraphics[width=0.45\textwidth]{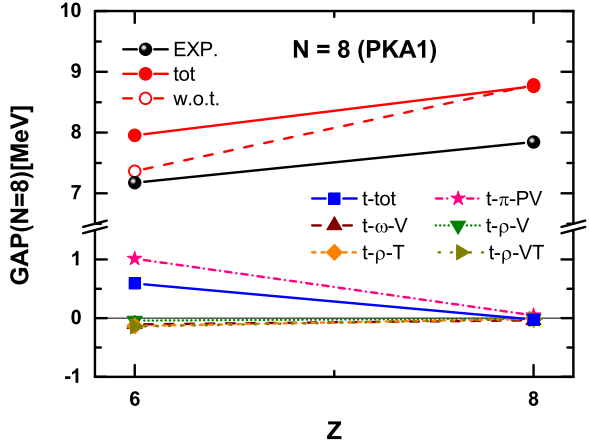}% Here is how to import EPS art
\caption{(Color online) Same as Fig.~\ref{Fig:gap-Z=8}, but for the neutron gap $N=8$ in the $N=8$ isotones as a function of proton number $Z$.}
\label{Fig:gap-N=8}
\end{figure}

The neutron gap $N=8$ is determined by the energy difference between the neutron spin-up $1d_{5/2}$ and spin-down $1p_{1/2}$ states.
The evolution of the $N=8$ gap from $^{14}$C to $^{16}$O is determined by the occupation of proton spin-down orbital $1p_{1/2}$.
As shown in Fig.~\ref{Fig:gap-N=8}, the net tensor effect of PKA1 decreases the $N=8$ gap from $^{12}$C to $^{16}$O by about $0.6$~MeV, and the empirical trend is reproduced well with this reduction.
Similar to the case of the proton gap $Z=8$, for the neutron gap $N=8$, the tensor contribution of the $\pi\text{-PV}$ coupling dominates over all other couplings, but partly canceled by those of the $\rho\text{-T}$ and $\rho\text{-VT}$ couplings.

Finally, it is interesting to point out that here the net tensor effect on the $Z=8$ gap from $^{16}$O to $^{22}$O is about $1$~MeV by PKA1 in the RHF scheme.
In contrast, this tensor effect reaches around $2.5$~MeV by SLy5$_{\text{wT}}$ and even around $4$~MeV by GT2 in the Skyrme and Gogny theories, respectively \cite{Moreno-Torres2010Phys.Rev.C81.064327}.
From $^{14}$C to $^{16}$O, the net tensor effect of PKA1 decreases the $N=8$ gap by about $0.6$~MeV, while the corresponding values are about $1.2$~MeV and $3$~MeV in the Skyrme SLy5$_{\text{wT}}$ and Gogny GT2 calculations, respectively.
This may imply that the $\pi$-PV coupling in PKA1 is somewhat too weak, which can be kept in mind for the future developments of the relativistic energy density functionals.

\subsubsection{$Z=20$ isotopes and $N=20$ isotones}

\begin{figure}
\includegraphics[width=0.45\textwidth]{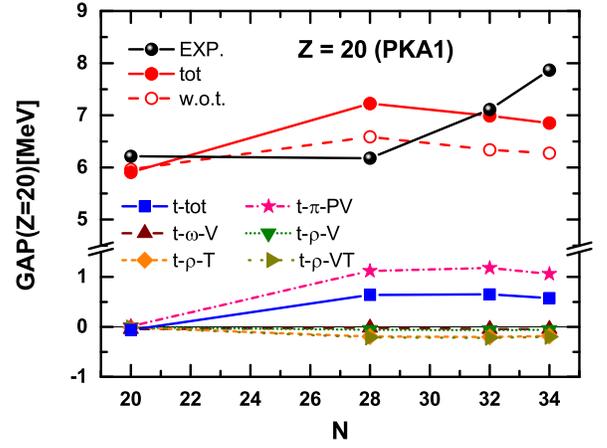}% Here is how to import EPS art
\caption{(Color online) Same as Fig.~\ref{Fig:gap-Z=8}, but for the proton gap $Z=20$ in the Ca isotopes.}\label{Fig:gap-Z=20}
\end{figure}

To clarify the effects of the tensor force on the proton gap $Z=20$, we calculate the Ca isotopes $^{40}\text{Ca}$, $^{48}\text{Ca}$, $^{52}\text{Ca}$, and $^{54}\text{Ca}$.
According to our calculations, the $Z=20$ gap in these isotopes are all determined by the single-particle energies of the spin-up $1f_{7/2}$ and spin-down $1d_{3/2}$ states.
In $^{48}\text{Ca}$, the neutron orbital $1f_{7/2}$ is fully occupied, and the $Z=20$ gap is expected to be enhanced by the tensor effects comparing with $^{40}\text{Ca}$.
As shown in Fig.~\ref{Fig:gap-Z=20}, our calculation gives an enhanced gap and the total tensor contribution also increases the gap.
In $^{52}\text{Ca}$, another neutron spin-up orbital, $2p_{3/2}$, is occupied, and thus it is expected to further enhance the $Z=20$ gap.
Our calculations show such an enhancement but the slope is very small.
From $^{52}\text{Ca}$ to $^{54}\text{Ca}$, the neutron spin-down orbital $2p_{1/2}$ is occupied and it is expected to weaken the $Z=20$ gap, which is consistent with the present calculated results.
Decomposed into each coupling, same as the cases of $Z=8$ and $N=8$, the tensor contribution of the $\pi\text{-PV}$ coupling is dominant and partially canceled by those of the other couplings, especially the $\rho\text{-T}$ and $\rho\text{-VT}$ ones.

\begin{figure}
\includegraphics[width=0.45\textwidth]{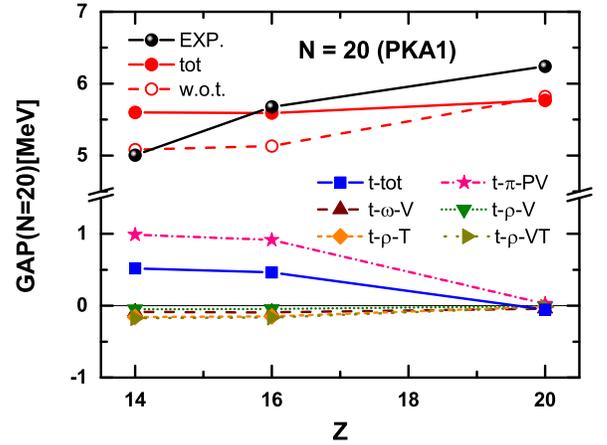}% Here is how to import EPS art
\caption{(Color online) Same as Fig.~\ref{Fig:gap-N=8}, but for the neutron gap $N=20$ in the $N=20$ isotones.}\label{Fig:gap-N=20}
\end{figure}

For the neutron gap $N=20$, we perform the RHF calculations for $^{34}\text{Si}$, $^{36}\text{S}$, and $^{40}\text{Ca}$, and show the corresponding results in Fig.~\ref{Fig:gap-N=20}.
The neutron gap $N=20$ in these nuclei are also determined by the $1f_{7/2}$ and $1d_{3/2}$ states.
From $^{34}\text{Si}$ to $^{36}\text{S}$, the $N=20$ gap keeps almost constant, and the tensor force does not present any remarkable effect.
This is because the two protons occupy only the $2s_{1/2}$ state and the $s$ orbitals give no tensor contribution \cite{Otsuka2005Phys.Rev.Lett.95.232502}.
From $^{36}\text{S}$ to $^{40}\text{Ca}$, the net tensor effect decreases the $N=20$ gap by around $0.5$~MeV as the protons
occupy the $1d_{3/2}$ state.
Finally at $^{40}\text{Ca}$, all the tensor contributions are basically vanishing because both neutrons and protons are spin saturated.

It is seen in Figs.~\ref{Fig:gap-Z=20} and \ref{Fig:gap-N=20} that, on the one hand, the present results show quite different behaviors from the empirical trend on the gap evolutions.
On the other hand, the present tensor effects coincide with every details of those by the Skyrme SLy5$_{\text{wT}}$ calculations \cite{Moreno-Torres2010Phys.Rev.C81.064327}, although the amplitude is somewhat smaller.

\subsubsection{$Z=28$ isotopes and $N=28$ isotones}

\begin{figure}
\includegraphics[width=0.45\textwidth]{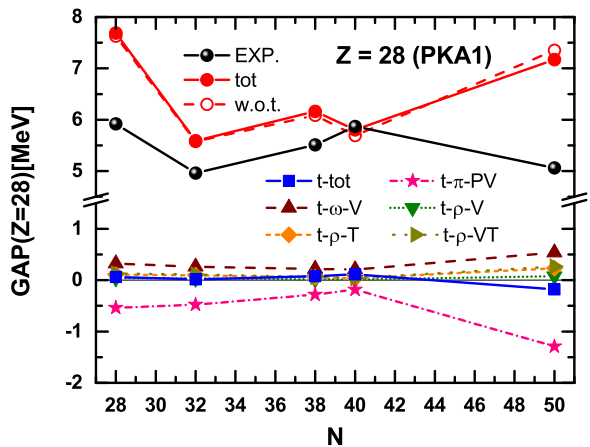}% Here is how to import EPS art
\caption{(Color online) Same as Fig.~\ref{Fig:gap-Z=8}, but for the proton gap $Z=28$ in the Ni isotopes.}\label{Fig:gap-Z=28}
\end{figure}

For the proton gap $Z=28$, we perform the RHF calculations for the Ni isotopes $^{56}\text{Ni}$, $^{60}\text{Ni}$, $^{66}\text{Ni}$, $^{68}\text{Ni}$, and $^{78}\text{Ni}$.
The $Z=28$ gap is determined by the proton $2p_{3/2}$ and $1f_{7/2}$ states from $^{56}\text{Ni}$ up to $^{68}\text{Ni}$, but by the $1f_{5/2}$ and $1f_{7/2}$ states for $^{78}\text{Ni}$.
The corresponding results are shown in Fig.~\ref{Fig:gap-Z=28}.
It is seen that the empirical trend of the gap evolution is followed from $^{56}\text{Ni}$ to $^{66}\text{Ni}$ but not further.
For the net tensor effect, it is noted that the $2p_{3/2}$ and $1f_{7/2}$ states are both $j_>$ states, and thus the tensor interactions act for the two states in the same direction.
As a result, the net tensor effect on the gap evolution is not profound at all up to $^{68}\text{Ni}$.
The $Z=28$ gap is then determined by the $1f$ spin doublets in $^{78}\text{Ni}$.
As a result, a visible but not large tensor effect is seen from $^{68}\text{Ni}$ to $^{78}\text{Ni}$.

Another important point is that the protons in the Ni isotopes are not spin saturated.
This makes it possible for the $\omega\text{-V}$ coupling to present considerable tensor contribution to the $Z=28$ gap.
As seen in Fig.~\ref{Fig:gap-Z=28}, its contributions are up to around $0.5$~MeV, which is comparable with those from the $\pi\text{-PV}$ coupling.
Nevertheless, because of the isoscalar nature of the $\omega\text{-V}$ coupling, its tensor contributions remain almost the same along the isotopes with respect to the change of the neutron number, as long as the single-particle configurations remain the same, i.e., from $^{56}\text{Ni}$ to $^{68}\text{Ni}$.

\begin{figure}
\includegraphics[width=0.45\textwidth]{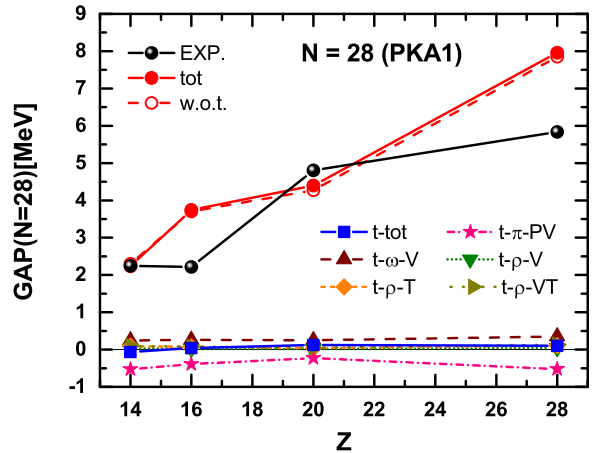}% Here is how to import EPS art
\caption{(Color online) Same as Fig.~\ref{Fig:gap-N=8}, but for the neutron gap $N=28$ in the $N=28$ isotones.}\label{Fig:gap-N=28}
\end{figure}

For the neutron gap $N=28$, we perform the RHF calculations for the isotones $^{42}\text{Si}$, $^{44}\text{S}$, $^{48}\text{Ca}$, and $^{56}\text{Ni}$.
The $N=28$ gap is determined by the neutron $2p_{3/2}$ and $1f_{7/2}$ orbitals for these considered nuclei.
The corresponding results are shown in Fig.~\ref{Fig:gap-N=28}.
It is interesting to see that although the overall increasing trend from $^{42}\text{Si}$ to $^{56}\text{Ni}$ can be reproduced, the detailed evolution at each sub-shell closure shows difference.
Nevertheless, this is not because of the tensor effect.
Since both the $2p_{3/2}$ and $1f_{7/2}$ states are the spin-up states, the tensor interactions act for the two states in the same direction, and thus the net tensor effects are almost invisible.

It is also interesting to point out that for the cases of the $Z=28$ and $N=28$ gaps, the tensor effects in the present results are substantially different from those by the Skyrme SLy5$_{\text{wT}}$ and Gogny GT2 calculations \cite{Moreno-Torres2010Phys.Rev.C81.064327}.
But one conclusion is in common: The $Z=28$ and $N=28$ gaps do not seem suitable for identifying the tensor effects.

\section{Summary and Perspectives}\label{Sec:summary}

We have identified the tensor force up to the $1/M^2$ order in each meson-nucleon coupling in the RHF theory, by the non-relativistic reduction for the relativistic two-body interactions.
It is found that all the couplings, except the $\sigma${-S} one, give rise to the tensor force, which is in agreement with the realistic Bonn nucleon-nucleon interactions in the one-boson-exchange picture.
The sum rule of the two-body matrix elements of tensor force has been also verified.

On the one hand, taking the nuclei $^{48}$Ca and $^{208}$Pb as examples, we have found that the tensor contributions to nuclear binding energies are in general tiny.
The tensor contribution from the $\pi$-PV coupling dominates and makes nuclei less bound, whereas all the other meson-nucleon couplings give opposite contributions.
In particular, with the effective interaction PKA1, not only the $\rho$-T but also $\omega$-V and $\rho$-VT couplings show substantial tensor contributions, and these contributions largely cancel out the $\pi$-PV one.
As a result, it is very difficult to determine the proper strengths of tensor force by fitting to the nuclear masses.

On the other hand, taking the isotopes and isotones $Z,\,N= 8,\,20$, and $28$ as examples, we have found that the tensor contributions to the evolutions of the magic gaps are much more profound.
Similar to the case of binding energy, here the $\pi$-PV tensor contribution is dominant and partially canceled by the $\rho$-T and $\rho$-VT ones.
The $\omega$-V coupling does not participate in the proton-neutron channel.
With the newly-developed formalism in this work, we are eventually able to make fair and quantitative comparisons with the corresponding results with and without tensor in the non-relativistic Skyrme and Gogny calculations.
The present results show the same conclusions by the non-relativistic theories in Ref.~\cite{Moreno-Torres2010Phys.Rev.C81.064327} that the $Z,\,N = 8$ and $20$ gaps are the candidates for constraining the tensor strengths, but the $Z,\,N = 28$ gaps are not.
Moreover, it is found that the total tensor effect in the effective interaction PKA1 is weaker than those in the Skyrme SLy5$_{\text{wT}}$ and Gogny GT2 effective interactions.

With the present formalism, we are able to further quantitatively evaluate the tensor contributions, from each meson-nucleon coupling in the relativistic framework, to a variety of nuclear ground-state and excited-state properties.
Those properties sensitive to the tensor force can be selected, and then they can serve as efficient constraints for the strengths of the tensor force in return.
In the non-relativistic framework, the sensitivities of the tensor force in the excitation energies of the $0^-$ states \cite{Anguiano2011Phys.Rev.C83.064306}, the electric and magnetic multipole responses \cite{Cao2009Phys.Rev.C80.064304}, the Gamow-Teller \cite{Bai2009Phys.Lett.B675.28} and spin-dipole \cite{Bai2010Phys.Rev.Lett.105.072501} resonances, and the $\beta$-decay half-lives \cite{Minato2013Phys.Rev.Lett.110.122501} have been investigated.
Very recently, \textit{ab initio} relativistic Brueckner-Hartree-Fock calculations \cite{Shen2016Chin.Phys.Lett.33.102103,Shen2017Phys.Rev.C96.014316, Shen2018Phys.Lett.B781.227} based on the realistic nucleon-nucleon interactions showed a systematic and specific pattern in the evolution of spin-orbit splittings in neutron drops \cite{SHEN2018Phys.Lett.B778.344, Shen2018Phys.Rev.C97.054312}.
It was also shown that the tensor force plays a critical role in reproducing this pattern, and the tensor force in the existing effective interactions in the RHF theory seems not strong enough.
In addition, it is found that the form factors in the meson-nucleon couplings play important roles in the RHF theory \cite{Hu2010Phys.Lett.B687.271,  Hu2010Eur.Phys.J.A43.323}.
All these aspects will promote the developments of the nuclear density functional theory in the near future.

\begin{acknowledgments}
The authors are grateful to Dr.~Li Juan Jiang and Dr.~Jia Jie Li for the helpful discussions.
This work was partially supported by the Natural Science Foundation of China under Grant No.~11675065,
the JSPS Grant-in-Aid for Early-Career Scientists under Grant No.~18K13549,
and the JSPS-NSFC Bilateral Program for Joint Research Project on Nuclear mass and life for unravelling mysteries of the r-process.
Z.W. acknowledges the scholarship of China Scholarship Council,
and H.L. thanks the RIKEN iTHEMS program.
Z.W. is also grateful to Professor Takashi Nakatsukasa for help and support during his studying in the University of Tsukuba as a joint Ph.D. student.
\end{acknowledgments}

\appendix

\begin{widetext}
\section{Details of non-relativistic reduction in the zero-density limit}\label{App:Reduction}

In this Appendix, we will show the detailed derivations for the non-relativistic reduced two-body interactions $\hat{\mathcal{V}}_{0,\phi}$ in the zero-density limit, and, in particular, identify their tensor components $\hat{\mathcal{V}}^{\text{t}}_{0,\phi}$ for each meson-nucleon coupling.

First of all, the interaction vertices used in Eq.~\eqref{eq:Vphi_abcd} read
\begin{subequations}\label{vertexp}
  \begin{align}
    \Gamma_{\sigma\text{-S}}(1,2) = &\,-[g_\sigma]_1[g_\sigma]_2,\\
    \Gamma_{\omega\text{-V}}(1,2) = &\,[g_\omega\gamma_\mu]_1 [g_\omega\gamma^\mu]_2,\\
    \Gamma_{\rho\text{-V}}(1,2) = &\,[g_\rho\gamma_\mu\vec{\tau}]_1 \cdot [g_\rho\gamma^\mu\vec{\tau}]_2,\\
    \Gamma_{\rho\text{-T}}(1,2) = &\,\frac{1}{4M^2}[f_\rho q^i\sigma_{\mu i}\vec{\tau}]_1
                                    \cdot [f_\rho q_j  \sigma^{\mu j}\vec{\tau}]_2,\\
    \Gamma_{\rho\text{-VT}}(1,2) = &\,\frac{i}{2M}[f_\rho\sigma^{\mu i }q_i\vec{\tau}]_1\cdot[ g_\rho\gamma_{\mu}\vec{\tau}]_2 -(1 \leftrightarrow 2),\\
    \Gamma_{\pi\text{-PV}}(1,2) = &\,-\left[\frac{f_{\pi}}{m_\pi}\gamma^i{q}_i\gamma_5\vec{\tau}\right]_1 \cdot
                                       \left[\frac{f_{\pi}}{m_\pi}\gamma^j{q}_j\gamma_5\vec{\tau}\right]_2.
  \end{align}
\end{subequations}
They are obtained by applying the interaction vertices in Eq.~\eqref{eq:vertex} on the plane waves \eqref{eq:Plane} and neglecting the retardation effect.

From Eqs.~\eqref{eq:Plane} and \eqref{eq:upa}, the upper components of the plane waves in the zero-density limit read
\begin{equation}\label{eq:upan}
  {\xi}_{\bm{p}_a}(\bm{r}) =\,\mathcal{C}_{p_a} \chi_\frac{1}{2}(s_a)\chi_\frac{1}{2}(\tau_a) \, e^{i \bm p_a \cdot \bm r},
\end{equation}
with $\mathcal{C}_{p_a} \equiv \sqrt{(M+\varepsilon_a)/(2\varepsilon_a)}$.

Inserting the interaction vertices~\eqref{vertexp} into Eq.~\eqref{eq:Vphi_abcd}, and keeping in mind that the upper components~\eqref{eq:upan} serve as the bra and ket in the r.h.s. of Eq.~\eqref{eq:NRdef}, one can obtain the non-relativistic reduced interactions $\mathcal{\hat V}_{0,\phi}$ for each meson-nucleon coupling.
Here we will give the key steps of derivations, and expand the $\mathcal{\hat V}_{0,\phi}$ up to the $1/M^2$ order.
We define $\bm{q} \equiv \bm{p}_a - \bm{p}_c = \bm{p}_d - \bm{p}_b$,  $\bm k \equiv(\bm p_a+\bm p_c)/2 $, and $\bm k'\equiv (\bm p_b+\bm p_d)/2$.
For simplicity, the spin and isospin spinors as well as the isospin operator $\vec\tau(1) \cdot \vec\tau(2)$ for the isovector mesons are not shown explicitly here.

For the $\pi\text{-PV}$ coupling,
\begin{align}
  & -\frac{f^2_{\pi}}{m^2_\pi}\frac{1}{m_\pi^2+\bm{q}^2} (\mathcal{C}_{p_a}\mathcal{C}_{p_b}\mathcal{C}_{p_c}\mathcal{C}_{p_d})^{-1} [\bar u_a\bm\gamma\cdot \bm q\gamma_5 u_c]_1
  [\bar u_b\bm\gamma \cdot\bm q\gamma_5 u_d]_2\nonumber\\
=\,&-\frac{f^2_{\pi}}{m^2_\pi}\frac{1}{m_\pi^2+\bm{q}^2}
 \left[
 \left(
     \begin{array}{cc}
     1 & -\frac{\bm{\sigma}\cdot\bm{p}_a}{2M}  \\
     \end{array}
   \right)
   \left(
   \begin{array}{cc}
    \bm{\sigma}\cdot \bm q  & 0\\
    0 & -\bm{\sigma}\cdot \bm q\\
    \end{array}
   \right)
    \left(
   \begin{array}{c}
    1 \\
    \frac{\bm{\sigma}\cdot\bm{p}_c}{2M} \\
     \end{array}
  \right)
  \right]_1
  \left[
   \left(
     \begin{array}{cc}
     1 & -\frac{\bm{\sigma}\cdot\bm{p}_b}{2M}  \\
     \end{array}
   \right)
   \left(
   \begin{array}{cc}
    \bm{\sigma}\cdot \bm q  & 0\\
    0 & -\bm{\sigma}\cdot \bm q\\
    \end{array}
   \right)
   \left(
   \begin{array}{c}
    1 \\
    \frac{\bm{\sigma}\cdot\bm{p}_d}{2M} \\
     \end{array}
   \right)
   \right]_2\nonumber\\
=\,&-\frac{f^2_{\pi}}{m^2_\pi}\frac{1}{m_\pi^2+\bm{q}^2}(\bm{\sigma}_1\cdot \bm q)(\bm{\sigma}_2\cdot \bm q).
\end{align}
In the one-boson-exchange picture \cite{Machleidt1989Adv.Nucl.Phys.19.189}, ${f^2_{\pi}}/{m^2_\pi} = {g^2_{\pi}}/{4M^2}$, i.e., this order is regarded as $O(1/M^2)$.

For the $\sigma\text{-S}$ coupling,
\begin{align}
  & -g_\sigma^2\frac{1}{m_\sigma^2+\bm{q}^2} (\mathcal{C}_{p_a}\mathcal{C}_{p_b}\mathcal{C}_{p_c}\mathcal{C}_{p_d})^{-1} [\bar u_au_c]_1[\bar u_bu_d]_2\nonumber\\
=\,&-g_\sigma^2\frac{1}{m_\sigma^2+\bm{q}^2}
   \left[
   \left(
     \begin{array}{cc}
     1 & - \frac{\bm{\sigma}\cdot\bm{p}_a}{2M}  \\
     \end{array}
   \right)
   \left(
   \begin{array}{c}
    1 \\
    \frac{\bm{\sigma}\cdot\bm{p}_c}{2M} \\
     \end{array}
  \right)
  \right]_1
  \left[
   \left(
     \begin{array}{cc}
     1 & - \frac{\bm{\sigma}\cdot\bm{p}_b}{2M}  \\
     \end{array}
   \right)
   \left(
   \begin{array}{c}
    1 \\
    \frac{\bm{\sigma}\cdot\bm{p}_d}{2M} \\
     \end{array}
  \right)
  \right]_2
  \nonumber\\
=\,&-g_\sigma^2\frac{1}{m_\sigma^2+\bm{q}^2}
   \left[1-\frac{(\bm{\sigma}\cdot{\frac{2\bm k + \bm q}{2}})(\bm{\sigma}\cdot{\frac{2\bm k - \bm q}{2}})}{4M^2}  \right]_1
   \left[1-\frac{(\bm{\sigma}\cdot{\frac{2\bm k' - \bm q}{2}})(\bm{\sigma}\cdot{\frac{2 \bm k' + \bm q}{2}})}{4M^2}\right]_2\nonumber\\
=\,&-g_\sigma^2\frac{1}{m_\sigma^2+\bm{q}^2}
   \left[1-\frac{\bm k^2+{\bm {k}}'^2-i\bm \sigma_1\cdot(\bm k \times \bm q)-i\bm \sigma_2\cdot(\bm q \times \bm k')}{4M^2}
   +\frac{\bm q^2}{8M^2} \right].
\end{align}
It should be noticed that there is no tensor component up to this order.

For the $\omega\text{-V}$ coupling, its time component is similar to the $\sigma\text{-S}$ coupling, which reads
\begin{align}
  &g_\omega^2\frac{1}{m_\omega^2+\bm{q}^2} (\mathcal{C}_{p_a}\mathcal{C}_{p_b}\mathcal{C}_{p_c}\mathcal{C}_{p_d})^{-1} [\bar u_a\gamma^0 u_c]_1[\bar u_b\gamma_0 u_d]_2\nonumber\\
=\,&g_\omega^2\frac{1}{m_\omega^2+\bm{q}^2}
   \left[1+\frac{\bm k^2+{\bm {k}}'^2-i\bm \sigma_1\cdot(\bm k \times \bm q)-i\bm \sigma_2\cdot(\bm q \times \bm k')}{4M^2}
   -\frac{\bm q^2}{8M^2} \right].
\end{align}
Its space component is as following,
\begin{align}
 & -g_\omega^2\frac{1}{m_\omega^2+\bm{q}^2} (\mathcal{C}_{p_a}\mathcal{C}_{p_b}\mathcal{C}_{p_c}\mathcal{C}_{p_d})^{-1} [\bar u_a\bm\gamma u_c]_1\cdot[\bar u_b\bm\gamma u_d]_2\nonumber\\\nonumber
=\,&-g_\omega^2\frac{1}{m_\omega^2+\bm{q}^2}
  \left[
   \left(
     \begin{array}{cc}
     1 & -\frac{\bm{\sigma}\cdot\bm{p}_a}{2M}  \\
     \end{array}
   \right)
   \left(
   \begin{array}{cc}
    0 & \bm{\sigma} \\
    -\bm{\sigma} & 0 \\
    \end{array}
   \right)
   \left(
   \begin{array}{c}
    1 \\
    \frac{\bm{\sigma}\cdot\bm{p}_c}{2M} \\
     \end{array}
  \right)\right]_1
  \cdot
   \left[ \left(
     \begin{array}{cc}
     1 & -\frac{\bm{\sigma}\cdot\bm{p}_b}{2M}  \\
     \end{array}
   \right)
   \left(
   \begin{array}{cc}
    0 & \bm{\sigma} \\
    -\bm{\sigma} & 0 \\
    \end{array}
   \right)
   \left(
   \begin{array}{c}
    1 \\
    \frac{\bm{\sigma}\cdot\bm{p}_d}{2M} \\
     \end{array}
   \right)
  \right]_2\nonumber\\
=\,&-g_\omega^2\frac{1}{m_\omega^2+\bm{q}^2}
  \left[\frac{(\bm \sigma\cdot\bm{p}_a)\bm{\sigma}+\bm{\sigma}(\bm \sigma\cdot\bm{p}_c)}{2M}\right]_1\cdot
  \left[\frac{(\bm \sigma\cdot\bm{p}_b)\bm{\sigma}+\bm{\sigma}(\bm \sigma\cdot\bm{p}_d)}{2M}\right]_2\nonumber\\
=\,&-\frac{g_\omega^2}{4M^2}\frac{1}{m_\omega^2+\bm{q}^2}
  \left[4\bm {k}\cdot \bm{k}'-2i(\bm{q}\times\bm\sigma_1)\cdot\bm k'
  +2i\bm k\cdot(\bm{q}\times\bm\sigma_2)+(\bm\sigma_1\cdot\bm\sigma_2)\bm{q}^2-(\bm\sigma_1\cdot\bm{q})(\bm\sigma_2\cdot\bm{q})
  \right].
\end{align}

For the $\rho\text{-V}$ coupling, it is similar to $\omega\text{-V}$ coupling except for the isospin part.

For the $\rho\text{-T}$ coupling, with
\begin{align}
 q^i\sigma_{0 i} = \,q_j\sigma^{0 j}
=\,-i\left(
   \begin{array}{cc}
    0 & \bm q\cdot\bm{\sigma} \\
    \bm q\cdot\bm{\sigma} & 0 \\
    \end{array}
   \right),
\end{align}
we get its time component as
\begin{align}
%\begin{split}
  &\frac{f^2_\rho}{4M^2}\frac{1}{m_\rho^2+\bm{q}^2} (\mathcal{C}_{p_a}\mathcal{C}_{p_b}\mathcal{C}_{p_c}\mathcal{C}_{p_d})^{-1} [\bar u_aq^i\sigma_{0 i} u_c]_1
  [ \bar u_bq_j  \sigma^{0 j} u_d]_2\nonumber\\
=\,&-\frac{f^2_\rho}{4M^2}\frac{1}{m_\rho^2+\bm{q}^2}
  \left[
   \left(
     \begin{array}{cc}
     1 & -\frac{\bm{\sigma}\cdot\bm{p}_a}{2M}  \\
     \end{array}
   \right)
   \left(
   \begin{array}{cc}
    0 & \bm q\cdot\bm{\sigma} \\
    \bm q\cdot\bm{\sigma} & 0 \\
    \end{array}
   \right)
   \left(
   \begin{array}{c}
    1 \\
    \frac{\bm{\sigma}\cdot\bm{p}_c}{2M} \\
     \end{array}
   \right)
  \right]_1
  \left[
   \left(
     \begin{array}{cc}
     1 & -\frac{\bm{\sigma}\cdot\bm{p}_b}{2M}  \\
     \end{array}
   \right)
   \left(
   \begin{array}{cc}
    0 & \bm q\cdot\bm{\sigma} \\
    \bm q\cdot\bm{\sigma} & 0 \\
    \end{array}
   \right)
   \left(
   \begin{array}{c}
    1 \\
    \frac{\bm{\sigma}\cdot\bm{p}_d}{2M} \\
     \end{array}
   \right)
  \right]_2\nonumber\\
=\,& 0,
\end{align}
because the leading order is of $O(1/M^4)$ here.
With
\begin{align}
%\begin{split}
q^i\sigma_{k i} =\, - q_j\sigma^{k j}
=\,\left(
   \begin{array}{cc}
   \bm q\times\bm\sigma & 0 \\
    0 & \bm q\times\bm\sigma \\
    \end{array}
 \right)^k,
%%\end{split}
\end{align}
we get its space component as
\begin{align}
%\begin{split}
&\frac{f^2_\rho}{4M^2}\frac{1}{m_\rho^2+\bm{q}^2} (\mathcal{C}_{p_a}\mathcal{C}_{p_b}\mathcal{C}_{p_c}\mathcal{C}_{p_d})^{-1} [\bar u_aq^i\sigma_{k i} u_c]_1
  [ \bar u_bq_j  \sigma^{k j} u_d]_2\nonumber\\
=\,&-\frac{f^2_\rho }{4M^2}\frac{1}{m_\rho^2+\bm{q}^2}
  \left[
   \bar u_a
   \left(
   \begin{array}{cc}
   \bm q\times\bm\sigma & 0 \\
    0 & \bm q\times\bm\sigma \\
    \end{array}
   \right)
   u_c
   \right]_1\cdot
  \left[
  \bar u_b
   \left(
   \begin{array}{cc}
   \bm q\times\bm\sigma & 0 \\
    0 & \bm q\times\bm\sigma \\
    \end{array}
   \right)
   u_d
  \right]_2\nonumber\\
=\,&-\frac{f^2_\rho}{4M^2}\frac{1}{m_\rho^2+\bm{q}^2}
   \left[\bm q\times\bm{\sigma}  -\frac{(\bm{\sigma}\cdot\bm{p}_a)(\bm q\times\bm{\sigma})(\bm{\sigma}\cdot\bm{p}_c)}{4M^2}
   \right]_1\cdot
   \left[\bm q\times\bm{\sigma}-\frac{(\bm{\sigma}\cdot\bm{p}_b)(\bm q\times\bm{\sigma})(\bm{\sigma}\cdot\bm{p}_d)}{4M^2}
   \right]_2\nonumber\\
=\,&\frac{f^2_\rho}{4M^2}\frac{1}{m_\rho^2+\bm{q}^2}
  \left[(\bm{\sigma}_1\cdot\bm q)(\bm{\sigma}_2\cdot\bm q)-(\bm{\sigma}_1 \cdot\bm{\sigma}_2)\bm q^2  \right].
\end{align}
Following the derivations of $\rho\text{-V}$ and $\rho\text{-T}$ couplings, one can easily get the corresponding  two-body interaction matrix element of the time component of  $\rho\text{-VT}$ coupling,
\begin{align}%\label{exch-ome}
%\begin{split}
  &i\frac{f_\rho(1) g_\rho(2)}{2M}\frac{1}{m_\rho^2+\bm{q}^2} (\mathcal{C}_{p_a}\mathcal{C}_{p_b}\mathcal{C}_{p_c}\mathcal{C}_{p_d})^{-1} [\bar u_a\sigma^{0 i }q_i u_c]_1
  [\bar u_b  \gamma_{0} u_d]_2\nonumber\\
  &-i\frac{g_\rho(1) f_\rho(2)}{2M}\frac{1}{m_\rho^2+\bm{q}^2} (\mathcal{C}_{p_a}\mathcal{C}_{p_b}\mathcal{C}_{p_c}\mathcal{C}_{p_d})^{-1} [\bar u_a\gamma_{0} u_c]_1
  [\bar u_b  \sigma^{0 i }q_iu_d]_2\nonumber\\
=\,&\frac{f_\rho(1) g_\rho(2)}{4M^2}\frac{1}{m_\rho^2+\bm{q}^2}
  \left[-\bm{q}^2+2i\bm{\sigma}_1\cdot(\bm{q}\times\bm {k})\right]-\frac{g_\rho(1) f_\rho(2)}{4M^2}\frac{1}{m_\rho^2+\bm{q}^2}
  \left[\bm{q}^2+2i\bm{\sigma}_2\cdot(\bm{q}\times\bm {k}')\right],
\end{align}
and the corresponding  two-body interaction matrix element of the space component of  $\rho\text{-VT}$ coupling,
\begin{align}%\label{exch-ome}
 &i\frac{f_\rho(1) g_\rho(2)}{2M}\frac{1}{m_\rho^2+\bm{q}^2} (\mathcal{C}_{p_a}\mathcal{C}_{p_b}\mathcal{C}_{p_c}\mathcal{C}_{p_d})^{-1} [\bar u_a\sigma^{j i }q_i u_c]_1
  [\bar u_b  \gamma_{j} u_d]_2\nonumber\\
 &-i\frac{g_\rho(1) f_\rho(2)}{2M}\frac{1}{m_\rho^2+\bm{q}^2}
 (\mathcal{C}_{p_a}\mathcal{C}_{p_b}\mathcal{C}_{p_c}\mathcal{C}_{p_d})^{-1} [\bar u_a\gamma_{j} u_c]_1  [\bar u_b  \sigma^{j i }q_i u_d]_2\nonumber\\
=\,&i\frac{f_\rho(1) g_\rho(2)}{4M^2}\frac{1}{m_\rho^2+\bm{q}^2}
   \left(\bm q\times\bm{\sigma}\right)_1\cdot
   \left(2\bm k' +i\bm{q}\times\bm\sigma\right)_2-i\frac{g_\rho(1) f_\rho(2)}{4M^2}\frac{1}{m_\rho^2+\bm{q}^2}
  \left(2\bm k-i\bm{q}\times\bm\sigma\right)_1 \cdot\left(\bm q\times\bm{\sigma}\right)_2\nonumber\\
=\,&\frac{f_\rho(1) g_\rho(2)}{4M^2}\frac{1}{m_\rho^2+\bm{q}^2}
  \left[-2i\bm{\sigma}_1\cdot(\bm{q}\times\bm {k}') +(\bm{\sigma}_1\cdot\bm q)(\bm{\sigma}_2\cdot\bm q)
  -(\bm{\sigma}_1 \cdot\bm{\sigma}_2)\bm q^2 \right]\nonumber\\
  &+\frac{g_\rho(1) f_\rho(2)}{4M^2}\frac{1}{m_\rho^2+\bm{q}^2}
  \left[2i\bm{\sigma}_2\cdot(\bm{q}\times\bm {k})+( \bm{\sigma}_1\cdot\bm q)(\bm{\sigma}_2\cdot\bm q)
  -(\bm{\sigma}_1 \cdot\bm{\sigma}_2)\bm q^2 \right].%\nonumber\\
\end{align}

In short, up to the $1/M^2$ order, the non-relativistic reduced two-body interactions $\mathcal{\hat V}_{0,\phi}$ in the zero-density limit read
\begin{subequations}\label{Vabcd2}
\begin{align}
 \mathcal{\hat V}_{0,\sigma\text{-S}}
=\,&-g_\sigma(1)g_\sigma(2)\frac{1}{m_\sigma^2+\bm{q}^2}
   \left[1-\frac{\bm k^2+{\bm {k}}'^2-i\bm \sigma_1\cdot(\bm k \times \bm q)-i\bm \sigma_2\cdot(\bm q \times \bm k')}{4M^2}
   +\frac{\bm q^2}{8M^2} \right],\\
\mathcal{\hat V}_{0,\omega\text{-V}}
=\,&+g_\omega(1)g_\omega(2)\frac{1}{m_\omega^2+\bm{q}^2}
   \left[1+\frac{\bm k^2+{\bm {k}}'^2-i\bm \sigma_1\cdot(\bm k \times \bm q)-i\bm \sigma_2\cdot(\bm q \times \bm k')}{4M^2}
   -\frac{\bm q^2}{8M^2}   \right]\nonumber\\
  &-\frac{g_\omega(1)g_\omega(2)}{4M^2}\frac{1}{m_\omega^2+\bm{q}^2}
  \left[4\bm {k}\cdot \bm{k}'+2i\bm\sigma_1\cdot(\bm{q}\times\bm k')
  -2i\bm\sigma_2\cdot(\bm{q}\times\bm k)+\frac{2}{3}(\bm\sigma_1\cdot\bm\sigma_2)\bm{q}^2-S_{12}\right],\\
\mathcal{\hat V}_{0,\rho\text{-V}}
=\,&+\vec{\tau}(1)\cdot\vec{\tau}(2)g_\rho(1)g_\rho(2)\frac{1}{m_\rho^2+\bm{q}^2}
   \left[1+\frac{\bm k^2+{\bm {k}}'^2-i\bm \sigma_1\cdot(\bm k \times \bm q)-i\bm \sigma_2\cdot(\bm q \times \bm k')}{4M^2}
   -\frac{\bm q^2}{8M^2}   \right]\nonumber\\
  &-\vec{\tau}(1)\cdot\vec{\tau}(2)\frac{g_\rho(1)g_\rho(2)}{4M^2}\frac{1}{m_\rho^2+\bm{q}^2}
  \left[4\bm {k}\cdot \bm{k}'+2i\bm\sigma_1\cdot(\bm{q}\times\bm k')
  -2i\bm\sigma_2\cdot(\bm{q}\times\bm k)+\frac{2}{3}(\bm\sigma_1\cdot\bm\sigma_2)\bm{q}^2-S_{12}\right],\\
  \mathcal{\hat V}_{0,\pi\text{-PV}} =\,&-\vec{\tau}(1)\cdot\vec{\tau}(2)\frac{f_{\pi}(1)f_{\pi}(2)}{m^2_\pi}\frac{1}{m_\pi^2+\bm{q}^2}
  \left[ S_{12}+ \frac{1}{3}(\bm\sigma_1\cdot\bm\sigma_2)  \bm q^2\right],\\
  \mathcal{\hat V}_{0,\rho\text{-T}}
=\,&+\vec{\tau}(1)\cdot\vec{\tau}(2)\frac{f_\rho(1) f_\rho(2)}{4M^2}\frac{1}{m_\rho^2+\bm{q}^2}
  \left[  S_{12}-\frac{2}{3}(\bm\sigma_1\cdot\bm\sigma_2)\bm q^2\right],\\
  \mathcal{\hat V}_{0,\rho\text{-VT}}
=\,&+\vec{\tau}(1)\cdot\vec{\tau}(2)\frac{f_\rho(1) g_\rho(2)}{4M^2}\frac{1}{m_\rho^2+\bm{q}^2}
  \left[-\bm{q}^2+2i\bm{\sigma}_1\cdot(\bm{q}\times\bm {k})-2i\bm{\sigma}_1\cdot(\bm{q}\times\bm {k}')
  -\frac{2}{3}(\bm\sigma_1\cdot\bm\sigma_2)\bm q^2 +S_{12}\right]\nonumber\\
 &+\vec{\tau}(1)\cdot\vec{\tau}(2)\frac{g_\rho(1) f_\rho(2)}{4M^2}\frac{1}{m_\rho^2+\bm{q}^2}
  \left[-\bm{q}^2-2i\bm{\sigma}_2\cdot(\bm{q}\times\bm {k}')+2i\bm{\sigma}_2\cdot(\bm{q}\times\bm {k})
  -\frac{2}{3}(\bm\sigma_1\cdot\bm\sigma_2)\bm q^2 +S_{12}\right].
\end{align}
\end{subequations}
Note that by transfering these results to the center-of-mass frame ($\bm{k}' = -\bm{k}$), these expressions are consistent with Eqs.~(A17)--(A19) in Ref.~\cite{Machleidt1989Adv.Nucl.Phys.19.189} for the bare Bonn interactions.

\section{Non-relativistic reduction with finite density}\label{App:Reduction*}

In this Appendix, we will show the non-relativistic reduced two-body interactions $\hat{\mathcal{V}}_{\phi}$ in the case of finite density. The general strategy is quite similar to that used in Appendix~\ref{App:Reduction}, but starting with the Dirac spinor Eq.~\eqref{eq:upa*} with the starred quantities instead.

\begin{figure}
    \includegraphics[width=0.9\textwidth]{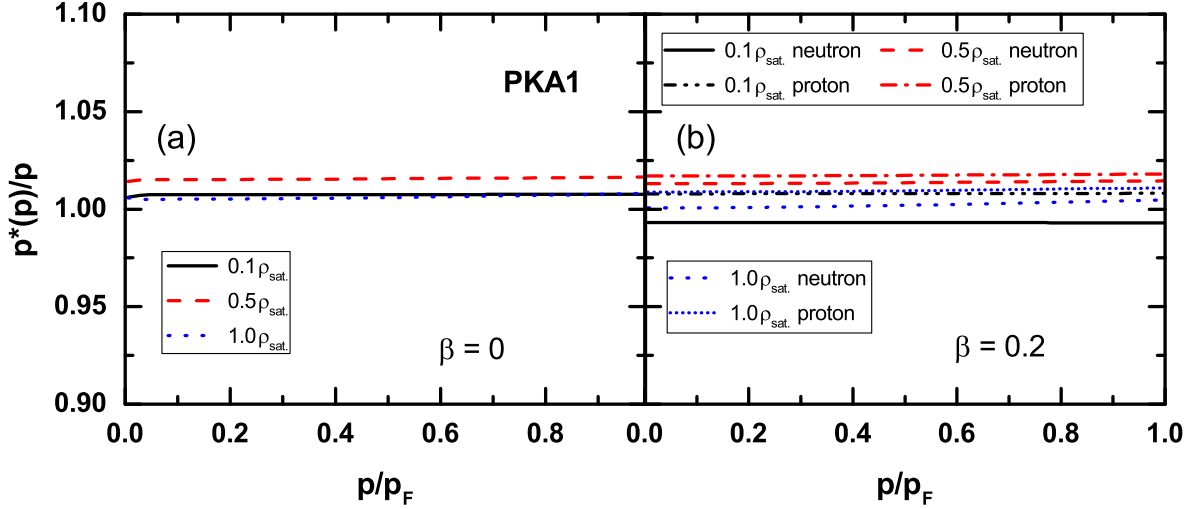}% Here is how to import EPS art
    \caption{Ratio between the starred momentum and its undressed counterpart $\bm p^*/\bm p$ as a function of $p/p_F$ for the (a) symmetric ($\beta=0$) and (b) asymmetric ($\beta=0.2$) nuclear matter with $\rho = 0.1\rho_{\rm sat.}$, $0.5\rho_{\rm sat.}$, and $\rho_{\rm sat.}$ calculated by RHF with PKA1.}
    \label{fig:ps}
\end{figure}

\begin{figure}
    \includegraphics[width=0.9\textwidth]{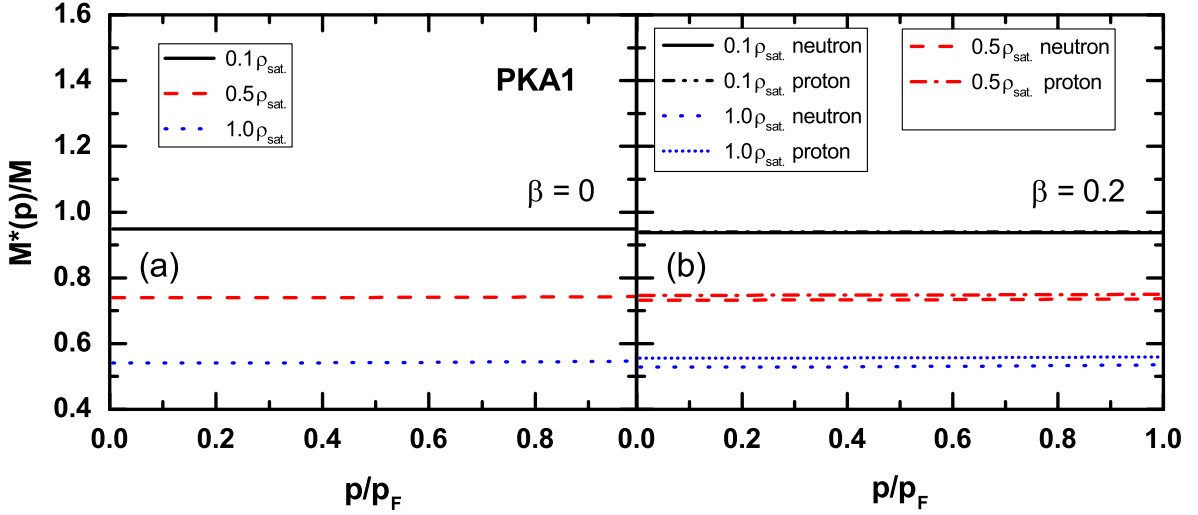}% Here is how to import EPS art
    \caption{Ratio between the Dirac mass and nucleon mass $M^*/M$ as a function of $p/p_F$ for the (a) symmetric ($\beta=0$) and (b) asymmetric ($\beta=0.2$) nuclear matter with $\rho = 0.1\rho_{\rm sat.}$, $0.5\rho_{\rm sat.}$, and $\rho_{\rm sat.}$ calculated by RHF with PKA1.}
    \label{fig:ms}
\end{figure}

During the derivations, we further replace the starred momentum $\bm p^*$ by the undressed one $\bm p$ and adopt the same Dirac mass $M^*$ for the two states at one vertex, as justified below.
The final results of $\hat{\mathcal{V}}_{\phi}$ up to the $1/M^2$ order read
\begin{subequations}\label{VALL}
\begin{align}
  \hat{\mathcal V}_{\sigma\text{-S}}
=\,&-g_\sigma(1)g_\sigma(2)\frac{1}{m_\sigma^2+\bm{q}^2}
   \left[1-\frac{1}{4}\frac{4\bm k^2-\bm q^2-4i\bm \sigma_1\cdot(\bm k \times \bm q)}{4M^*(1)M^*(1)}
   -\frac{1}{4}\frac{4 {\bm k'}^2-\bm q^2-4i\bm \sigma_2\cdot(\bm q \times \bm k')   }{4M^*(2)M^*(2)}\right],\\
\hat{\mathcal V}_{\omega\text{-V}}
=\,&+g_\omega(1)g_\omega(2)\frac{1}{m_\omega^2+\bm{q}^2}
   \left[1+\frac{1}{4}\frac{4\bm k^2-\bm q^2-4i\bm \sigma_1\cdot(\bm k \times \bm q)}{4M^*(1)M^*(1)}
   +\frac{1}{4}\frac{4{\bm k'}^2-\bm q^2-4i\bm \sigma_2\cdot(\bm q \times \bm k')}{4M^*(2)M^*(2)}\right]\nonumber\\
   &-\frac{g_\omega(1)g_\omega(2)}{4M^*(1)M^*(2)}\frac{1}{m_\omega^2+\bm{q}^2}
   \left[4\bm {k}\bm k'+2i\bm\sigma_1\cdot(\bm{q}\times\bm k')
   -2i\bm\sigma_2\cdot(\bm{q}\times\bm k)+\frac{2}{3}(\bm\sigma_1\cdot\bm\sigma_2)\bm{q}^2-S_{12}\right],\\
\hat{\mathcal V}_{\rho\text{-V}}
=\,&+\vec{\tau}(1)\cdot\vec{\tau}(2)g_\rho(1)g_\rho(2)\frac{1}{m_\rho^2+\bm{q}^2}
   \left[1+\frac{1}{4}\frac{4\bm k^2-\bm q^2-4i\bm \sigma_1\cdot(\bm k \times \bm q)}{4M^*(1)M^*(1)}
   +\frac{1}{4}\frac{4{\bm k'}^2-\bm q^2-4i\bm \sigma_2\cdot(\bm q \times \bm k')}{4M^*(2)M^*(2)}\right]\nonumber\\
   &-\vec{\tau}(1)\cdot\vec{\tau}(2)\frac{g_\rho(1)g_\rho(2)}{4M^*(1)M^*(2)}\frac{1}{m_\rho^2+\bm{q}^2}
   \left[4\bm {k}\bm k'+2i\bm\sigma_1\cdot(\bm{q}\times\bm k')
   -2i\bm\sigma_2\cdot(\bm{q}\times\bm k)+\frac{2}{3}(\bm\sigma_1\cdot\bm\sigma_2)\bm{q}^2-S_{12}\right],\\
  \hat{\mathcal V}_{\pi\text{-PV}}
 =\,&-\vec{\tau}(1)\cdot\vec{\tau}(2)\frac{f_{\pi}(1)f_{\pi}(2)}{m^2_\pi}\frac{1}{m_\pi^2+\bm{q}^2}
   \left[ S_{12}+ \frac{1}{3}(\bm\sigma_1\cdot\bm\sigma_2)  \bm q^2\right],\label{eq:v_pipv}\\
  \hat{\mathcal V}_{\rho\text{-T}}
=\,&+\vec{\tau}(1)\cdot\vec{\tau}(2)\frac{f_\rho(1) f_\rho(2)}{4M^2}\frac{1}{m_\rho^2+\bm{q}^2}
   \left[S_{12}- \frac{2}{3}(\bm\sigma_1\cdot\bm\sigma_2)\bm q^2\right],\label{eq:v_rhot}\\
  \hat{\mathcal { V}}_{\rho\text{-VT}}
=\,&+\vec{\tau}(1)\cdot\vec{\tau}(2)\frac{f_\rho(1) g_\rho(2)}{4MM^*(1)}\frac{1}{m_\rho^2+\bm{q}^2}
  \left[-\bm{q}^2+2i\bm{\sigma}_1\cdot(\bm{q}\times\bm {k})\right]\nonumber\\
  &+\vec{\tau}(1)\cdot\vec{\tau}(2)\frac{f_\rho(1) g_\rho(2)}{4MM^*(2)}\frac{1}{m_\rho^2+\bm{q}^2}
  \left[-2i\bm{\sigma}_1\cdot(\bm{q}\times\bm {k}')  -\frac{2}{3}(\bm\sigma_1\cdot\bm\sigma_2)\bm q^2 +S_{12}\right]\nonumber\\
 &+\vec{\tau}(1)\cdot\vec{\tau}(2)\frac{g_\rho(1) f_\rho(2)}{4MM^*(2)}\frac{1}{m_\rho^2+\bm{q}^2}
  \left[-\bm{q}^2-2i\bm{\sigma}_2\cdot(\bm{q}\times\bm {k}')\right]\nonumber\\
 & +\vec{\tau}(1)\cdot\vec{\tau}(2)\frac{g_\rho(1) f_\rho(2)}{4MM^*(1)}\frac{1}{m_\rho^2+\bm{q}^2}
  \left[+2i\bm{\sigma}_2\cdot(\bm{q}\times\bm {k})
  -\frac{2}{3}(\bm\sigma_1\cdot\bm\sigma_2)\bm q^2 +S_{12}\right].
\end{align}
\end{subequations}

In Fig.~\ref{fig:ps}, we show the ratio between the starred momentum and its undressed counterpart $\bm p^*/\bm p$ as a function of $p/p_F$ for the symmetric ($\beta=0$) and asymmetric ($\beta=0.2$) nuclear matter with $\rho = 0.1\rho_{\rm sat.}$, $0.5\rho_{\rm sat.}$, and $\rho_{\rm sat.}$.
It is seen that in all these representative cases the adopted approximation $\bm p^* \approx \bm p$ introduces less than $2\%$ errors.

In Fig.~\ref{fig:ms}, we show the momentum dependence of the Dirac mass $M^*$ for the symmetric ($\beta=0$) and asymmetric ($\beta=0.2$) nuclear matter with $\rho = 0.1\rho_{\rm sat.}$, $0.5\rho_{\rm sat.}$, and $\rho_{\rm sat.}$.
It is also seen that the momentum dependence of the Dirac mass $M^*$ is rather weak.
Therefore, it is reasonable to adopt the same value of $M^*$ for the two states at one vertex.
In practice, the value of $M^*$ is chosen as the one with the Fermi momentum $M^*(p_F)$.

\section{Evaluation of two-body interaction matrix elements of tensor force }\label{APP:Matrix_tensor}

In this Appendix, we will show the alternative ways to carry out the integral in Eq.~\eqref{eq:VTabbasum} for the tensor contribution to the two-body interaction matrix elements.

Since there are only tensor and central terms without any other distracting terms, up to the leading order, in the $\pi$-PV and $\rho$-T couplings, it inspires us to evaluate the  $V^\text{t}_{\pi\text{-PV},\alpha\beta\beta\alpha}$ and $V^\text{t}_{\rho\text{-T}, \alpha\beta\beta\alpha}$ indirectly by excluding the contribution of the central term from the whole two-body interaction matrix elements of the $\pi$-PV and $\rho$-T channels \cite{Long2008Europhys.Lett.82.12001}.
The central terms in Eqs.~\eqref{eq:v_pipv} and \eqref{eq:v_rhot} can be divided into two parts, denoted as the zero-range (ZR) and finite-range (FR) parts, respectively, as follows,
\begin{align}
    (\bm\sigma_1\cdot\bm\sigma_2)\frac{\bm q^2}{m_\phi^2+\bm{q}^2}
    =\,(\bm\sigma_1\cdot\bm\sigma_2)\left(1-\frac{m_\phi^2}{m_\phi^2+\bm{q}^2}\right),
\end{align}
and its Fourier transformation gives the presentation in the coordinate space,
\begin{align}\label{Fourierrhot}
(\bm\sigma_1\cdot\bm\sigma_2)\left[\delta(\bm r_1-\bm r_2)-\frac{m_\phi^2}{4\pi}
\frac{e^{-m_\phi |\bm r_1-\bm r_2|}}{ |\bm r_1-\bm r_2|}\right].
\end{align}
Thus, the evaluation of the tensor contributions in the $\pi$-PV and $\rho$-T channels is relatively easy.
Based on that, we can actually find two different ways to evaluate the tensor contributions $V_{\phi, \alpha\beta\beta\alpha}^{\text{t}}$ of each meson-nucleon coupling in the pseudovector (PV) and tensor (T) forms.
It is confirmed that the numerical results by these two different ways are all identical to each other.

\subsection{Pseudovector form}

We denote the tensor contributions evaluated in the PV form as $V^\text{t,PV}_{\phi,\alpha\beta\beta\alpha}$, which can be expressed as
\begin{align}
 V^\text{t,PV}_{\phi,\alpha\beta\beta\alpha}
=\,V_{\phi, \alpha\beta\beta\alpha}^{\text{non,PV}}-V_{\phi, \alpha\beta\beta\alpha}^{\text{ZR,PV}}
 -V_{\phi, \alpha\beta\beta\alpha}^{\text{FR,PV}},
 \end{align}
with
\begin{subequations}
 \begin{align}
  V_{\phi, \alpha\beta\beta\alpha}^{\text{non,PV}}
 =\,& \iint d{\bm r_1}\, d\bm r_2\,\mathscr{F}_\phi(1,2){\xi}_\alpha^\dagger(\bm{r}_1){\xi}_\beta^\dagger(\bm{r}_2)
   \left[\bm\sigma\cdot\bm\nabla\right]_1\left[ \bm\sigma\cdot\bm\nabla\right]_2 \frac{1}{4\pi}\frac{e^{-m_\phi|\bm{x}_1-\bm{x}_2|}}{|\bm{x}_1-\bm{x}_2|}{\xi}_{\beta}(\bm{r}_1)\xi_{\alpha}(\bm{r}_2),
\\
  V_{\phi, \alpha\beta\beta\alpha}^{\text{ZR,PV}}
 =\,& \frac{1}{3}\iint d{\bm r_1}\, d\bm r_2\,\mathscr{F}_\phi(1,2)
   {\xi}_\alpha^\dagger(\bm{r}_1){\xi}_\beta^\dagger(\bm{r}_2)(\bm\sigma_1\cdot\bm\sigma_2)\delta(\bm r_1-\bm r_2) {\xi}_{\beta}(\bm{r}_1)\xi_{\alpha}(\bm{r}_2),
\\
  V_{\phi, \alpha\beta\beta\alpha}^{\text{FR,PV}}
 =\,& -\frac{1}{3}\frac{m_\phi^2}{4\pi}\iint d{\bm r_1}\, d\bm r_2\,\mathscr{F}_\phi(1,2)
   {\xi}_\alpha^\dagger(\bm{r}_1){\xi}_\beta^\dagger(\bm{r}_2)(\bm\sigma_1\cdot\bm\sigma_2)
   \frac{e^{-m_\phi |\bm r_1-\bm r_2|}}{ |\bm r_1-\bm r_2|} {\xi}_{\beta}(\bm{r}_1) \xi_{\alpha}(\bm{r}_2).
\end{align}
\end{subequations}

These three terms can be further expressed as
\begin{subequations}
 \begin{align}
 \sum_{m_\alpha m_\beta} V_{\phi, \alpha\beta\beta\alpha}^{\text{non,PV}}
 =\,& \frac{\hat j_a^2\hat j_b^2}{4\pi}
   \left\{\iint dr_1\,dr_2\, \frac{\mathscr{F}_\phi(1,2)}{2r_1^2}\delta(r_1-r_2)(G_aG_b)_1(G_aG_b)_2
   -m_\phi^2\sum_L'' \hat L^{-4} \left(C_{j_a\frac{1}{2}j_b-\frac{1}{2}}^{L0}\right)^2 \sum_{L_1L_2}^{L\pm1} i^{L_2-L_1} \right. \nonumber\\
 &\qquad \left.\times \iint dr_1 \,dr_2\,\mathscr{F}_\phi(1,2)\left[  \left(\kappa_{ab}+\beta_{LL_1}\right) G_aG_b \right]_1 R_{L_1L_2}(m_\phi;r_1,r_2) \left[\left(\kappa_{ab}+\beta_{LL_2}\right) G_aG_b \right]_2 \right\},\\
  \sum_{m_\alpha m_\beta} V_{\phi, \alpha\beta\beta\alpha}^{\text{ZR,PV}}
 =\,& \frac{\hat j_a^2\hat j_b^2}{4\pi} \iint dr_1\,dr_2\,  \frac{\mathscr{F}_\phi(1,2)}{2 r_1^2}\delta(r_1-r_2)(G_aG_b)_1(G_aG_b)_2 ,\\
   \sum_{m_\alpha m_\beta} V_{\phi, \alpha\beta\beta\alpha}^{\text{FR,PV}}
 =\,& -\frac{m_\phi^2}{3}\frac{\hat j_a^2\hat j_b^2}{4\pi} \sum_L'\iint dr_1\,dr_2\, \mathscr{F}_\phi(1,2)R_{LL}(m_\phi;r_1,r_2)\nonumber\\
  &~~~~~~~~~~~~~~~~~~~~~~~~~~~~~~~~~~
  \times(G_aG_b)_1(G_aG_b)_2 \left[2\left(C_{l_a0l_b0}^{L0}\right)^2-\left(C_{j_a\frac{1}{2}j_b-\frac{1}{2}}^{L0}\right)^2\right],
  \end{align}
\end{subequations}
where $\kappa_{ab}=\kappa_a+\kappa_b$ and
\begin{align}
\beta_{LL_1}
=\,\left\{
 \begin{array}{ll}
 -L & \mbox{for}\quad L_1=L-1, \\
 L+1 & \mbox{for}\quad L_1=L+1.
 \end{array}
 \right.
\end{align}
Note that the isospin operators in $\mathscr{F}_\phi(1,2)$ in Table~\ref{Table:F}, $1$ or $\vec{\tau}\cdot\vec{\tau}$, here become the isospin factors
\begin{align}
\left\{
 \begin{array}{ll}
 \delta_{\tau_a\tau_b} & \mbox{for}\quad \sigma,\, \omega, \\
 2-\delta_{\tau_a\tau_b} & \mbox{for}\quad \rho,\, \pi.
 \end{array}
 \right.
\end{align}
The definition of $R_{L_1L_2}$ reads
 \begin{equation}
  R_{L_1L_2}(m_\phi;r_1,r_2)\equiv\,{m_\phi\sqrt\frac{1}{z_1z_2}}\left[I_{L_1+\frac{1}{2}}(z_1)K_{L_2+\frac{1}{2}}(z_2)\theta(z_2-z_1)
                                 +K_{L_1+\frac{1}{2}}(z_1)I_{L_2+\frac{1}{2}}(z_2)\theta(z_1-z_2)\right],
\end{equation}
with $z=m_\phi r$, $I$ and $K$ the spherical Bessel functions, and $\theta$ the step function.
The summation $\sum\limits '_L$ ($\sum\limits ''_L$) means $L+l_a+l_b$ must be even (odd).

\subsection{Tensor form}

We denote the tensor contributions evaluated in the T form as $V^\text{t,T}_{\phi,\alpha\beta\beta\alpha}$, which can be expressed as
\begin{align}
 V^\text{t,T}_{\phi,\alpha\beta\beta\alpha}
=\,V_{\phi, \alpha\beta\beta\alpha}^{\text{non,T}}-V_{\phi, \alpha\beta\beta\alpha}^{\text{ZR,T}}
 -V_{\phi, \alpha\beta\beta\alpha}^{\text{FR,T}},
\end{align}
with
\begin{subequations}
\begin{align}
 V_{\phi, \alpha\beta\beta\alpha}^{\text{non,T}}
=&\, \frac{1}{4}\iint d{\bm r_1}\, d\bm r_2\,\mathscr{F}_\phi(1,2)\sum_{\mu=\pm 1,0}(-1)^\mu
  {\xi}_\alpha^\dagger(\bm{r}_1){\xi}_\beta^\dagger(\bm{r}_2)\left[[\sigma_\mu, \bm\sigma]\cdot\bm\nabla\right]_1
  \left[ [\sigma_{-\mu}, \bm\sigma]\cdot\bm\nabla\right]_2 \frac{1}{4\pi}\frac{e^{-m_\phi|\bm{x}_1-\bm{x}_2|}}{|\bm{x}_1-\bm{x}_2|}{\xi}_{\beta}(\bm{r}_1)\xi_{\alpha}(\bm{r}_2),
\\
 V_{\phi, \alpha\beta\beta\alpha}^{\text{ZR,T}}
=&\, -\frac{2}{3}\iint d{\bm r_1}\, d\bm r_2\,\mathscr{F}_\phi(1,2)
  {\xi}_\alpha^\dagger(\bm{r}_1){\xi}_\beta^\dagger(\bm{r}_2)(\bm\sigma_1\cdot\bm\sigma_2)\delta(\bm r_1-\bm r_2) {\xi}_{\beta}(\bm{r}_1) \xi_{\alpha}(\bm{r}_2),
\\
 V_{\phi, \alpha\beta\beta\alpha}^{\text{FR,T}}
=&\, \frac{2}{3}\frac{m_\phi^2}{4\pi}\iint d{\bm r_1} \,d\bm r_2\,\mathscr{F}_\phi(1,2)
  {\xi}_\alpha^\dagger(\bm{r}_1){\xi}_\beta^\dagger(\bm{r}_2)(\bm\sigma_1\cdot\bm\sigma_2)
  \frac{e^{-m_\phi |\bm r_1-\bm r_2|}}{ |\bm r_1-\bm r_2|} {\xi}_{\beta}(\bm{r}_1) \xi_{\alpha}(\bm{r}_2).
\end{align}
\end{subequations}
Here $[\sigma_\mu, \bm\sigma]$ and $[\sigma_{-\mu}, \bm\sigma]$ are commutators.

These three terms can be further expressed as
\begin{subequations}
\begin{align}
 \sum_{m_\alpha m_\beta} V_{\phi, \alpha\beta\beta\alpha}^{\text{non,T}}
=\,&-6{m_\phi^2}
 \frac{\hat j_a^2 \hat j_b^2}{4\pi} \sum_L'' \sum_{\mathcal J} \hat{\mathcal J}^{-2} \left(C_{j_a\frac{1}{2}j_b-\frac{1}{2}}^{\mathcal J 0}\right)^2 \sum_{L_1L_2}^{L\pm1} f_{L\mathcal J}^{L_1} f_{L\mathcal J}^{L_2} \nonumber\\
 &\times \iint dr_1\, dr_2\, \mathscr{F}_\phi(1,2)\left( B_{\mathcal JL_1}^{ab} G_aG_b \right)_1 \left[-R_{L_1L_2}(m_\phi;r_1,r_2)+\frac{1}{m_\phi^2r_1^2}\delta(r_1-r_2)\right]
  \left( B_{\mathcal JL_2}^{ab} G_aG_b \right)_2,\\
 \sum_{m_\alpha m_\beta} V_{\phi, \alpha\beta\beta\alpha}^{\text{ZR,T}}
=\,& - \frac{\hat j_a^2 \hat j_b^2}{4\pi} \iint dr_1\, dr_2\, \,\mathscr{F}_\phi(1,2)\frac{1}{r_1^2}\delta(r_1-r_2)(G_aG_b)_1(G_aG_b)_2,\\
 \sum_{m_\alpha m_\beta} V_{\phi, \alpha\beta\beta\alpha}^{\text{FR,T}}
=\,&\frac{2}{3}\frac{\hat{j}_a^2\hat{j}_b^2}{4\pi}m_\phi^2\sum_{L}'\iint dr_1 \, dr_2\,\mathscr{F}_\phi(1,2)
  R_{LL}(m_\phi,r_1,r_2)\nonumber\\
&~~~~~~~~~~~~~~~~~~~~~~~~~~~~~~~~~~~~~~~~
\times(G_aG_b)_1(G_aG_b)_2 \left[2\left(C_{l_a0l_b0}^{L0}\right)^2-\left(C_{j_a\frac{1}{2}j_b-\frac{1}{2}}^{L0}\right)^2\right],
\end{align}
\end{subequations}
where
\begin{align}
 f_{L\mathcal J}^{L_1}
=\,&(-1)^{L_1}\hat L C_{L010}^{L_10}\left\{\begin{matrix} L_1&L&1\\ 1&1&\mathcal J \end{matrix}\right\},
\end{align}
and
\begin{equation}
 B_{\mathcal J L}^{ab}
=\,\left\{\begin{array}{ll} (-1)^{j_a+l_a+\frac{1}{2}} \frac{\kappa_{ab} + \mathcal J +1}{\sqrt{\mathcal J +1}} & \mbox{for}\quad \mathcal J=L-1,\\
\hat{\mathcal J}\frac{\hat j_a^2 +(-1)^{j_a+j_b-\mathcal J }\hat j_b^2}{2\sqrt{\mathcal J \left(\mathcal J +1\right)}} & \mbox{for}\quad \mathcal J=L,\\
(-1)^{j_a+l_a+\frac{1}{2}} \frac{\kappa_{ab} - \mathcal J }{\sqrt{\mathcal J }} & \mbox{for}\quad \mathcal J=L+1.
\end{array} \right.
\end{equation}
From the coding point of view, the T form is more complicated than the PV form.

\end{widetext}

%\newpage %Just because of unusual number of tables stacked at end

%\bibliography{ZHWANG}% Produces the bibliography via BibTeX.

\begin{thebibliography}{93}%
\makeatletter
\providecommand \@ifxundefined [1]{%
 \@ifx{#1\undefined}
}%
\providecommand \@ifnum [1]{%
 \ifnum #1\expandafter \@firstoftwo
 \else \expandafter \@secondoftwo
 \fi
}%
\providecommand \@ifx [1]{%
 \ifx #1\expandafter \@firstoftwo
 \else \expandafter \@secondoftwo
 \fi
}%
\providecommand \natexlab [1]{#1}%
\providecommand \enquote  [1]{``#1''}%
\providecommand \bibnamefont  [1]{#1}%
\providecommand \bibfnamefont [1]{#1}%
\providecommand \citenamefont [1]{#1}%
\providecommand \href@noop [0]{\@secondoftwo}%
\providecommand \href [0]{\begingroup \@sanitize@url \@href}%
\providecommand \@href[1]{\@@startlink{#1}\@@href}%
\providecommand \@@href[1]{\endgroup#1\@@endlink}%
\providecommand \@sanitize@url [0]{\catcode `\\12\catcode `\$12\catcode
  `\&12\catcode `\#12\catcode `\^12\catcode `\_12\catcode `\%12\relax}%
\providecommand \@@startlink[1]{}%
\providecommand \@@endlink[0]{}%
\providecommand \url  [0]{\begingroup\@sanitize@url \@url }%
\providecommand \@url [1]{\endgroup\@href {#1}{\urlprefix }}%
\providecommand \urlprefix  [0]{URL }%
\providecommand \Eprint [0]{\href }%
\providecommand \doibase [0]{http://dx.doi.org/}%
\providecommand \selectlanguage [0]{\@gobble}%
\providecommand \bibinfo  [0]{\@secondoftwo}%
\providecommand \bibfield  [0]{\@secondoftwo}%
\providecommand \translation [1]{[#1]}%
\providecommand \BibitemOpen [0]{}%
\providecommand \bibitemStop [0]{}%
\providecommand \bibitemNoStop [0]{.\EOS\space}%
\providecommand \EOS [0]{\spacefactor3000\relax}%
\providecommand \BibitemShut  [1]{\csname bibitem#1\endcsname}%
\let\auto@bib@innerbib\@empty
%</preamble>
\bibitem [{\citenamefont
  {Yukawa}(1935)}]{YUKAWA1935Proc.Phys.Math.Soc.Japan17.48}%
  \BibitemOpen
  \bibfield  {author} {\bibinfo {author} {\bibfnamefont {H.}~\bibnamefont
  {Yukawa}},\ }\href {\doibase 10.11429/ppmsj1919.17.0_48} {\bibfield
  {journal} {\bibinfo  {journal} {Proc. Phys. Math. Soc. Japan}\ }\textbf
  {\bibinfo {volume} {17}},\ \bibinfo {pages} {48} (\bibinfo {year}
  {1935})}\BibitemShut {NoStop}%
\bibitem [{\citenamefont {Fayache}\ \emph {et~al.}(1997)\citenamefont
  {Fayache}, \citenamefont {Zamick},\ and\ \citenamefont
  {Castel}}]{Fayache1997Phys.Rep.290.201}%
  \BibitemOpen
  \bibfield  {author} {\bibinfo {author} {\bibfnamefont {M.~S.}\ \bibnamefont
  {Fayache}}, \bibinfo {author} {\bibfnamefont {L.}~\bibnamefont {Zamick}}, \
  and\ \bibinfo {author} {\bibfnamefont {B.}~\bibnamefont {Castel}},\ }\href
  {\doibase https://doi.org/10.1016/S0370-1573(97)00013-6} {\bibfield
  {journal} {\bibinfo  {journal} {Phys. Rep.}\ }\textbf {\bibinfo {volume}
  {290}},\ \bibinfo {pages} {201} (\bibinfo {year} {1997})}\BibitemShut
  {NoStop}%
\bibitem [{\citenamefont {Sagawa}\ and\ \citenamefont
  {Col\`o}(2014)}]{Sagawa2014Prog.Part.Nucl.Phys.76.76}%
  \BibitemOpen
  \bibfield  {author} {\bibinfo {author} {\bibfnamefont {H.}~\bibnamefont
  {Sagawa}}\ and\ \bibinfo {author} {\bibfnamefont {G.}~\bibnamefont
  {Col\`o}},\ }\href {\doibase https://doi.org/10.1016/j.ppnp.2014.01.006}
  {\bibfield  {journal} {\bibinfo  {journal} {Prog. Part. Nucl. Phys.}\
  }\textbf {\bibinfo {volume} {76}},\ \bibinfo {pages} {76} (\bibinfo {year}
  {2014})}\BibitemShut {NoStop}%
\bibitem [{\citenamefont {Rarita}\ and\ \citenamefont
  {Schwinger}(1941)}]{Rarita1941Phys.Rev.59.436}%
  \BibitemOpen
  \bibfield  {author} {\bibinfo {author} {\bibfnamefont {W.}~\bibnamefont
  {Rarita}}\ and\ \bibinfo {author} {\bibfnamefont {J.}~\bibnamefont
  {Schwinger}},\ }\href {\doibase 10.1103/PhysRev.59.436} {\bibfield  {journal}
  {\bibinfo  {journal} {Phys. Rev.}\ }\textbf {\bibinfo {volume} {59}},\
  \bibinfo {pages} {436} (\bibinfo {year} {1941})}\BibitemShut {NoStop}%
\bibitem [{\citenamefont {Gerjuoy}\ and\ \citenamefont
  {Schwinger}(1942)}]{Gerjuoy1942Phys.Rev.61.138}%
  \BibitemOpen
  \bibfield  {author} {\bibinfo {author} {\bibfnamefont {E.}~\bibnamefont
  {Gerjuoy}}\ and\ \bibinfo {author} {\bibfnamefont {J.}~\bibnamefont
  {Schwinger}},\ }\href {\doibase 10.1103/PhysRev.61.138} {\bibfield  {journal}
  {\bibinfo  {journal} {Phys. Rev.}\ }\textbf {\bibinfo {volume} {61}},\
  \bibinfo {pages} {138} (\bibinfo {year} {1942})}\BibitemShut {NoStop}%
\bibitem [{\citenamefont {Sorlin}\ and\ \citenamefont
  {Porquet}(2008)}]{Sorlin2008Prog.Part.Nucl.Phys.61.602}%
  \BibitemOpen
  \bibfield  {author} {\bibinfo {author} {\bibfnamefont {O.}~\bibnamefont
  {Sorlin}}\ and\ \bibinfo {author} {\bibfnamefont {M.~G.}\ \bibnamefont
  {Porquet}},\ }\href {\doibase https://doi.org/10.1016/j.ppnp.2008.05.001}
  {\bibfield  {journal} {\bibinfo  {journal} {Prog. Part. Nucl. Phys.}\
  }\textbf {\bibinfo {volume} {61}},\ \bibinfo {pages} {602} (\bibinfo {year}
  {2008})}\BibitemShut {NoStop}%
\bibitem [{\citenamefont {Wienholtz}\ \emph {et~al.}(2013)\citenamefont
  {Wienholtz} \emph {et~al.}}]{Wienholtz2013Nature498.346}%
  \BibitemOpen
  \bibfield  {author} {\bibinfo {author} {\bibfnamefont {F.}~\bibnamefont
  {Wienholtz}} \emph {et~al.},\ }\href {\doibase 10.1038/nature12226}
  {\bibfield  {journal} {\bibinfo  {journal} {Nature}\ }\textbf {\bibinfo
  {volume} {498}},\ \bibinfo {pages} {346} (\bibinfo {year}
  {2013})}\BibitemShut {NoStop}%
\bibitem [{\citenamefont {Steppenbeck}\ \emph {et~al.}(2013)\citenamefont
  {Steppenbeck} \emph {et~al.}}]{Steppenbeck2013Nature502.207}%
  \BibitemOpen
  \bibfield  {author} {\bibinfo {author} {\bibfnamefont {D.}~\bibnamefont
  {Steppenbeck}} \emph {et~al.},\ }\href {\doibase 10.1038/nature12522}
  {\bibfield  {journal} {\bibinfo  {journal} {Nature}\ }\textbf {\bibinfo
  {volume} {502}},\ \bibinfo {pages} {207} (\bibinfo {year}
  {2013})}\BibitemShut {NoStop}%
\bibitem [{\citenamefont {Otsuka}\ \emph {et~al.}(2001)\citenamefont {Otsuka},
  \citenamefont {Fujimoto}, \citenamefont {Utsuno}, \citenamefont {Brown},
  \citenamefont {Honma},\ and\ \citenamefont
  {Mizusaki}}]{Otsuka2001Phys.Rev.Lett.87.082502}%
  \BibitemOpen
  \bibfield  {author} {\bibinfo {author} {\bibfnamefont {T.}~\bibnamefont
  {Otsuka}}, \bibinfo {author} {\bibfnamefont {R.}~\bibnamefont {Fujimoto}},
  \bibinfo {author} {\bibfnamefont {Y.}~\bibnamefont {Utsuno}}, \bibinfo
  {author} {\bibfnamefont {B.~A.}\ \bibnamefont {Brown}}, \bibinfo {author}
  {\bibfnamefont {M.}~\bibnamefont {Honma}}, \ and\ \bibinfo {author}
  {\bibfnamefont {T.}~\bibnamefont {Mizusaki}},\ }\href {\doibase
  10.1103/PhysRevLett.87.082502} {\bibfield  {journal} {\bibinfo  {journal}
  {Phys. Rev. Lett.}\ }\textbf {\bibinfo {volume} {87}},\ \bibinfo {pages}
  {082502} (\bibinfo {year} {2001})}\BibitemShut {NoStop}%
\bibitem [{\citenamefont {Otsuka}\ \emph {et~al.}(2005)\citenamefont {Otsuka},
  \citenamefont {Suzuki}, \citenamefont {Fujimoto}, \citenamefont {Grawe},\
  and\ \citenamefont {Akaishi}}]{Otsuka2005Phys.Rev.Lett.95.232502}%
  \BibitemOpen
  \bibfield  {author} {\bibinfo {author} {\bibfnamefont {T.}~\bibnamefont
  {Otsuka}}, \bibinfo {author} {\bibfnamefont {T.}~\bibnamefont {Suzuki}},
  \bibinfo {author} {\bibfnamefont {R.}~\bibnamefont {Fujimoto}}, \bibinfo
  {author} {\bibfnamefont {H.}~\bibnamefont {Grawe}}, \ and\ \bibinfo {author}
  {\bibfnamefont {Y.}~\bibnamefont {Akaishi}},\ }\href {\doibase
  10.1103/PhysRevLett.95.232502} {\bibfield  {journal} {\bibinfo  {journal}
  {Phys. Rev. Lett.}\ }\textbf {\bibinfo {volume} {95}},\ \bibinfo {pages}
  {232502} (\bibinfo {year} {2005})}\BibitemShut {NoStop}%
\bibitem [{\citenamefont {Otsuka}\ \emph {et~al.}(2006)\citenamefont {Otsuka},
  \citenamefont {Matsuo},\ and\ \citenamefont
  {Abe}}]{Otsuka2006Phys.Rev.Lett.97.162501}%
  \BibitemOpen
  \bibfield  {author} {\bibinfo {author} {\bibfnamefont {T.}~\bibnamefont
  {Otsuka}}, \bibinfo {author} {\bibfnamefont {T.}~\bibnamefont {Matsuo}}, \
  and\ \bibinfo {author} {\bibfnamefont {D.}~\bibnamefont {Abe}},\ }\href
  {\doibase 10.1103/PhysRevLett.97.162501} {\bibfield  {journal} {\bibinfo
  {journal} {Phys. Rev. Lett.}\ }\textbf {\bibinfo {volume} {97}},\ \bibinfo
  {pages} {162501} (\bibinfo {year} {2006})}\BibitemShut {NoStop}%
\bibitem [{\citenamefont {Otsuka}\ \emph {et~al.}(2010)\citenamefont {Otsuka},
  \citenamefont {Suzuki}, \citenamefont {Honma}, \citenamefont {Utsuno},
  \citenamefont {Tsunoda}, \citenamefont {Tsukiyama},\ and\ \citenamefont
  {Hjorth-Jensen}}]{Otsuka2010Phys.Rev.Lett.104.012501}%
  \BibitemOpen
  \bibfield  {author} {\bibinfo {author} {\bibfnamefont {T.}~\bibnamefont
  {Otsuka}}, \bibinfo {author} {\bibfnamefont {T.}~\bibnamefont {Suzuki}},
  \bibinfo {author} {\bibfnamefont {M.}~\bibnamefont {Honma}}, \bibinfo
  {author} {\bibfnamefont {Y.}~\bibnamefont {Utsuno}}, \bibinfo {author}
  {\bibfnamefont {N.}~\bibnamefont {Tsunoda}}, \bibinfo {author} {\bibfnamefont
  {K.}~\bibnamefont {Tsukiyama}}, \ and\ \bibinfo {author} {\bibfnamefont
  {M.}~\bibnamefont {Hjorth-Jensen}},\ }\href {\doibase
  10.1103/PhysRevLett.104.012501} {\bibfield  {journal} {\bibinfo  {journal}
  {Phys. Rev. Lett.}\ }\textbf {\bibinfo {volume} {104}},\ \bibinfo {pages}
  {012501} (\bibinfo {year} {2010})}\BibitemShut {NoStop}%
\bibitem [{\citenamefont {Bender}\ \emph {et~al.}(2003)\citenamefont {Bender},
  \citenamefont {Heenen},\ and\ \citenamefont
  {Reinhard}}]{Bender2003Rev.Mod.Phys.75.121}%
  \BibitemOpen
  \bibfield  {author} {\bibinfo {author} {\bibfnamefont {M.}~\bibnamefont
  {Bender}}, \bibinfo {author} {\bibfnamefont {P.~H.}\ \bibnamefont {Heenen}},
  \ and\ \bibinfo {author} {\bibfnamefont {P.~G.}\ \bibnamefont {Reinhard}},\
  }\href {\doibase 10.1103/RevModPhys.75.121} {\bibfield  {journal} {\bibinfo
  {journal} {Rev. Mod. Phys.}\ }\textbf {\bibinfo {volume} {75}},\ \bibinfo
  {pages} {121} (\bibinfo {year} {2003})}\BibitemShut {NoStop}%
\bibitem [{\citenamefont {Meng}\ \emph {et~al.}(2006)\citenamefont {Meng},
  \citenamefont {Toki}, \citenamefont {Zhou}, \citenamefont {Zhang},
  \citenamefont {Long},\ and\ \citenamefont
  {Geng}}]{Meng2006Prog.Part.Nucl.Phys.57.470}%
  \BibitemOpen
  \bibfield  {author} {\bibinfo {author} {\bibfnamefont {J.}~\bibnamefont
  {Meng}}, \bibinfo {author} {\bibfnamefont {H.}~\bibnamefont {Toki}}, \bibinfo
  {author} {\bibfnamefont {S.~G.}\ \bibnamefont {Zhou}}, \bibinfo {author}
  {\bibfnamefont {S.~Q.}\ \bibnamefont {Zhang}}, \bibinfo {author}
  {\bibfnamefont {W.~H.}\ \bibnamefont {Long}}, \ and\ \bibinfo {author}
  {\bibfnamefont {L.~S.}\ \bibnamefont {Geng}},\ }\href {\doibase
  https://doi.org/10.1016/j.ppnp.2005.06.001} {\bibfield  {journal} {\bibinfo
  {journal} {Prog. Part. Nucl. Phys.}\ }\textbf {\bibinfo {volume} {57}},\
  \bibinfo {pages} {470} (\bibinfo {year} {2006})}\BibitemShut {NoStop}%
\bibitem [{\citenamefont {Nakatsukasa}\ \emph {et~al.}(2016)\citenamefont
  {Nakatsukasa}, \citenamefont {Matsuyanagi}, \citenamefont {Matsuo},\ and\
  \citenamefont {Yabana}}]{Nakatsukasa2016Rev.Mod.Phys.88.045004}%
  \BibitemOpen
  \bibfield  {author} {\bibinfo {author} {\bibfnamefont {T.}~\bibnamefont
  {Nakatsukasa}}, \bibinfo {author} {\bibfnamefont {K.}~\bibnamefont
  {Matsuyanagi}}, \bibinfo {author} {\bibfnamefont {M.}~\bibnamefont {Matsuo}},
  \ and\ \bibinfo {author} {\bibfnamefont {K.}~\bibnamefont {Yabana}},\ }\href
  {\doibase 10.1103/RevModPhys.88.045004} {\bibfield  {journal} {\bibinfo
  {journal} {Rev. Mod. Phys.}\ }\textbf {\bibinfo {volume} {88}},\ \bibinfo
  {pages} {045004} (\bibinfo {year} {2016})}\BibitemShut {NoStop}%
\bibitem [{\citenamefont {Xia}\ \emph {et~al.}(2018)\citenamefont {Xia} \emph
  {et~al.}}]{Xia2018At.DataNucl.DataTables121.1}%
  \BibitemOpen
  \bibfield  {author} {\bibinfo {author} {\bibfnamefont {X.~W.}\ \bibnamefont
  {Xia}} \emph {et~al.},\ }\href {\doibase 10.1016/j.adt.2017.09.001}
  {\bibfield  {journal} {\bibinfo  {journal} {At. Data Nucl. Data Tables}\
  }\textbf {\bibinfo {volume} {121--122}},\ \bibinfo {pages} {1} (\bibinfo
  {year} {2018})}\BibitemShut {NoStop}%
\bibitem [{\citenamefont {Stancu}\ \emph {et~al.}(1977)\citenamefont {Stancu},
  \citenamefont {Brink},\ and\ \citenamefont
  {Flocard}}]{Stancu1977Phys.Lett.B68.108}%
  \BibitemOpen
  \bibfield  {author} {\bibinfo {author} {\bibfnamefont {F.}~\bibnamefont
  {Stancu}}, \bibinfo {author} {\bibfnamefont {D.~M.}\ \bibnamefont {Brink}}, \
  and\ \bibinfo {author} {\bibfnamefont {H.}~\bibnamefont {Flocard}},\ }\href
  {\doibase https://doi.org/10.1016/0370-2693(77)90178-2} {\bibfield  {journal}
  {\bibinfo  {journal} {Phys. Lett. B}\ }\textbf {\bibinfo {volume} {68}},\
  \bibinfo {pages} {108} (\bibinfo {year} {1977})}\BibitemShut {NoStop}%
\bibitem [{\citenamefont {Skyrme}(1958)}]{Skyrme1958NuclearPhys.9.615}%
  \BibitemOpen
  \bibfield  {author} {\bibinfo {author} {\bibfnamefont {T.~H.~R.}\
  \bibnamefont {Skyrme}},\ }\href {\doibase
  https://doi.org/10.1016/0029-5582(58)90345-6} {\bibfield  {journal} {\bibinfo
   {journal} {Nuclear Phys.}\ }\textbf {\bibinfo {volume} {9}},\ \bibinfo
  {pages} {615} (\bibinfo {year} {1958})}\BibitemShut {NoStop}%
\bibitem [{\citenamefont {Decharg\'e}\ and\ \citenamefont
  {Gogny}(1980)}]{Decharge1980Phys.Rev.C21.1568}%
  \BibitemOpen
  \bibfield  {author} {\bibinfo {author} {\bibfnamefont {J.}~\bibnamefont
  {Decharg\'e}}\ and\ \bibinfo {author} {\bibfnamefont {D.}~\bibnamefont
  {Gogny}},\ }\href {\doibase 10.1103/PhysRevC.21.1568} {\bibfield  {journal}
  {\bibinfo  {journal} {Phys. Rev. C}\ }\textbf {\bibinfo {volume} {21}},\
  \bibinfo {pages} {1568} (\bibinfo {year} {1980})}\BibitemShut {NoStop}%
\bibitem [{\citenamefont {Berger}\ \emph {et~al.}(1991)\citenamefont {Berger},
  \citenamefont {Girod},\ and\ \citenamefont
  {Gogny}}]{Berger1991Comput.Phys.Commun.63.365}%
  \BibitemOpen
  \bibfield  {author} {\bibinfo {author} {\bibfnamefont {J.~F.}\ \bibnamefont
  {Berger}}, \bibinfo {author} {\bibfnamefont {M.}~\bibnamefont {Girod}}, \
  and\ \bibinfo {author} {\bibfnamefont {D.}~\bibnamefont {Gogny}},\ }\href
  {\doibase https://doi.org/10.1016/0010-4655(91)90263-K} {\bibfield  {journal}
  {\bibinfo  {journal} {Comput. Phys. Commun.}\ }\textbf {\bibinfo {volume}
  {63}},\ \bibinfo {pages} {365} (\bibinfo {year} {1991})}\BibitemShut
  {NoStop}%
\bibitem [{\citenamefont {Walecka}(1974)}]{Walecka1974Ann.Phys.83.491}%
  \BibitemOpen
  \bibfield  {author} {\bibinfo {author} {\bibfnamefont {J.~D.}\ \bibnamefont
  {Walecka}},\ }\href {\doibase https://doi.org/10.1016/0003-4916(74)90208-5}
  {\bibfield  {journal} {\bibinfo  {journal} {Ann. Phys.}\ }\textbf {\bibinfo
  {volume} {83}},\ \bibinfo {pages} {491} (\bibinfo {year} {1974})}\BibitemShut
  {NoStop}%
\bibitem [{\citenamefont {Ring}(1996)}]{RING1996PROG.PART.NUCL.PHYS.37.193}%
  \BibitemOpen
  \bibfield  {author} {\bibinfo {author} {\bibfnamefont {P.}~\bibnamefont
  {Ring}},\ }\href {\doibase https://doi.org/10.1016/0146-6410(96)00054-3}
  {\bibfield  {journal} {\bibinfo  {journal} {Prog. Part. Nucl. Phys.}\
  }\textbf {\bibinfo {volume} {37}},\ \bibinfo {pages} {193} (\bibinfo {year}
  {1996})}\BibitemShut {NoStop}%
\bibitem [{\citenamefont {Nik\v{s}i\'c}\ \emph {et~al.}(2011)\citenamefont
  {Nik\v{s}i\'c}, \citenamefont {Vretenar},\ and\ \citenamefont
  {Ring}}]{NIKSIC2011PROG.PART.NUCL.PHYS.66.519}%
  \BibitemOpen
  \bibfield  {author} {\bibinfo {author} {\bibfnamefont {T.}~\bibnamefont
  {Nik\v{s}i\'c}}, \bibinfo {author} {\bibfnamefont {D.}~\bibnamefont
  {Vretenar}}, \ and\ \bibinfo {author} {\bibfnamefont {P.}~\bibnamefont
  {Ring}},\ }\href {\doibase https://doi.org/10.1016/j.ppnp.2011.01.055}
  {\bibfield  {journal} {\bibinfo  {journal} {Prog. Part. Nucl. Phys.}\
  }\textbf {\bibinfo {volume} {66}},\ \bibinfo {pages} {519} (\bibinfo {year}
  {2011})}\BibitemShut {NoStop}%
\bibitem [{\citenamefont {Vretenar}\ \emph {et~al.}(2005)\citenamefont
  {Vretenar}, \citenamefont {Afanasjev}, \citenamefont {Lalazissis},\ and\
  \citenamefont {Ring}}]{VRETENAR2005PHYS.REP.409.101}%
  \BibitemOpen
  \bibfield  {author} {\bibinfo {author} {\bibfnamefont {D.}~\bibnamefont
  {Vretenar}}, \bibinfo {author} {\bibfnamefont {A.~V.}\ \bibnamefont
  {Afanasjev}}, \bibinfo {author} {\bibfnamefont {G.~A.}\ \bibnamefont
  {Lalazissis}}, \ and\ \bibinfo {author} {\bibfnamefont {P.}~\bibnamefont
  {Ring}},\ }\href {\doibase https://doi.org/10.1016/j.physrep.2004.10.001}
  {\bibfield  {journal} {\bibinfo  {journal} {Phys. Rep.}\ }\textbf {\bibinfo
  {volume} {409}},\ \bibinfo {pages} {101} (\bibinfo {year}
  {2005})}\BibitemShut {NoStop}%
\bibitem [{\citenamefont {Meng}(2016)}]{meng2016}%
  \BibitemOpen
  \bibinfo {editor} {\bibfnamefont {J.}~\bibnamefont {Meng}},\ ed.,\ \href@noop
  {} {\emph {\bibinfo {title} {Relativistic Density Functional for Nuclear
  Structure}}},\ \bibinfo {series} {International Review of Nuclear Physics},
  Vol.~\bibinfo {volume} {10}\ (\bibinfo  {publisher} {World Scientific,
  Singapore},\ \bibinfo {year} {2016})\BibitemShut {NoStop}%
\bibitem [{\citenamefont {Kamada}\ \emph {et~al.}(2001)\citenamefont {Kamada}
  \emph {et~al.}}]{Kamada2001Phys.Rev.C64.044001}%
  \BibitemOpen
  \bibfield  {author} {\bibinfo {author} {\bibfnamefont {H.}~\bibnamefont
  {Kamada}} \emph {et~al.},\ }\href {\doibase 10.1103/PhysRevC.64.044001}
  {\bibfield  {journal} {\bibinfo  {journal} {Phys. Rev. C}\ }\textbf {\bibinfo
  {volume} {64}},\ \bibinfo {pages} {044001} (\bibinfo {year}
  {2001})}\BibitemShut {NoStop}%
\bibitem [{\citenamefont {Myo}\ \emph {et~al.}(2005)\citenamefont {Myo},
  \citenamefont {Kat\=o},\ and\ \citenamefont
  {Ikeda}}]{Myo2005Prog.Theor.Phys.113.763}%
  \BibitemOpen
  \bibfield  {author} {\bibinfo {author} {\bibfnamefont {T.}~\bibnamefont
  {Myo}}, \bibinfo {author} {\bibfnamefont {K.}~\bibnamefont {Kat\=o}}, \ and\
  \bibinfo {author} {\bibfnamefont {K.}~\bibnamefont {Ikeda}},\ }\href
  {\doibase 10.1143/PTP.113.763} {\bibfield  {journal} {\bibinfo  {journal}
  {Prog. Theor. Phys.}\ }\textbf {\bibinfo {volume} {113}},\ \bibinfo {pages}
  {763} (\bibinfo {year} {2005})}\BibitemShut {NoStop}%
\bibitem [{\citenamefont {Myo}\ \emph {et~al.}(2007)\citenamefont {Myo},
  \citenamefont {Kat\=o}, \citenamefont {Toki},\ and\ \citenamefont
  {Ikeda}}]{Myo2007Phys.Rev.C76.024305}%
  \BibitemOpen
  \bibfield  {author} {\bibinfo {author} {\bibfnamefont {T.}~\bibnamefont
  {Myo}}, \bibinfo {author} {\bibfnamefont {K.}~\bibnamefont {Kat\=o}},
  \bibinfo {author} {\bibfnamefont {H.}~\bibnamefont {Toki}}, \ and\ \bibinfo
  {author} {\bibfnamefont {K.}~\bibnamefont {Ikeda}},\ }\href {\doibase
  10.1103/PhysRevC.76.024305} {\bibfield  {journal} {\bibinfo  {journal} {Phys.
  Rev. C}\ }\textbf {\bibinfo {volume} {76}},\ \bibinfo {pages} {024305}
  (\bibinfo {year} {2007})}\BibitemShut {NoStop}%
\bibitem [{\citenamefont {Myo}\ \emph {et~al.}(2017{\natexlab{a}})\citenamefont
  {Myo}, \citenamefont {Toki}, \citenamefont {Ikeda}, \citenamefont
  {Horiuchi},\ and\ \citenamefont {Suhara}}]{Myo2017Phys.Lett.B769.213}%
  \BibitemOpen
  \bibfield  {author} {\bibinfo {author} {\bibfnamefont {T.}~\bibnamefont
  {Myo}}, \bibinfo {author} {\bibfnamefont {H.}~\bibnamefont {Toki}}, \bibinfo
  {author} {\bibfnamefont {K.}~\bibnamefont {Ikeda}}, \bibinfo {author}
  {\bibfnamefont {H.}~\bibnamefont {Horiuchi}}, \ and\ \bibinfo {author}
  {\bibfnamefont {T.}~\bibnamefont {Suhara}},\ }\href {\doibase
  10.1016/j.physletb.2017.03.059} {\bibfield  {journal} {\bibinfo  {journal}
  {Phys. Lett. B}\ }\textbf {\bibinfo {volume} {769}},\ \bibinfo {pages} {213}
  (\bibinfo {year} {2017}{\natexlab{a}})}\BibitemShut {NoStop}%
\bibitem [{\citenamefont {Myo}\ \emph {et~al.}(2017{\natexlab{b}})\citenamefont
  {Myo}, \citenamefont {Toki}, \citenamefont {Ikeda}, \citenamefont
  {Horiuchi},\ and\ \citenamefont {Suhara}}]{Myo2017Phys.Rev.C95.044314}%
  \BibitemOpen
  \bibfield  {author} {\bibinfo {author} {\bibfnamefont {T.}~\bibnamefont
  {Myo}}, \bibinfo {author} {\bibfnamefont {H.}~\bibnamefont {Toki}}, \bibinfo
  {author} {\bibfnamefont {K.}~\bibnamefont {Ikeda}}, \bibinfo {author}
  {\bibfnamefont {H.}~\bibnamefont {Horiuchi}}, \ and\ \bibinfo {author}
  {\bibfnamefont {T.}~\bibnamefont {Suhara}},\ }\href {\doibase
  10.1103/PhysRevC.95.044314} {\bibfield  {journal} {\bibinfo  {journal} {Phys.
  Rev. C}\ }\textbf {\bibinfo {volume} {95}},\ \bibinfo {pages} {044314}
  (\bibinfo {year} {2017}{\natexlab{b}})}\BibitemShut {NoStop}%
\bibitem [{\citenamefont {Myo}\ \emph {et~al.}(2017{\natexlab{c}})\citenamefont
  {Myo}, \citenamefont {Toki}, \citenamefont {Ikeda}, \citenamefont {Horiuchi},
  \citenamefont {Suhara}, \citenamefont {Lyu}, \citenamefont {Isaka},\ and\
  \citenamefont {Yamada}}]{Myo2017Prog.Theor.Exp.Phys.2017.111D01}%
  \BibitemOpen
  \bibfield  {author} {\bibinfo {author} {\bibfnamefont {T.}~\bibnamefont
  {Myo}}, \bibinfo {author} {\bibfnamefont {H.}~\bibnamefont {Toki}}, \bibinfo
  {author} {\bibfnamefont {K.}~\bibnamefont {Ikeda}}, \bibinfo {author}
  {\bibfnamefont {H.}~\bibnamefont {Horiuchi}}, \bibinfo {author}
  {\bibfnamefont {T.}~\bibnamefont {Suhara}}, \bibinfo {author} {\bibfnamefont
  {M.}~\bibnamefont {Lyu}}, \bibinfo {author} {\bibfnamefont {M.}~\bibnamefont
  {Isaka}}, \ and\ \bibinfo {author} {\bibfnamefont {T.}~\bibnamefont
  {Yamada}},\ }\href {\doibase 10.1093/ptep/ptx143} {\bibfield  {journal}
  {\bibinfo  {journal} {Prog. Theor. Exp. Phys.}\ }\textbf {\bibinfo {volume}
  {2017}},\ \bibinfo {pages} {111D01} (\bibinfo {year}
  {2017}{\natexlab{c}})}\BibitemShut {NoStop}%
\bibitem [{\citenamefont {Schiffer}\ \emph {et~al.}(2004)\citenamefont
  {Schiffer} \emph {et~al.}}]{Schiffer2004Phys.Rev.Lett.92.162501}%
  \BibitemOpen
  \bibfield  {author} {\bibinfo {author} {\bibfnamefont {J.~P.}\ \bibnamefont
  {Schiffer}} \emph {et~al.},\ }\href {\doibase 10.1103/PhysRevLett.92.162501}
  {\bibfield  {journal} {\bibinfo  {journal} {Phys. Rev. Lett.}\ }\textbf
  {\bibinfo {volume} {92}},\ \bibinfo {pages} {162501} (\bibinfo {year}
  {2004})}\BibitemShut {NoStop}%
\bibitem [{\citenamefont {Cottle}\ and\ \citenamefont
  {Kemper}(1998)}]{Cottle1998Phys.Rev.C58.3761}%
  \BibitemOpen
  \bibfield  {author} {\bibinfo {author} {\bibfnamefont {P.~D.}\ \bibnamefont
  {Cottle}}\ and\ \bibinfo {author} {\bibfnamefont {K.~W.}\ \bibnamefont
  {Kemper}},\ }\href {\doibase 10.1103/PhysRevC.58.3761} {\bibfield  {journal}
  {\bibinfo  {journal} {Phys. Rev. C}\ }\textbf {\bibinfo {volume} {58}},\
  \bibinfo {pages} {3761} (\bibinfo {year} {1998})}\BibitemShut {NoStop}%
\bibitem [{\citenamefont {Brown}\ \emph {et~al.}(2006)\citenamefont {Brown},
  \citenamefont {Duguet}, \citenamefont {Otsuka}, \citenamefont {Abe},\ and\
  \citenamefont {Suzuki}}]{Brown2006Phys.Rev.C74.061303}%
  \BibitemOpen
  \bibfield  {author} {\bibinfo {author} {\bibfnamefont {B.~A.}\ \bibnamefont
  {Brown}}, \bibinfo {author} {\bibfnamefont {T.}~\bibnamefont {Duguet}},
  \bibinfo {author} {\bibfnamefont {T.}~\bibnamefont {Otsuka}}, \bibinfo
  {author} {\bibfnamefont {D.}~\bibnamefont {Abe}}, \ and\ \bibinfo {author}
  {\bibfnamefont {T.}~\bibnamefont {Suzuki}},\ }\href {\doibase
  10.1103/PhysRevC.74.061303} {\bibfield  {journal} {\bibinfo  {journal} {Phys.
  Rev. C}\ }\textbf {\bibinfo {volume} {74}},\ \bibinfo {pages} {061303}
  (\bibinfo {year} {2006})}\BibitemShut {NoStop}%
\bibitem [{\citenamefont {Brink}\ and\ \citenamefont
  {Stancu}(2007)}]{Brink2007Phys.Rev.C75.064311}%
  \BibitemOpen
  \bibfield  {author} {\bibinfo {author} {\bibfnamefont {D.~M.}\ \bibnamefont
  {Brink}}\ and\ \bibinfo {author} {\bibfnamefont {F.}~\bibnamefont {Stancu}},\
  }\href {\doibase 10.1103/PhysRevC.75.064311} {\bibfield  {journal} {\bibinfo
  {journal} {Phys. Rev. C}\ }\textbf {\bibinfo {volume} {75}},\ \bibinfo
  {pages} {064311} (\bibinfo {year} {2007})}\BibitemShut {NoStop}%
\bibitem [{\citenamefont {Col\`o}\ \emph {et~al.}(2007)\citenamefont {Col\`o},
  \citenamefont {Sagawa}, \citenamefont {Fracasso},\ and\ \citenamefont
  {Bortignon}}]{Colo2007Phys.Lett.B646.227}%
  \BibitemOpen
  \bibfield  {author} {\bibinfo {author} {\bibfnamefont {G.}~\bibnamefont
  {Col\`o}}, \bibinfo {author} {\bibfnamefont {H.}~\bibnamefont {Sagawa}},
  \bibinfo {author} {\bibfnamefont {S.}~\bibnamefont {Fracasso}}, \ and\
  \bibinfo {author} {\bibfnamefont {P.~F.}\ \bibnamefont {Bortignon}},\ }\href
  {\doibase https://doi.org/10.1016/j.physletb.2007.01.033} {\bibfield
  {journal} {\bibinfo  {journal} {Phys. Lett. B}\ }\textbf {\bibinfo {volume}
  {646}},\ \bibinfo {pages} {227} (\bibinfo {year} {2007})}\BibitemShut
  {NoStop}%
\bibitem [{\citenamefont {Sugimoto}\ \emph {et~al.}(2007)\citenamefont
  {Sugimoto}, \citenamefont {Toki},\ and\ \citenamefont
  {Ikeda}}]{Sugimoto2007Phys.Rev.C76.054310}%
  \BibitemOpen
  \bibfield  {author} {\bibinfo {author} {\bibfnamefont {S.}~\bibnamefont
  {Sugimoto}}, \bibinfo {author} {\bibfnamefont {H.}~\bibnamefont {Toki}}, \
  and\ \bibinfo {author} {\bibfnamefont {K.}~\bibnamefont {Ikeda}},\ }\href
  {\doibase 10.1103/PhysRevC.76.054310} {\bibfield  {journal} {\bibinfo
  {journal} {Phys. Rev. C}\ }\textbf {\bibinfo {volume} {76}},\ \bibinfo
  {pages} {054310} (\bibinfo {year} {2007})}\BibitemShut {NoStop}%
\bibitem [{\citenamefont {Zou}\ \emph {et~al.}(2008)\citenamefont {Zou},
  \citenamefont {Col\`o}, \citenamefont {Ma}, \citenamefont {Sagawa},\ and\
  \citenamefont {Bortignon}}]{Zou2008Phys.Rev.C77.014314}%
  \BibitemOpen
  \bibfield  {author} {\bibinfo {author} {\bibfnamefont {W.}~\bibnamefont
  {Zou}}, \bibinfo {author} {\bibfnamefont {G.}~\bibnamefont {Col\`o}},
  \bibinfo {author} {\bibfnamefont {Z.~Y.}\ \bibnamefont {Ma}}, \bibinfo
  {author} {\bibfnamefont {H.}~\bibnamefont {Sagawa}}, \ and\ \bibinfo {author}
  {\bibfnamefont {P.~F.}\ \bibnamefont {Bortignon}},\ }\href {\doibase
  10.1103/PhysRevC.77.014314} {\bibfield  {journal} {\bibinfo  {journal} {Phys.
  Rev. C}\ }\textbf {\bibinfo {volume} {77}},\ \bibinfo {pages} {014314}
  (\bibinfo {year} {2008})}\BibitemShut {NoStop}%
\bibitem [{\citenamefont {Anguiano}\ \emph {et~al.}(2012)\citenamefont
  {Anguiano}, \citenamefont {Grasso}, \citenamefont {Co'}, \citenamefont
  {De~Donno},\ and\ \citenamefont {Lallena}}]{Anguiano2012Phys.Rev.C86.054302}%
  \BibitemOpen
  \bibfield  {author} {\bibinfo {author} {\bibfnamefont {M.}~\bibnamefont
  {Anguiano}}, \bibinfo {author} {\bibfnamefont {M.}~\bibnamefont {Grasso}},
  \bibinfo {author} {\bibfnamefont {G.}~\bibnamefont {Co'}}, \bibinfo {author}
  {\bibfnamefont {V.}~\bibnamefont {De~Donno}}, \ and\ \bibinfo {author}
  {\bibfnamefont {A.~M.}\ \bibnamefont {Lallena}},\ }\href {\doibase
  10.1103/PhysRevC.86.054302} {\bibfield  {journal} {\bibinfo  {journal} {Phys.
  Rev. C}\ }\textbf {\bibinfo {volume} {86}},\ \bibinfo {pages} {054302}
  (\bibinfo {year} {2012})}\BibitemShut {NoStop}%
\bibitem [{\citenamefont {Nakada}(2013)}]{Nakada2013Phys.Rev.C87.014336}%
  \BibitemOpen
  \bibfield  {author} {\bibinfo {author} {\bibfnamefont {H.}~\bibnamefont
  {Nakada}},\ }\href {\doibase 10.1103/PhysRevC.87.014336} {\bibfield
  {journal} {\bibinfo  {journal} {Phys. Rev. C}\ }\textbf {\bibinfo {volume}
  {87}},\ \bibinfo {pages} {014336} (\bibinfo {year} {2013})}\BibitemShut
  {NoStop}%
\bibitem [{\citenamefont {Long}\ \emph {et~al.}(2007)\citenamefont {Long},
  \citenamefont {Sagawa}, \citenamefont {Van~Giai},\ and\ \citenamefont
  {Meng}}]{Long2007Phys.Rev.C76.034314}%
  \BibitemOpen
  \bibfield  {author} {\bibinfo {author} {\bibfnamefont {W.~H.}\ \bibnamefont
  {Long}}, \bibinfo {author} {\bibfnamefont {H.}~\bibnamefont {Sagawa}},
  \bibinfo {author} {\bibfnamefont {N.}~\bibnamefont {Van~Giai}}, \ and\
  \bibinfo {author} {\bibfnamefont {J.}~\bibnamefont {Meng}},\ }\href {\doibase
  10.1103/PhysRevC.76.034314} {\bibfield  {journal} {\bibinfo  {journal} {Phys.
  Rev. C}\ }\textbf {\bibinfo {volume} {76}},\ \bibinfo {pages} {034314}
  (\bibinfo {year} {2007})}\BibitemShut {NoStop}%
\bibitem [{\citenamefont {Long}\ \emph {et~al.}(2008)\citenamefont {Long},
  \citenamefont {Sagawa}, \citenamefont {Meng},\ and\ \citenamefont
  {Van~Giai}}]{Long2008Europhys.Lett.82.12001}%
  \BibitemOpen
  \bibfield  {author} {\bibinfo {author} {\bibfnamefont {W.~H.}\ \bibnamefont
  {Long}}, \bibinfo {author} {\bibfnamefont {H.}~\bibnamefont {Sagawa}},
  \bibinfo {author} {\bibfnamefont {J.}~\bibnamefont {Meng}}, \ and\ \bibinfo
  {author} {\bibfnamefont {N.}~\bibnamefont {Van~Giai}},\ }\href
  {http://stacks.iop.org/0295-5075/82/i=1/a=12001} {\bibfield  {journal}
  {\bibinfo  {journal} {Europhys. Lett.}\ }\textbf {\bibinfo {volume} {82}},\
  \bibinfo {pages} {12001} (\bibinfo {year} {2008})}\BibitemShut {NoStop}%
\bibitem [{\citenamefont {Wang}\ \emph {et~al.}(2013)\citenamefont {Wang},
  \citenamefont {Dong},\ and\ \citenamefont
  {Long}}]{Wang2013Phys.Rev.C87.047301}%
  \BibitemOpen
  \bibfield  {author} {\bibinfo {author} {\bibfnamefont {L.~J.}\ \bibnamefont
  {Wang}}, \bibinfo {author} {\bibfnamefont {J.~M.}\ \bibnamefont {Dong}}, \
  and\ \bibinfo {author} {\bibfnamefont {W.~H.}\ \bibnamefont {Long}},\ }\href
  {\doibase 10.1103/PhysRevC.87.047301} {\bibfield  {journal} {\bibinfo
  {journal} {Phys. Rev. C}\ }\textbf {\bibinfo {volume} {87}},\ \bibinfo
  {pages} {047301} (\bibinfo {year} {2013})}\BibitemShut {NoStop}%
\bibitem [{\citenamefont {Marcos}\ \emph {et~al.}(2014)\citenamefont {Marcos},
  \citenamefont {L{\'{o}}pez-Quelle}, \citenamefont {Niembro},\ and\
  \citenamefont {Savushkin}}]{Marcos2014Phys.At.Nucl.77.299}%
  \BibitemOpen
  \bibfield  {author} {\bibinfo {author} {\bibfnamefont {S.}~\bibnamefont
  {Marcos}}, \bibinfo {author} {\bibfnamefont {M.}~\bibnamefont
  {L{\'{o}}pez-Quelle}}, \bibinfo {author} {\bibfnamefont {R.}~\bibnamefont
  {Niembro}}, \ and\ \bibinfo {author} {\bibfnamefont {L.~N.}\ \bibnamefont
  {Savushkin}},\ }\href {\doibase 10.1134/s1063778814020136} {\bibfield
  {journal} {\bibinfo  {journal} {Phys. At. Nucl.}\ }\textbf {\bibinfo {volume}
  {77}},\ \bibinfo {pages} {299} (\bibinfo {year} {2014})}\BibitemShut
  {NoStop}%
\bibitem [{\citenamefont {L\'opez-Quelle}\ \emph {et~al.}(2018)\citenamefont
  {L\'opez-Quelle}, \citenamefont {Marcos}, \citenamefont {Niembro},\ and\
  \citenamefont {Savushkin}}]{Lopez-Quelle2018Nucl.Phys.A971.149}%
  \BibitemOpen
  \bibfield  {author} {\bibinfo {author} {\bibfnamefont {M.}~\bibnamefont
  {L\'opez-Quelle}}, \bibinfo {author} {\bibfnamefont {S.}~\bibnamefont
  {Marcos}}, \bibinfo {author} {\bibfnamefont {R.}~\bibnamefont {Niembro}}, \
  and\ \bibinfo {author} {\bibfnamefont {L.~N.}\ \bibnamefont {Savushkin}},\
  }\href {\doibase https://doi.org/10.1016/j.nuclphysa.2018.01.012} {\bibfield
  {journal} {\bibinfo  {journal} {Nucl. Phys. A}\ }\textbf {\bibinfo {volume}
  {971}},\ \bibinfo {pages} {149} (\bibinfo {year} {2018})}\BibitemShut
  {NoStop}%
\bibitem [{\citenamefont {Tarpanov}\ \emph {et~al.}(2008)\citenamefont
  {Tarpanov}, \citenamefont {Liang}, \citenamefont {Van~Giai},\ and\
  \citenamefont {Stoyanov}}]{Tarpanov2008Phys.Rev.C77.054316}%
  \BibitemOpen
  \bibfield  {author} {\bibinfo {author} {\bibfnamefont {D.}~\bibnamefont
  {Tarpanov}}, \bibinfo {author} {\bibfnamefont {H.~Z.}\ \bibnamefont {Liang}},
  \bibinfo {author} {\bibfnamefont {N.}~\bibnamefont {Van~Giai}}, \ and\
  \bibinfo {author} {\bibfnamefont {C.}~\bibnamefont {Stoyanov}},\ }\href
  {\doibase 10.1103/PhysRevC.77.054316} {\bibfield  {journal} {\bibinfo
  {journal} {Phys. Rev. C}\ }\textbf {\bibinfo {volume} {77}},\ \bibinfo
  {pages} {054316} (\bibinfo {year} {2008})}\BibitemShut {NoStop}%
\bibitem [{\citenamefont {Moreno-Torres}\ \emph {et~al.}(2010)\citenamefont
  {Moreno-Torres}, \citenamefont {Grasso}, \citenamefont {Liang}, \citenamefont
  {De~Donno}, \citenamefont {Anguiano},\ and\ \citenamefont
  {Van~Giai}}]{Moreno-Torres2010Phys.Rev.C81.064327}%
  \BibitemOpen
  \bibfield  {author} {\bibinfo {author} {\bibfnamefont {M.}~\bibnamefont
  {Moreno-Torres}}, \bibinfo {author} {\bibfnamefont {M.}~\bibnamefont
  {Grasso}}, \bibinfo {author} {\bibfnamefont {H.~Z.}\ \bibnamefont {Liang}},
  \bibinfo {author} {\bibfnamefont {V.}~\bibnamefont {De~Donno}}, \bibinfo
  {author} {\bibfnamefont {M.}~\bibnamefont {Anguiano}}, \ and\ \bibinfo
  {author} {\bibfnamefont {N.}~\bibnamefont {Van~Giai}},\ }\href {\doibase
  10.1103/PhysRevC.81.064327} {\bibfield  {journal} {\bibinfo  {journal} {Phys.
  Rev. C}\ }\textbf {\bibinfo {volume} {81}},\ \bibinfo {pages} {064327}
  (\bibinfo {year} {2010})}\BibitemShut {NoStop}%
\bibitem [{\citenamefont {Lesinski}\ \emph {et~al.}(2007)\citenamefont
  {Lesinski}, \citenamefont {Bender}, \citenamefont {Bennaceur}, \citenamefont
  {Duguet},\ and\ \citenamefont {Meyer}}]{Lesinski2007Phys.Rev.C76.014312}%
  \BibitemOpen
  \bibfield  {author} {\bibinfo {author} {\bibfnamefont {T.}~\bibnamefont
  {Lesinski}}, \bibinfo {author} {\bibfnamefont {M.}~\bibnamefont {Bender}},
  \bibinfo {author} {\bibfnamefont {K.}~\bibnamefont {Bennaceur}}, \bibinfo
  {author} {\bibfnamefont {T.}~\bibnamefont {Duguet}}, \ and\ \bibinfo {author}
  {\bibfnamefont {J.}~\bibnamefont {Meyer}},\ }\href {\doibase
  10.1103/PhysRevC.76.014312} {\bibfield  {journal} {\bibinfo  {journal} {Phys.
  Rev. C}\ }\textbf {\bibinfo {volume} {76}},\ \bibinfo {pages} {014312}
  (\bibinfo {year} {2007})}\BibitemShut {NoStop}%
\bibitem [{\citenamefont {Bouyssy}\ \emph {et~al.}(1987)\citenamefont
  {Bouyssy}, \citenamefont {Mathiot}, \citenamefont {Van~Giai},\ and\
  \citenamefont {Marcos}}]{Bouyssy1987Phys.Rev.C36.380}%
  \BibitemOpen
  \bibfield  {author} {\bibinfo {author} {\bibfnamefont {A.}~\bibnamefont
  {Bouyssy}}, \bibinfo {author} {\bibfnamefont {J.~F.}\ \bibnamefont
  {Mathiot}}, \bibinfo {author} {\bibfnamefont {N.}~\bibnamefont {Van~Giai}}, \
  and\ \bibinfo {author} {\bibfnamefont {S.}~\bibnamefont {Marcos}},\ }\href
  {\doibase 10.1103/PhysRevC.36.380} {\bibfield  {journal} {\bibinfo  {journal}
  {Phys. Rev. C}\ }\textbf {\bibinfo {volume} {36}},\ \bibinfo {pages} {380}
  (\bibinfo {year} {1987})}\BibitemShut {NoStop}%
\bibitem [{\citenamefont {Long}\ \emph
  {et~al.}(2006{\natexlab{a}})\citenamefont {Long}, \citenamefont {Van~Giai},\
  and\ \citenamefont {Meng}}]{Long2006Phys.Lett.B640.150}%
  \BibitemOpen
  \bibfield  {author} {\bibinfo {author} {\bibfnamefont {W.~H.}\ \bibnamefont
  {Long}}, \bibinfo {author} {\bibfnamefont {N.}~\bibnamefont {Van~Giai}}, \
  and\ \bibinfo {author} {\bibfnamefont {J.}~\bibnamefont {Meng}},\ }\href
  {\doibase https://doi.org/10.1016/j.physletb.2006.07.064} {\bibfield
  {journal} {\bibinfo  {journal} {Phys. Lett. B}\ }\textbf {\bibinfo {volume}
  {640}},\ \bibinfo {pages} {150} (\bibinfo {year}
  {2006}{\natexlab{a}})}\BibitemShut {NoStop}%
\bibitem [{\citenamefont {Long}(2005)}]{Long2005PhD.Thesis}%
  \BibitemOpen
  \bibfield  {author} {\bibinfo {author} {\bibfnamefont {W.~H.}\ \bibnamefont
  {Long}},\ }\emph {\bibinfo {title} {Relativistic Hartree-Fock approach with
  density-dependent meson-nucleon couplings}},\ \href@noop {} {Ph.D. thesis},\
  \bibinfo  {school} {Peking University, China and Universit{\'e} Paris
  Sud-Paris XI, France,} (\bibinfo {year} {2005})\BibitemShut {NoStop}%
\bibitem [{\citenamefont {Sun}\ \emph {et~al.}(2008)\citenamefont {Sun},
  \citenamefont {Long}, \citenamefont {Meng},\ and\ \citenamefont
  {Lombardo}}]{Sun2008Phys.Rev.C78.065805}%
  \BibitemOpen
  \bibfield  {author} {\bibinfo {author} {\bibfnamefont {B.~Y.}\ \bibnamefont
  {Sun}}, \bibinfo {author} {\bibfnamefont {W.~H.}\ \bibnamefont {Long}},
  \bibinfo {author} {\bibfnamefont {J.}~\bibnamefont {Meng}}, \ and\ \bibinfo
  {author} {\bibfnamefont {U.}~\bibnamefont {Lombardo}},\ }\href {\doibase
  10.1103/PhysRevC.78.065805} {\bibfield  {journal} {\bibinfo  {journal} {Phys.
  Rev. C}\ }\textbf {\bibinfo {volume} {78}},\ \bibinfo {pages} {065805}
  (\bibinfo {year} {2008})}\BibitemShut {NoStop}%
\bibitem [{\citenamefont {Zhao}\ \emph {et~al.}(2015)\citenamefont {Zhao},
  \citenamefont {Sun},\ and\ \citenamefont
  {Long}}]{Zhao2015J.Phys.G:Nucl.Part.Phys.42.095101}%
  \BibitemOpen
  \bibfield  {author} {\bibinfo {author} {\bibfnamefont {Q.}~\bibnamefont
  {Zhao}}, \bibinfo {author} {\bibfnamefont {B.~Y.}\ \bibnamefont {Sun}}, \
  and\ \bibinfo {author} {\bibfnamefont {W.~H.}\ \bibnamefont {Long}},\ }\href
  {http://stacks.iop.org/0954-3899/42/i=9/a=095101} {\bibfield  {journal}
  {\bibinfo  {journal} {J. Phys. G: Nucl. Part. Phys.}\ }\textbf {\bibinfo
  {volume} {42}},\ \bibinfo {pages} {095101} (\bibinfo {year}
  {2015})}\BibitemShut {NoStop}%
\bibitem [{\citenamefont {Liu}\ \emph {et~al.}(2018)\citenamefont {Liu},
  \citenamefont {Qian}, \citenamefont {Xing}, \citenamefont {Niu},\ and\
  \citenamefont {Sun}}]{Liu2018Phys.Rev.C97.025801}%
  \BibitemOpen
  \bibfield  {author} {\bibinfo {author} {\bibfnamefont {Z.~W.}\ \bibnamefont
  {Liu}}, \bibinfo {author} {\bibfnamefont {Z.}~\bibnamefont {Qian}}, \bibinfo
  {author} {\bibfnamefont {R.~Y.}\ \bibnamefont {Xing}}, \bibinfo {author}
  {\bibfnamefont {J.~R.}\ \bibnamefont {Niu}}, \ and\ \bibinfo {author}
  {\bibfnamefont {B.~Y.}\ \bibnamefont {Sun}},\ }\href {\doibase
  10.1103/PhysRevC.97.025801} {\bibfield  {journal} {\bibinfo  {journal} {Phys.
  Rev. C}\ }\textbf {\bibinfo {volume} {97}},\ \bibinfo {pages} {025801}
  (\bibinfo {year} {2018})}\BibitemShut {NoStop}%
\bibitem [{\citenamefont {Long}\ \emph
  {et~al.}(2006{\natexlab{b}})\citenamefont {Long}, \citenamefont {Sagawa},
  \citenamefont {Meng},\ and\ \citenamefont
  {Van~Giai}}]{LONG2006Phys.Lett.B639.242}%
  \BibitemOpen
  \bibfield  {author} {\bibinfo {author} {\bibfnamefont {W.~H.}\ \bibnamefont
  {Long}}, \bibinfo {author} {\bibfnamefont {H.}~\bibnamefont {Sagawa}},
  \bibinfo {author} {\bibfnamefont {J.}~\bibnamefont {Meng}}, \ and\ \bibinfo
  {author} {\bibfnamefont {N.}~\bibnamefont {Van~Giai}},\ }\href {\doibase
  https://doi.org/10.1016/j.physletb.2006.05.065} {\bibfield  {journal}
  {\bibinfo  {journal} {Phys. Lett. B}\ }\textbf {\bibinfo {volume} {639}},\
  \bibinfo {pages} {242} (\bibinfo {year} {2006}{\natexlab{b}})}\BibitemShut
  {NoStop}%
\bibitem [{\citenamefont {Liang}\ \emph {et~al.}(2010)\citenamefont {Liang},
  \citenamefont {Long}, \citenamefont {Meng},\ and\ \citenamefont
  {Van~Giai}}]{Liang2010Eur.Phys.J.A44.119}%
  \BibitemOpen
  \bibfield  {author} {\bibinfo {author} {\bibfnamefont {H.~Z.}\ \bibnamefont
  {Liang}}, \bibinfo {author} {\bibfnamefont {W.~H.}\ \bibnamefont {Long}},
  \bibinfo {author} {\bibfnamefont {J.}~\bibnamefont {Meng}}, \ and\ \bibinfo
  {author} {\bibfnamefont {N.}~\bibnamefont {Van~Giai}},\ }\href {\doibase
  10.1140/epja/i2010-10938-6} {\bibfield  {journal} {\bibinfo  {journal} {Eur.
  Phys. J. A}\ }\textbf {\bibinfo {volume} {44}},\ \bibinfo {pages} {119}
  (\bibinfo {year} {2010})}\BibitemShut {NoStop}%
\bibitem [{\citenamefont {Liang}\ \emph {et~al.}(2015)\citenamefont {Liang},
  \citenamefont {Meng},\ and\ \citenamefont {Zhou}}]{Liang2015Phys.Rep.570.1}%
  \BibitemOpen
  \bibfield  {author} {\bibinfo {author} {\bibfnamefont {H.~Z.}\ \bibnamefont
  {Liang}}, \bibinfo {author} {\bibfnamefont {J.}~\bibnamefont {Meng}}, \ and\
  \bibinfo {author} {\bibfnamefont {S.~G.}\ \bibnamefont {Zhou}},\ }\href
  {\doibase https://doi.org/10.1016/j.physrep.2014.12.005} {\bibfield
  {journal} {\bibinfo  {journal} {Phys. Rep.}\ }\textbf {\bibinfo {volume}
  {570}},\ \bibinfo {pages} {1} (\bibinfo {year} {2015})}\BibitemShut {NoStop}%
\bibitem [{\citenamefont {Long}\ \emph {et~al.}(2010)\citenamefont {Long},
  \citenamefont {Ring}, \citenamefont {Meng}, \citenamefont {Van~Giai},\ and\
  \citenamefont {Bertulani}}]{Long2010Phys.Rev.C81.031302}%
  \BibitemOpen
  \bibfield  {author} {\bibinfo {author} {\bibfnamefont {W.~H.}\ \bibnamefont
  {Long}}, \bibinfo {author} {\bibfnamefont {P.}~\bibnamefont {Ring}}, \bibinfo
  {author} {\bibfnamefont {J.}~\bibnamefont {Meng}}, \bibinfo {author}
  {\bibfnamefont {N.}~\bibnamefont {Van~Giai}}, \ and\ \bibinfo {author}
  {\bibfnamefont {C.~A.}\ \bibnamefont {Bertulani}},\ }\href {\doibase
  10.1103/PhysRevC.81.031302} {\bibfield  {journal} {\bibinfo  {journal} {Phys.
  Rev. C}\ }\textbf {\bibinfo {volume} {81}},\ \bibinfo {pages} {031302}
  (\bibinfo {year} {2010})}\BibitemShut {NoStop}%
\bibitem [{\citenamefont {Li}\ \emph {et~al.}(2016{\natexlab{a}})\citenamefont
  {Li}, \citenamefont {Long}, \citenamefont {Song},\ and\ \citenamefont
  {Zhao}}]{Li2016Phys.Rev.C93.054312}%
  \BibitemOpen
  \bibfield  {author} {\bibinfo {author} {\bibfnamefont {J.~J.}\ \bibnamefont
  {Li}}, \bibinfo {author} {\bibfnamefont {W.~H.}\ \bibnamefont {Long}},
  \bibinfo {author} {\bibfnamefont {J.~L.}\ \bibnamefont {Song}}, \ and\
  \bibinfo {author} {\bibfnamefont {Q.}~\bibnamefont {Zhao}},\ }\href {\doibase
  10.1103/PhysRevC.93.054312} {\bibfield  {journal} {\bibinfo  {journal} {Phys.
  Rev. C}\ }\textbf {\bibinfo {volume} {93}},\ \bibinfo {pages} {054312}
  (\bibinfo {year} {2016}{\natexlab{a}})}\BibitemShut {NoStop}%
\bibitem [{\citenamefont {Ebran}\ \emph {et~al.}(2011)\citenamefont {Ebran},
  \citenamefont {Khan}, \citenamefont {Pe\~na Arteaga},\ and\ \citenamefont
  {Vretenar}}]{Ebran2011Phys.Rev.C83.064323}%
  \BibitemOpen
  \bibfield  {author} {\bibinfo {author} {\bibfnamefont {J.~P.}\ \bibnamefont
  {Ebran}}, \bibinfo {author} {\bibfnamefont {E.}~\bibnamefont {Khan}},
  \bibinfo {author} {\bibfnamefont {D.}~\bibnamefont {Pe\~na Arteaga}}, \ and\
  \bibinfo {author} {\bibfnamefont {D.}~\bibnamefont {Vretenar}},\ }\href
  {\doibase 10.1103/PhysRevC.83.064323} {\bibfield  {journal} {\bibinfo
  {journal} {Phys. Rev. C}\ }\textbf {\bibinfo {volume} {83}},\ \bibinfo
  {pages} {064323} (\bibinfo {year} {2011})}\BibitemShut {NoStop}%
\bibitem [{\citenamefont {Li}\ \emph {et~al.}(2014)\citenamefont {Li},
  \citenamefont {Long}, \citenamefont {Margueron},\ and\ \citenamefont
  {Van~Giai}}]{LI2014Phys.Lett.B732.169}%
  \BibitemOpen
  \bibfield  {author} {\bibinfo {author} {\bibfnamefont {J.~J.}\ \bibnamefont
  {Li}}, \bibinfo {author} {\bibfnamefont {W.~H.}\ \bibnamefont {Long}},
  \bibinfo {author} {\bibfnamefont {J.}~\bibnamefont {Margueron}}, \ and\
  \bibinfo {author} {\bibfnamefont {N.}~\bibnamefont {Van~Giai}},\ }\href
  {\doibase https://doi.org/10.1016/j.physletb.2014.03.031} {\bibfield
  {journal} {\bibinfo  {journal} {Phys. Lett. B}\ }\textbf {\bibinfo {volume}
  {732}},\ \bibinfo {pages} {169} (\bibinfo {year} {2014})}\BibitemShut
  {NoStop}%
\bibitem [{\citenamefont {Li}\ \emph {et~al.}(2016{\natexlab{b}})\citenamefont
  {Li}, \citenamefont {Margueron}, \citenamefont {Long},\ and\ \citenamefont
  {Van~Giai}}]{Li2016Phys.Lett.B753.97}%
  \BibitemOpen
  \bibfield  {author} {\bibinfo {author} {\bibfnamefont {J.~J.}\ \bibnamefont
  {Li}}, \bibinfo {author} {\bibfnamefont {J.}~\bibnamefont {Margueron}},
  \bibinfo {author} {\bibfnamefont {W.~H.}\ \bibnamefont {Long}}, \ and\
  \bibinfo {author} {\bibfnamefont {N.}~\bibnamefont {Van~Giai}},\ }\href
  {\doibase https://doi.org/10.1016/j.physletb.2015.12.004} {\bibfield
  {journal} {\bibinfo  {journal} {Phys. Lett. B}\ }\textbf {\bibinfo {volume}
  {753}},\ \bibinfo {pages} {97} (\bibinfo {year}
  {2016}{\natexlab{b}})}\BibitemShut {NoStop}%
\bibitem [{\citenamefont {Li}\ \emph {et~al.}()\citenamefont {Li},
  \citenamefont {Long}, \citenamefont {Margueron},\ and\ \citenamefont
  {Giai}}]{Li2018arXiv:1807.10000v1}%
  \BibitemOpen
  \bibfield  {author} {\bibinfo {author} {\bibfnamefont {J.~J.}\ \bibnamefont
  {Li}}, \bibinfo {author} {\bibfnamefont {W.~H.}\ \bibnamefont {Long}},
  \bibinfo {author} {\bibfnamefont {J.}~\bibnamefont {Margueron}}, \ and\
  \bibinfo {author} {\bibfnamefont {N.~V.}\ \bibnamefont {Giai}},\ }\href@noop
  {} {\bibinfo  {journal} {arXiv:1807.10000 [nucl-th]}\ }\BibitemShut {NoStop}%
\bibitem [{\citenamefont {Liang}\ \emph {et~al.}(2009)\citenamefont {Liang},
  \citenamefont {Van~Giai},\ and\ \citenamefont
  {Meng}}]{Liang2009Phys.Rev.C79.064316}%
  \BibitemOpen
\bibfield  {journal} {  }\bibfield  {author} {\bibinfo {author} {\bibfnamefont
  {H.~Z.}\ \bibnamefont {Liang}}, \bibinfo {author} {\bibfnamefont
  {N.}~\bibnamefont {Van~Giai}}, \ and\ \bibinfo {author} {\bibfnamefont
  {J.}~\bibnamefont {Meng}},\ }\href {\doibase 10.1103/PhysRevC.79.064316}
  {\bibfield  {journal} {\bibinfo  {journal} {Phys. Rev. C}\ }\textbf {\bibinfo
  {volume} {79}},\ \bibinfo {eid} {064316} (\bibinfo {year}
  {2009})}\BibitemShut {NoStop}%
\bibitem [{\citenamefont {Gu}\ \emph {et~al.}(2013)\citenamefont {Gu},
  \citenamefont {Liang}, \citenamefont {Long}, \citenamefont {Van~Giai},\ and\
  \citenamefont {Meng}}]{Gu2013Phys.Rev.C87.041301R}%
  \BibitemOpen
  \bibfield  {author} {\bibinfo {author} {\bibfnamefont {H.~Q.}\ \bibnamefont
  {Gu}}, \bibinfo {author} {\bibfnamefont {H.~Z.}\ \bibnamefont {Liang}},
  \bibinfo {author} {\bibfnamefont {W.~H.}\ \bibnamefont {Long}}, \bibinfo
  {author} {\bibfnamefont {N.}~\bibnamefont {Van~Giai}}, \ and\ \bibinfo
  {author} {\bibfnamefont {J.}~\bibnamefont {Meng}},\ }\href {\doibase
  10.1103/PhysRevC.87.041301} {\bibfield  {journal} {\bibinfo  {journal} {Phys.
  Rev. C}\ }\textbf {\bibinfo {volume} {87}},\ \bibinfo {pages} {041301(R)}
  (\bibinfo {year} {2013})}\BibitemShut {NoStop}%
\bibitem [{\citenamefont {Liang}\ \emph {et~al.}(2008)\citenamefont {Liang},
  \citenamefont {Van~Giai},\ and\ \citenamefont
  {Meng}}]{Liang2008Phys.Rev.Lett.101.122502}%
  \BibitemOpen
  \bibfield  {author} {\bibinfo {author} {\bibfnamefont {H.~Z.}\ \bibnamefont
  {Liang}}, \bibinfo {author} {\bibfnamefont {N.}~\bibnamefont {Van~Giai}}, \
  and\ \bibinfo {author} {\bibfnamefont {J.}~\bibnamefont {Meng}},\ }\href
  {\doibase 10.1103/PhysRevLett.101.122502} {\bibfield  {journal} {\bibinfo
  {journal} {Phys. Rev. Lett.}\ }\textbf {\bibinfo {volume} {101}},\ \bibinfo
  {pages} {122502} (\bibinfo {year} {2008})}\BibitemShut {NoStop}%
\bibitem [{\citenamefont {Liang}\ \emph
  {et~al.}(2012{\natexlab{a}})\citenamefont {Liang}, \citenamefont {Zhao},\
  and\ \citenamefont {Meng}}]{Liang2012Phys.Rev.C85.064302}%
  \BibitemOpen
  \bibfield  {author} {\bibinfo {author} {\bibfnamefont {H.~Z.}\ \bibnamefont
  {Liang}}, \bibinfo {author} {\bibfnamefont {P.~W.}\ \bibnamefont {Zhao}}, \
  and\ \bibinfo {author} {\bibfnamefont {J.}~\bibnamefont {Meng}},\ }\href
  {\doibase 10.1103/PhysRevC.85.064302} {\bibfield  {journal} {\bibinfo
  {journal} {Phys. Rev. C}\ }\textbf {\bibinfo {volume} {85}},\ \bibinfo
  {pages} {064302} (\bibinfo {year} {2012}{\natexlab{a}})}\BibitemShut
  {NoStop}%
\bibitem [{\citenamefont {Liang}\ \emph
  {et~al.}(2012{\natexlab{b}})\citenamefont {Liang}, \citenamefont {Zhao},
  \citenamefont {Ring}, \citenamefont {Roca-Maza},\ and\ \citenamefont
  {Meng}}]{Liang2012Phys.Rev.C86.021302R}%
  \BibitemOpen
  \bibfield  {author} {\bibinfo {author} {\bibfnamefont {H.~Z.}\ \bibnamefont
  {Liang}}, \bibinfo {author} {\bibfnamefont {P.~W.}\ \bibnamefont {Zhao}},
  \bibinfo {author} {\bibfnamefont {P.}~\bibnamefont {Ring}}, \bibinfo {author}
  {\bibfnamefont {X.}~\bibnamefont {Roca-Maza}}, \ and\ \bibinfo {author}
  {\bibfnamefont {J.}~\bibnamefont {Meng}},\ }\href {\doibase
  10.1103/PhysRevC.86.021302} {\bibfield  {journal} {\bibinfo  {journal} {Phys.
  Rev. C}\ }\textbf {\bibinfo {volume} {86}},\ \bibinfo {pages} {021302(R)}
  (\bibinfo {year} {2012}{\natexlab{b}})}\BibitemShut {NoStop}%
\bibitem [{\citenamefont {Niu}\ \emph {et~al.}(2017)\citenamefont {Niu},
  \citenamefont {Niu}, \citenamefont {Liang}, \citenamefont {Long},\ and\
  \citenamefont {Meng}}]{Niu2017Phys.Rev.C95.044301}%
  \BibitemOpen
  \bibfield  {author} {\bibinfo {author} {\bibfnamefont {Z.~M.}\ \bibnamefont
  {Niu}}, \bibinfo {author} {\bibfnamefont {Y.~F.}\ \bibnamefont {Niu}},
  \bibinfo {author} {\bibfnamefont {H.~Z.}\ \bibnamefont {Liang}}, \bibinfo
  {author} {\bibfnamefont {W.~H.}\ \bibnamefont {Long}}, \ and\ \bibinfo
  {author} {\bibfnamefont {J.}~\bibnamefont {Meng}},\ }\href {\doibase
  10.1103/PhysRevC.95.044301} {\bibfield  {journal} {\bibinfo  {journal} {Phys.
  Rev. C}\ }\textbf {\bibinfo {volume} {95}},\ \bibinfo {pages} {044301}
  (\bibinfo {year} {2017})}\BibitemShut {NoStop}%
\bibitem [{\citenamefont {Niu}\ \emph {et~al.}(2013)\citenamefont {Niu},
  \citenamefont {Niu}, \citenamefont {Liang}, \citenamefont {Long},
  \citenamefont {Nik\v{s}i\'{c}}, \citenamefont {Vretenar},\ and\ \citenamefont
  {Meng}}]{NIU2013Phys.Lett.B723.172}%
  \BibitemOpen
  \bibfield  {author} {\bibinfo {author} {\bibfnamefont {Z.~M.}\ \bibnamefont
  {Niu}}, \bibinfo {author} {\bibfnamefont {Y.~F.}\ \bibnamefont {Niu}},
  \bibinfo {author} {\bibfnamefont {H.~Z.}\ \bibnamefont {Liang}}, \bibinfo
  {author} {\bibfnamefont {W.~H.}\ \bibnamefont {Long}}, \bibinfo {author}
  {\bibfnamefont {T.}~\bibnamefont {Nik\v{s}i\'{c}}}, \bibinfo {author}
  {\bibfnamefont {D.}~\bibnamefont {Vretenar}}, \ and\ \bibinfo {author}
  {\bibfnamefont {J.}~\bibnamefont {Meng}},\ }\href {\doibase
  https://doi.org/10.1016/j.physletb.2013.04.048} {\bibfield  {journal}
  {\bibinfo  {journal} {Phys. Lett. B}\ }\textbf {\bibinfo {volume} {723}},\
  \bibinfo {pages} {172} (\bibinfo {year} {2013})}\BibitemShut {NoStop}%
\bibitem [{\citenamefont {Long}\ \emph {et~al.}(2012)\citenamefont {Long},
  \citenamefont {Sun}, \citenamefont {Hagino},\ and\ \citenamefont
  {Sagawa}}]{Long2012Phys.Rev.C85.025806}%
  \BibitemOpen
  \bibfield  {author} {\bibinfo {author} {\bibfnamefont {W.~H.}\ \bibnamefont
  {Long}}, \bibinfo {author} {\bibfnamefont {B.~Y.}\ \bibnamefont {Sun}},
  \bibinfo {author} {\bibfnamefont {K.}~\bibnamefont {Hagino}}, \ and\ \bibinfo
  {author} {\bibfnamefont {H.}~\bibnamefont {Sagawa}},\ }\href {\doibase
  10.1103/PhysRevC.85.025806} {\bibfield  {journal} {\bibinfo  {journal} {Phys.
  Rev. C}\ }\textbf {\bibinfo {volume} {85}},\ \bibinfo {pages} {025806}
  (\bibinfo {year} {2012})}\BibitemShut {NoStop}%
\bibitem [{\citenamefont {Li}\ \emph {et~al.}(2018{\natexlab{a}})\citenamefont
  {Li}, \citenamefont {Sedrakian},\ and\ \citenamefont
  {Weber}}]{Li2018Phys.Lett.B783.234}%
  \BibitemOpen
  \bibfield  {author} {\bibinfo {author} {\bibfnamefont {J.~J.}\ \bibnamefont
  {Li}}, \bibinfo {author} {\bibfnamefont {A.}~\bibnamefont {Sedrakian}}, \
  and\ \bibinfo {author} {\bibfnamefont {F.}~\bibnamefont {Weber}},\ }\href
  {\doibase https://doi.org/10.1016/j.physletb.2018.06.051} {\bibfield
  {journal} {\bibinfo  {journal} {Phys. Lett. B}\ }\textbf {\bibinfo {volume}
  {783}},\ \bibinfo {pages} {234} (\bibinfo {year}
  {2018}{\natexlab{a}})}\BibitemShut {NoStop}%
\bibitem [{\citenamefont {Li}\ \emph {et~al.}(2018{\natexlab{b}})\citenamefont
  {Li}, \citenamefont {Long},\ and\ \citenamefont
  {Sedrakian}}]{Li2018Eur.Phys.J.A54.133}%
  \BibitemOpen
  \bibfield  {author} {\bibinfo {author} {\bibfnamefont {J.~J.}\ \bibnamefont
  {Li}}, \bibinfo {author} {\bibfnamefont {W.~H.}\ \bibnamefont {Long}}, \ and\
  \bibinfo {author} {\bibfnamefont {A.}~\bibnamefont {Sedrakian}},\ }\href
  {\doibase 10.1140/epja/i2018-12566-6} {\bibfield  {journal} {\bibinfo
  {journal} {Eur. Phys. J. A}\ }\textbf {\bibinfo {volume} {54}},\ \bibinfo
  {pages} {133} (\bibinfo {year} {2018}{\natexlab{b}})}\BibitemShut {NoStop}%
\bibitem [{\citenamefont {Jiang}\ \emph
  {et~al.}(2015{\natexlab{a}})\citenamefont {Jiang}, \citenamefont {Yang},
  \citenamefont {Sun}, \citenamefont {Long},\ and\ \citenamefont
  {Gu}}]{Jiang2015Phys.Rev.C91.034326}%
  \BibitemOpen
  \bibfield  {author} {\bibinfo {author} {\bibfnamefont {L.~J.}\ \bibnamefont
  {Jiang}}, \bibinfo {author} {\bibfnamefont {S.}~\bibnamefont {Yang}},
  \bibinfo {author} {\bibfnamefont {B.~Y.}\ \bibnamefont {Sun}}, \bibinfo
  {author} {\bibfnamefont {W.~H.}\ \bibnamefont {Long}}, \ and\ \bibinfo
  {author} {\bibfnamefont {H.~Q.}\ \bibnamefont {Gu}},\ }\href {\doibase
  10.1103/PhysRevC.91.034326} {\bibfield  {journal} {\bibinfo  {journal} {Phys.
  Rev. C}\ }\textbf {\bibinfo {volume} {91}},\ \bibinfo {pages} {034326}
  (\bibinfo {year} {2015}{\natexlab{a}})}\BibitemShut {NoStop}%
\bibitem [{\citenamefont {Jiang}\ \emph
  {et~al.}(2015{\natexlab{b}})\citenamefont {Jiang}, \citenamefont {Yang},
  \citenamefont {Dong},\ and\ \citenamefont
  {Long}}]{Jiang2015Phys.Rev.C91.025802}%
  \BibitemOpen
  \bibfield  {author} {\bibinfo {author} {\bibfnamefont {L.~J.}\ \bibnamefont
  {Jiang}}, \bibinfo {author} {\bibfnamefont {S.}~\bibnamefont {Yang}},
  \bibinfo {author} {\bibfnamefont {J.~M.}\ \bibnamefont {Dong}}, \ and\
  \bibinfo {author} {\bibfnamefont {W.~H.}\ \bibnamefont {Long}},\ }\href
  {\doibase 10.1103/PhysRevC.91.025802} {\bibfield  {journal} {\bibinfo
  {journal} {Phys. Rev. C}\ }\textbf {\bibinfo {volume} {91}},\ \bibinfo
  {pages} {025802} (\bibinfo {year} {2015}{\natexlab{b}})}\BibitemShut
  {NoStop}%
\bibitem [{\citenamefont {Zong}\ and\ \citenamefont
  {Sun}(2018)}]{Zong2018Chin.Phys.C42.024101}%
  \BibitemOpen
  \bibfield  {author} {\bibinfo {author} {\bibfnamefont {Y.~Y.}\ \bibnamefont
  {Zong}}\ and\ \bibinfo {author} {\bibfnamefont {B.~Y.}\ \bibnamefont {Sun}},\
  }\href {http://stacks.iop.org/1674-1137/42/i=2/a=024101} {\bibfield
  {journal} {\bibinfo  {journal} {Chin. Phys. C}\ }\textbf {\bibinfo {volume}
  {42}},\ \bibinfo {pages} {024101} (\bibinfo {year} {2018})}\BibitemShut
  {NoStop}%
\bibitem [{\citenamefont {Foldy}\ and\ \citenamefont
  {Wouthuysen}(1950)}]{Foldy1950Phys.Rev.78.29}%
  \BibitemOpen
  \bibfield  {author} {\bibinfo {author} {\bibfnamefont {L.~L.}\ \bibnamefont
  {Foldy}}\ and\ \bibinfo {author} {\bibfnamefont {S.~A.}\ \bibnamefont
  {Wouthuysen}},\ }\href {\doibase 10.1103/PhysRev.78.29} {\bibfield  {journal}
  {\bibinfo  {journal} {Phys. Rev.}\ }\textbf {\bibinfo {volume} {78}},\
  \bibinfo {pages} {29} (\bibinfo {year} {1950})}\BibitemShut {NoStop}%
\bibitem [{\citenamefont
  {Machleidt}(1989)}]{Machleidt1989Adv.Nucl.Phys.19.189}%
  \BibitemOpen
  \bibfield  {author} {\bibinfo {author} {\bibfnamefont {R.}~\bibnamefont
  {Machleidt}},\ }\href
  {https://link.springer.com/chapter/10.1007/978-1-4613-9907-0_2#citeas}
  {\bibfield  {journal} {\bibinfo  {journal} {Adv. Nucl. Phys.}\ }\textbf
  {\bibinfo {volume} {19}},\ \bibinfo {pages} {189} (\bibinfo {year}
  {1989})}\BibitemShut {NoStop}%
%%CITATION = ANUPB,19,189;%%
\bibitem [{\citenamefont {Lalazissis}\ \emph {et~al.}(2005)\citenamefont
  {Lalazissis}, \citenamefont {Nik\ifmmode \check{s}\else
  \v{s}\fi{}i\ifmmode~\acute{c}\else \'{c}\fi{}}, \citenamefont {Vretenar},\
  and\ \citenamefont {Ring}}]{Lalazissis2005Phys.Rev.C71.024312}%
  \BibitemOpen
  \bibfield  {author} {\bibinfo {author} {\bibfnamefont {G.~A.}\ \bibnamefont
  {Lalazissis}}, \bibinfo {author} {\bibfnamefont {T.}~\bibnamefont
  {Nik\ifmmode \check{s}\else \v{s}\fi{}i\ifmmode~\acute{c}\else \'{c}\fi{}}},
  \bibinfo {author} {\bibfnamefont {D.}~\bibnamefont {Vretenar}}, \ and\
  \bibinfo {author} {\bibfnamefont {P.}~\bibnamefont {Ring}},\ }\href {\doibase
  10.1103/PhysRevC.71.024312} {\bibfield  {journal} {\bibinfo  {journal} {Phys.
  Rev. C}\ }\textbf {\bibinfo {volume} {71}},\ \bibinfo {pages} {024312}
  (\bibinfo {year} {2005})}\BibitemShut {NoStop}%
\bibitem [{\citenamefont {Wang}\ \emph {et~al.}(2017)\citenamefont {Wang},
  \citenamefont {Audi}, \citenamefont {Kondev}, \citenamefont {Huang},
  \citenamefont {Naimi},\ and\ \citenamefont
  {Xu}}]{Wang2017Chin.Phys.C41.030003}%
  \BibitemOpen
  \bibfield  {author} {\bibinfo {author} {\bibfnamefont {M.}~\bibnamefont
  {Wang}}, \bibinfo {author} {\bibfnamefont {G.}~\bibnamefont {Audi}}, \bibinfo
  {author} {\bibfnamefont {F.~G.}\ \bibnamefont {Kondev}}, \bibinfo {author}
  {\bibfnamefont {W.~J.}\ \bibnamefont {Huang}}, \bibinfo {author}
  {\bibfnamefont {S.}~\bibnamefont {Naimi}}, \ and\ \bibinfo {author}
  {\bibfnamefont {X.}~\bibnamefont {Xu}},\ }\href {\doibase
  10.1088/1674-1137/41/3/030003} {\bibfield  {journal} {\bibinfo  {journal}
  {Chin. Phys. C}\ }\textbf {\bibinfo {volume} {41}},\ \bibinfo {pages}
  {030003} (\bibinfo {year} {2017})}\BibitemShut {NoStop}%
\bibitem [{\citenamefont {Goriely}\ \emph {et~al.}(2002)\citenamefont
  {Goriely}, \citenamefont {Samyn}, \citenamefont {Heenen}, \citenamefont
  {Pearson},\ and\ \citenamefont {Tondeur}}]{Goriely2002Phys.Rev.C66.024326}%
  \BibitemOpen
  \bibfield  {author} {\bibinfo {author} {\bibfnamefont {S.}~\bibnamefont
  {Goriely}}, \bibinfo {author} {\bibfnamefont {M.}~\bibnamefont {Samyn}},
  \bibinfo {author} {\bibfnamefont {P.~H.}\ \bibnamefont {Heenen}}, \bibinfo
  {author} {\bibfnamefont {J.~M.}\ \bibnamefont {Pearson}}, \ and\ \bibinfo
  {author} {\bibfnamefont {F.}~\bibnamefont {Tondeur}},\ }\href {\doibase
  10.1103/PhysRevC.66.024326} {\bibfield  {journal} {\bibinfo  {journal} {Phys.
  Rev. C}\ }\textbf {\bibinfo {volume} {66}},\ \bibinfo {pages} {024326}
  (\bibinfo {year} {2002})}\BibitemShut {NoStop}%
\bibitem [{\citenamefont {Anguiano}\ \emph {et~al.}(2011)\citenamefont
  {Anguiano}, \citenamefont {Co'}, \citenamefont {De~Donno},\ and\
  \citenamefont {Lallena}}]{Anguiano2011Phys.Rev.C83.064306}%
  \BibitemOpen
  \bibfield  {author} {\bibinfo {author} {\bibfnamefont {M.}~\bibnamefont
  {Anguiano}}, \bibinfo {author} {\bibfnamefont {G.}~\bibnamefont {Co'}},
  \bibinfo {author} {\bibfnamefont {V.}~\bibnamefont {De~Donno}}, \ and\
  \bibinfo {author} {\bibfnamefont {A.~M.}\ \bibnamefont {Lallena}},\ }\href
  {\doibase 10.1103/PhysRevC.83.064306} {\bibfield  {journal} {\bibinfo
  {journal} {Phys. Rev. C}\ }\textbf {\bibinfo {volume} {83}},\ \bibinfo
  {pages} {064306} (\bibinfo {year} {2011})}\BibitemShut {NoStop}%
\bibitem [{\citenamefont {Cao}\ \emph {et~al.}(2009)\citenamefont {Cao},
  \citenamefont {Col\`o}, \citenamefont {Sagawa}, \citenamefont {Bortignon},\
  and\ \citenamefont {Sciacchitano}}]{Cao2009Phys.Rev.C80.064304}%
  \BibitemOpen
  \bibfield  {author} {\bibinfo {author} {\bibfnamefont {L.~G.}\ \bibnamefont
  {Cao}}, \bibinfo {author} {\bibfnamefont {G.}~\bibnamefont {Col\`o}},
  \bibinfo {author} {\bibfnamefont {H.}~\bibnamefont {Sagawa}}, \bibinfo
  {author} {\bibfnamefont {P.~F.}\ \bibnamefont {Bortignon}}, \ and\ \bibinfo
  {author} {\bibfnamefont {L.}~\bibnamefont {Sciacchitano}},\ }\href {\doibase
  10.1103/PhysRevC.80.064304} {\bibfield  {journal} {\bibinfo  {journal} {Phys.
  Rev. C}\ }\textbf {\bibinfo {volume} {80}},\ \bibinfo {pages} {064304}
  (\bibinfo {year} {2009})}\BibitemShut {NoStop}%
\bibitem [{\citenamefont {Bai}\ \emph {et~al.}(2009)\citenamefont {Bai},
  \citenamefont {Sagawa}, \citenamefont {Zhang}, \citenamefont {Zhang},
  \citenamefont {Col\`o},\ and\ \citenamefont {Xu}}]{Bai2009Phys.Lett.B675.28}%
  \BibitemOpen
  \bibfield  {author} {\bibinfo {author} {\bibfnamefont {C.~L.}\ \bibnamefont
  {Bai}}, \bibinfo {author} {\bibfnamefont {H.}~\bibnamefont {Sagawa}},
  \bibinfo {author} {\bibfnamefont {H.~Q.}\ \bibnamefont {Zhang}}, \bibinfo
  {author} {\bibfnamefont {X.~Z.}\ \bibnamefont {Zhang}}, \bibinfo {author}
  {\bibfnamefont {G.}~\bibnamefont {Col\`o}}, \ and\ \bibinfo {author}
  {\bibfnamefont {F.~R.}\ \bibnamefont {Xu}},\ }\href {\doibase
  https://doi.org/10.1016/j.physletb.2009.03.077} {\bibfield  {journal}
  {\bibinfo  {journal} {Phys. Lett. B}\ }\textbf {\bibinfo {volume} {675}},\
  \bibinfo {pages} {28} (\bibinfo {year} {2009})}\BibitemShut {NoStop}%
\bibitem [{\citenamefont {Bai}\ \emph {et~al.}(2010)\citenamefont {Bai},
  \citenamefont {Zhang}, \citenamefont {Sagawa}, \citenamefont {Zhang},
  \citenamefont {Col\`o},\ and\ \citenamefont
  {Xu}}]{Bai2010Phys.Rev.Lett.105.072501}%
  \BibitemOpen
  \bibfield  {author} {\bibinfo {author} {\bibfnamefont {C.~L.}\ \bibnamefont
  {Bai}}, \bibinfo {author} {\bibfnamefont {H.~Q.}\ \bibnamefont {Zhang}},
  \bibinfo {author} {\bibfnamefont {H.}~\bibnamefont {Sagawa}}, \bibinfo
  {author} {\bibfnamefont {X.~Z.}\ \bibnamefont {Zhang}}, \bibinfo {author}
  {\bibfnamefont {G.}~\bibnamefont {Col\`o}}, \ and\ \bibinfo {author}
  {\bibfnamefont {F.~R.}\ \bibnamefont {Xu}},\ }\href {\doibase
  10.1103/PhysRevLett.105.072501} {\bibfield  {journal} {\bibinfo  {journal}
  {Phys. Rev. Lett.}\ }\textbf {\bibinfo {volume} {105}},\ \bibinfo {pages}
  {072501} (\bibinfo {year} {2010})}\BibitemShut {NoStop}%
\bibitem [{\citenamefont {Minato}\ and\ \citenamefont
  {Bai}(2013)}]{Minato2013Phys.Rev.Lett.110.122501}%
  \BibitemOpen
  \bibfield  {author} {\bibinfo {author} {\bibfnamefont {F.}~\bibnamefont
  {Minato}}\ and\ \bibinfo {author} {\bibfnamefont {C.~L.}\ \bibnamefont
  {Bai}},\ }\href {\doibase 10.1103/PhysRevLett.110.122501} {\bibfield
  {journal} {\bibinfo  {journal} {Phys. Rev. Lett.}\ }\textbf {\bibinfo
  {volume} {110}},\ \bibinfo {pages} {122501} (\bibinfo {year}
  {2013})}\BibitemShut {NoStop}%
\bibitem [{\citenamefont {Shen}\ \emph {et~al.}(2016)\citenamefont {Shen},
  \citenamefont {Hu}, \citenamefont {Liang}, \citenamefont {Meng},
  \citenamefont {Ring},\ and\ \citenamefont
  {Zhang}}]{Shen2016Chin.Phys.Lett.33.102103}%
  \BibitemOpen
  \bibfield  {author} {\bibinfo {author} {\bibfnamefont {S.~H.}\ \bibnamefont
  {Shen}}, \bibinfo {author} {\bibfnamefont {J.~N.}\ \bibnamefont {Hu}},
  \bibinfo {author} {\bibfnamefont {H.~Z.}\ \bibnamefont {Liang}}, \bibinfo
  {author} {\bibfnamefont {J.}~\bibnamefont {Meng}}, \bibinfo {author}
  {\bibfnamefont {P.}~\bibnamefont {Ring}}, \ and\ \bibinfo {author}
  {\bibfnamefont {S.~Q.}\ \bibnamefont {Zhang}},\ }\href
  {http://stacks.iop.org/0256-307X/33/i=10/a=102103} {\bibfield  {journal}
  {\bibinfo  {journal} {Chin. Phys. Lett.}\ }\textbf {\bibinfo {volume} {33}},\
  \bibinfo {pages} {102103} (\bibinfo {year} {2016})}\BibitemShut {NoStop}%
\bibitem [{\citenamefont {Shen}\ \emph {et~al.}(2017)\citenamefont {Shen},
  \citenamefont {Liang}, \citenamefont {Meng}, \citenamefont {Ring},\ and\
  \citenamefont {Zhang}}]{Shen2017Phys.Rev.C96.014316}%
  \BibitemOpen
  \bibfield  {author} {\bibinfo {author} {\bibfnamefont {S.~H.}\ \bibnamefont
  {Shen}}, \bibinfo {author} {\bibfnamefont {H.~Z.}\ \bibnamefont {Liang}},
  \bibinfo {author} {\bibfnamefont {J.}~\bibnamefont {Meng}}, \bibinfo {author}
  {\bibfnamefont {P.}~\bibnamefont {Ring}}, \ and\ \bibinfo {author}
  {\bibfnamefont {S.~Q.}\ \bibnamefont {Zhang}},\ }\href {\doibase
  10.1103/PhysRevC.96.014316} {\bibfield  {journal} {\bibinfo  {journal} {Phys.
  Rev. C}\ }\textbf {\bibinfo {volume} {96}},\ \bibinfo {pages} {014316}
  (\bibinfo {year} {2017})}\BibitemShut {NoStop}%
\bibitem [{\citenamefont {Shen}\ \emph
  {et~al.}(2018{\natexlab{a}})\citenamefont {Shen}, \citenamefont {Liang},
  \citenamefont {Meng}, \citenamefont {Ring},\ and\ \citenamefont
  {Zhang}}]{Shen2018Phys.Lett.B781.227}%
  \BibitemOpen
  \bibfield  {author} {\bibinfo {author} {\bibfnamefont {S.~H.}\ \bibnamefont
  {Shen}}, \bibinfo {author} {\bibfnamefont {H.~Z.}\ \bibnamefont {Liang}},
  \bibinfo {author} {\bibfnamefont {J.}~\bibnamefont {Meng}}, \bibinfo {author}
  {\bibfnamefont {P.}~\bibnamefont {Ring}}, \ and\ \bibinfo {author}
  {\bibfnamefont {S.~Q.}\ \bibnamefont {Zhang}},\ }\href {\doibase
  https://doi.org/10.1016/j.physletb.2018.03.080} {\bibfield  {journal}
  {\bibinfo  {journal} {Phys. Lett. B}\ }\textbf {\bibinfo {volume} {781}},\
  \bibinfo {pages} {227} (\bibinfo {year} {2018}{\natexlab{a}})}\BibitemShut
  {NoStop}%
\bibitem [{\citenamefont {Shen}\ \emph
  {et~al.}(2018{\natexlab{b}})\citenamefont {Shen}, \citenamefont {Liang},
  \citenamefont {Meng}, \citenamefont {Ring},\ and\ \citenamefont
  {Zhang}}]{SHEN2018Phys.Lett.B778.344}%
  \BibitemOpen
  \bibfield  {author} {\bibinfo {author} {\bibfnamefont {S.~H.}\ \bibnamefont
  {Shen}}, \bibinfo {author} {\bibfnamefont {H.~Z.}\ \bibnamefont {Liang}},
  \bibinfo {author} {\bibfnamefont {J.}~\bibnamefont {Meng}}, \bibinfo {author}
  {\bibfnamefont {P.}~\bibnamefont {Ring}}, \ and\ \bibinfo {author}
  {\bibfnamefont {S.~Q.}\ \bibnamefont {Zhang}},\ }\href {\doibase
  10.1016/j.physletb.2018.01.058} {\bibfield  {journal} {\bibinfo  {journal}
  {Phys. Lett. B}\ }\textbf {\bibinfo {volume} {778}},\ \bibinfo {pages} {344}
  (\bibinfo {year} {2018}{\natexlab{b}})}\BibitemShut {NoStop}%
\bibitem [{\citenamefont {Shen}\ \emph
  {et~al.}(2018{\natexlab{c}})\citenamefont {Shen}, \citenamefont {Liang},
  \citenamefont {Meng}, \citenamefont {Ring},\ and\ \citenamefont
  {Zhang}}]{Shen2018Phys.Rev.C97.054312}%
  \BibitemOpen
  \bibfield  {author} {\bibinfo {author} {\bibfnamefont {S.~H.}\ \bibnamefont
  {Shen}}, \bibinfo {author} {\bibfnamefont {H.~Z.}\ \bibnamefont {Liang}},
  \bibinfo {author} {\bibfnamefont {J.}~\bibnamefont {Meng}}, \bibinfo {author}
  {\bibfnamefont {P.}~\bibnamefont {Ring}}, \ and\ \bibinfo {author}
  {\bibfnamefont {S.~Q.}\ \bibnamefont {Zhang}},\ }\href {\doibase
  10.1103/PhysRevC.97.054312} {\bibfield  {journal} {\bibinfo  {journal} {Phys.
  Rev. C}\ }\textbf {\bibinfo {volume} {97}},\ \bibinfo {pages} {054312}
  (\bibinfo {year} {2018}{\natexlab{c}})}\BibitemShut {NoStop}%
\bibitem [{\citenamefont {Hu}\ \emph {et~al.}(2010{\natexlab{a}})\citenamefont
  {Hu}, \citenamefont {Toki}, \citenamefont {Wen},\ and\ \citenamefont
  {Shen}}]{Hu2010Phys.Lett.B687.271}%
  \BibitemOpen
  \bibfield  {author} {\bibinfo {author} {\bibfnamefont {J.}~\bibnamefont
  {Hu}}, \bibinfo {author} {\bibfnamefont {H.}~\bibnamefont {Toki}}, \bibinfo
  {author} {\bibfnamefont {W.}~\bibnamefont {Wen}}, \ and\ \bibinfo {author}
  {\bibfnamefont {H.}~\bibnamefont {Shen}},\ }\href {\doibase
  https://doi.org/10.1016/j.physletb.2010.03.027} {\bibfield  {journal}
  {\bibinfo  {journal} {Phys. Lett. B}\ }\textbf {\bibinfo {volume} {687}},\
  \bibinfo {pages} {271} (\bibinfo {year} {2010}{\natexlab{a}})}\BibitemShut
  {NoStop}%
\bibitem [{\citenamefont {Hu}\ \emph {et~al.}(2010{\natexlab{b}})\citenamefont
  {Hu}, \citenamefont {Toki}, \citenamefont {Wen},\ and\ \citenamefont
  {Shen}}]{Hu2010Eur.Phys.J.A43.323}%
  \BibitemOpen
  \bibfield  {author} {\bibinfo {author} {\bibfnamefont {J.}~\bibnamefont
  {Hu}}, \bibinfo {author} {\bibfnamefont {H.}~\bibnamefont {Toki}}, \bibinfo
  {author} {\bibfnamefont {W.}~\bibnamefont {Wen}}, \ and\ \bibinfo {author}
  {\bibfnamefont {H.}~\bibnamefont {Shen}},\ }\href {\doibase
  10.1140/epja/i2010-10917-y} {\bibfield  {journal} {\bibinfo  {journal} {Eur.
  Phys. J. A}\ }\textbf {\bibinfo {volume} {43}},\ \bibinfo {pages} {323}
  (\bibinfo {year} {2010}{\natexlab{b}})}\BibitemShut {NoStop}%
\end{thebibliography}


\end{document}